# The harm of class imbalance corrections for risk prediction models: illustration and simulation using logistic regression

van den Goorbergh R[1], van Smeden M[1], Timmerman D,[2,3] Van Calster B[2,4,5]

1 Julius Center for Health Sciences and Primary Care, UMC Utrecht, Utrecht University, Utrecht, Netherlands

2 Department of Development and Regeneration, KU Leuven, Leuven, Belgium

3 Department of Obstetrics and Gynecology, University Hospitals Leuven, Leuven, Belgium

4 Department of Biomedical Data Sciences, Leiden University Medical Center, Leiden, Netherlands

5 EPI-center, KU Leuven, Leuven, Belgium

Word count main text: 3148

**Abstract**

Methods to correct class imbalance, i.e. imbalance between the frequency of outcome events and non-events, are receiving increasing interest for developing prediction models. We examined the effect of imbalance correction on the performance of standard and penalized (ridge) logistic regression models in terms of discrimination, calibration, and classification. We examined random undersampling, random oversampling and SMOTE using Monte Carlo simulations and a case study on ovarian cancer diagnosis. The results indicated that all imbalance correction methods led to poor calibration (strong overestimation of the probability to belong to the minority class), but not to better discrimination in terms of the area under the receiver operating characteristic curve. Imbalance correction improved classification in terms of sensitivity and specificity, but similar results were obtained by shifting the probability threshold instead. Our study shows that outcome imbalance is not a problem in itself, and that imbalance correction may even worsen model performance.

**Introduction**

When developing clinical prediction models for a binary outcome, the percentage of individuals with the event of interest (i.e. the event fraction) is often much lower than 50%. When the frequency of individuals with and without the event is unequal, the term 'class imbalance' is often used.[1] Class imbalance is has been identified as a problem for the development of prediction models, in particularly when the interest is in the classification of individuals into a high risk versus low risk group ('classifier').[1-3] Commonly suggested solutions to address class imbalance include some form of resampling to create an artificially balanced dataset for model training. Common approaches are random undersampling (RUS), random oversampling (ROS), and SMOTE (Synthetic Minority Oversampling Technique).[2-5]

The classification accuracy of a model that classifies individuals into a high risk vs low risk group is defined as the percentage of individuals that are either true positive (individuals that have the event and are either correctly classified as high risk) or true negative (individuals that do not have the event and are correctly classified as low risk). To illustrate the possible impact of class imbalance, consider a simple model that classifies everyone as low risk. Such a classifier yields a classification accuracy of 50% if the event fraction is 50% (balanced), but a classification accuracy of 99% if the event fraction is 1% (highly imbalanced). That imbalanced datasets can easily lead to high classification accuracy is often labeled as problematic. For instance, He and Garcia write "we find that classifiers tend to provide a severely imbalanced degree of accuracy, with the majority class having close to 100 percent accuracy and the minority class having accuracies of 0-10 percent, for instance".[2] Fernandez and colleagues write "the truth is that classifiers ... tend to have great accuracy for the majority class while obtaining poor results (closer to 0%) for the minority class".[3]

We argue that the class imbalance is not a pervasive problem for prediction model development. First, the problem is specific to the classification accuracy measure. The limitations of focusing on classification accuracy as a measure of predictive performance is well known.[6,7] Second, if we consider models that produce estimated probabilities of the event of interest, an adjustment of the classification threshold probability can be used to ensure adequate classification performance (i.e. probability threshold to classify individuals as high risk does not have to be 0.5).[8] A probability threshold to select individuals for a given treatment implies certain misclassification costs and should be determined using clinical considerations.[8] If we use a probability threshold of 0.1 to classify individuals as high risk and suggest a specific treatment, this means that we accept to treat up to 10 individuals in order to treat 1 individual with the event: we accept up to 9 false positives, or unnecessary treatments, per true positive.[9-11] As Birch and colleagues write, models should be able to accommodate differing attitudes regarding misclassification costs.[12] The problem then shifts from class imbalance to probability calibration: the model's probability estimates should be reliable in order to make optimal decisions. This raises the question how class imbalance methods affect calibration.

In this study, we investigate the performance of standard and penalized logistic regression models developed in datasets with class imbalance. We hypothesize (1) that imbalance correction methods distort model calibration by leading to probability estimates that are too high, and (2) that shifting the probability threshold has similar impact on sensitivity and specificity as the use of imbalance correction methods.

**Methods**

*Imbalance correction methods and logistic regression models*

We examine RUS, ROS, and SMOTE, three common approaches to correct for class imbalance.[2-5] As lower sample size is well known to increase the risk of overfitting, we anticipated that RUS would require a larger sample size to perform well.[13-15] Prediction models were developed using standard maximum likelihood logistic regression (SLR) and using penalized logistic regression with the ridge (or L2) penalty (Ridge).[16] The lambda hyperparameter was tuned using a grid search based on 10-fold cross-validation.[17] See Supplementary Note for details.

*Case study: estimating the probability of ovarian cancer*

For illustration, we developed prediction models to estimate the risk of ovarian malignancy in premenopausal women presenting with at least one adnexal (ovarian, para-ovarian, or tubal) tumor. Prediction models for ovarian cancer diagnosis could be used to decide whether to operate and by whom (e.g. whether referral to an experienced gynecological oncologist is warranted or not). We use data women who were recruited consecutively across three waves (1999-2005, 2005-2007, and 2009-2012) of the International Ovarian Tumor Analysis (IOTA) study.[18,19] We have ethics approval for secondary use of these data for methodological/statistical research (Research Ethics Committee University Hospitals KU Leuven, S64709). The study only included patients who were operated on, such that the reference standard could be based on histology. Borderline malignant tumors were considered malignant. Overall, 5914 patients were recruited across the three waves, of which 3369 premenopausal patients between 18 and 59 years. The prevalence of malignancy was 20% (658/3369), reflecting moderate imbalance.

We used the following predictors: age of the patient (years), maximum diameter of the lesion (mm), and number of papillary structures (ordinal variable with values 0 to 4; 4 referring to four or more papillary structures). To investigate performance of all models in combination with the different imbalance solutions, the data was first split up into a training set and a test set using a 4:1 ratio. This yielded a training dataset of size 2695 (518 events), and a test dataset of size 674 (140 events). The training set was either unadjusted or pre-processed using RUS, ROS or SMOTE, resulting in four different datasets on which models were fitted: $D_{unadjusted}$; $D_{RUS}$; $D_{ROS}$ and $D_{SMOTE}$. Subsequently, prediction models were developed using SLR and Ridge, resulting in 4 (datasets) x 2 (algorithms) = 8 different models. The continuous predictor variables were modeled using restricted cubic splines with 3 knots to address potential nonlinearity. The resulting models were applied to the test set to obtain the following model performance in terms of the area under the ROC curve, accuracy, sensitivity, specificity, calibration intercept and slope, flexible calibration curves, and Net Benefit (Table 1).[10,11,20,21] For classification, the 'default' risk threshold of 0.5 was used as well as a risk threshold of 0.192 (518/2695, prevalence of malignancy in the training dataset) when class imbalance was not corrected.

*Monte Carlo simulation study*

We used the ADEMP (aim, data, estimands, methods, performance) guideline to design and report the simulation study.[22]

*Aim*. The aim of this study was to investigate the impact of class imbalance corrections on model performance in terms of discrimination, calibration and classification.

*Data generating mechanism*. Twenty-four scenarios were investigated by a varying the following simulation factors: original training set size ($N$) (2500 or 5000), number of predictors ($p$) (3, 6, 12 or 24), and outcome event fraction (0.3, 0.1, 0.01). The values for $p$ and the event fraction reflect common situations for clinical prediction models.[23] A sample size of 2500 will include 25 events on average when the event fraction is 1%. Smaller values for $N$ may hence lead to computational issues. Candidate predictor variables were drawn from a multivariate standard normal distribution with zero correlation between predictors. Then, the outcome probability of each case was computed by applying a logistic function to the generated predictors. The coefficients of this function were approximated numerically for each scenario (Supplementary Note), such that the predictors were of equal strength, the c-statistic of the data generating model was approximately 0.75, and the outcome prevalence expected in accordance with the simulation condition. The outcome variable was sampled from a binomial distribution.

*Estimands/targets of analysis*. The focus is on discrimination, calibration, and classification performance of the fitted models on a large out-of-sample dataset.

*Methods*. For each generated development data set, four prediction model development datasets were created: $D_{unadjusted}$, $D_{RUS}$, $D_{ROS}$ and $D_{SMOTE}$. On each of these data sets, SLR and Ridge models were fit. This resulted in 8 different prediction models per simulation scenario. Because we anticipated imbalance correction would lead to overestimation of probabilities (i.e. that the model intercept would be too high), we also implemented a logistic re-calibration approach for the models developed on $D_{RUS}$, $D_{ROS}$ and $D_{SMOTE}$, resulting in another 6 models.[24] This re-calibration was done by fitting a logistic regression model on the development dataset with the logit of the estimated probabilities from the initial model as an offset variable and the intercept as the only free parameter:

$$log\left(\frac{\pi_{i,recalibrated}}{1-\pi_{i,recalibrated}}\right) = a + log\left(\frac{\hat{\pi}_i}{1-\hat{\pi}_i}\right),$$

For each scenario, 2,000 simulation runs were performed. In each run, a newly simulated training dataset was used. To evaluate the performance of the resulting models for a given scenario, a single test set per scenario was simulated with size *N* = 100,000 using the same data generating mechanism.

*Performance metrics*. We applied each model on its respective test set, and calculated the AUROC, accuracy, sensitivity, specificity, calibration intercept and slope. To convert the estimated probabilities into a dichotomous prediction, a default risk threshold of 0.5 was used. For models trained on unadjusted development datasets, we also used a threshold that is equal to the true event fraction. The primary metric was the calibration intercept.[20,21]

*Software and error handling.* All analyses were performed using R version 3.6.2 ([www.R-project.org](www.R-project.org)). The simulation study was performed on a high-performance computing facility running on a Linux-based Operating System (CentOS7). To fit the regression models, the R packages `stat` and `glmnet` version 4.0-2 were used.[25,26] To implement SMOTE and simulate data from a multivariate normal distribution, we respectively used the `smotefamily` (Siriseriwan W. *smotefamily: A Collection of Oversampling Techniques for Class Imbalance Problem Based on SMOTE*. 2019. R package version 1.3.1) and the `MASS` version 7.3-51.5 R packages.[25] The code is available via https://github.com/benvancalster/classimb_calibration.

Errors in the generation of the development data sets and estimation of the models were closely monitored (details in Supplementary Note).[27] A summary of the data sets in which data separation occurred is given in in Table S1.

**Results**

*Case study*

There was little variation in discrimination across algorithms and imbalance correction methods, with average AUROC of 0.79 to 0.80 (Table S2). The calibration curves indicate that all imbalance correction methods had strong impact on calibration, yielding strongly overestimated probability estimates after imbalance correction but not without correction (Figure 1). This is confirmed by the calibration intercepts: these were 0.06 (95% CI -0.16 to 0.26) for SLR and 0.05 (-0.16 to 0.26) for Ridge on training data without imbalance correction, but varied between -1.32 (-1.54 to -1.11; SMOTE followed by SLR) and -1.50 (-1.72 to -1.28; RUS followed by SLR) when using imbalance corrections. The calibration slope was closest to the target value of 1 for models based on unadjusted data (0.99 for SLR, 1.03 for Ridge) and lowest (i.e. worst) for models after RUS (0.85 for SLR, 0.93 for Ridge). When using the 0.5 probability threshold on models trained on unadjusted data, specificity (96% for SLR and Ridge) was clearly higher than sensitivity (31% for SLR, 29% for Ridge). As expected, sensitivity increased and specificity increased by changing the classification threshold for models based on unadjusted data or using the 0.5 threshold for models after imbalance correction (sensitivities between 69% and 75%, specificities between 74% and 78%).

Our results also show that the overestimation of the probability of a malignancy for models that were trained on imbalance corrected datasets could lead to overtreatment: too many individuals would exceed a given probability threshold and would be selected for treatment (for instance, referral to specialized gynecologic oncology centers for surgery). This is reflected in the Net Benefit measurement of clinical utility (Figure 2). The decision curves show that models trained on imbalance corrected datasets had strongly reduced clinical utility, even to the extent that the Net Benefit was negative when using a probability threshold of 0.3 or higher to select individuals for treatment.

*Simulation study*

The simulation results did not provide evidence that imbalance correction methods systematically improved the AUROC compared to developing models on the original

(imbalanced) training data (Figures 3 and S2-6, Table S3). The median AUROC of models trained on unadjusted data was never lower than the median AUROC of models after RUS, ROS, or SMOTE. For RUS, the median AUROC was often lower, with larger differences when event fraction was lower, training set size was lower, and number of predictors was higher.

Training models on imbalance corrected datasets resulted in severe overestimation of the estimated probabilities as evidenced by the negative calibration intercepts (Figures 4 and S7-11, Table S4). Models trained on unadjusted data had median calibration intercepts between -0.05 and 0.03. Imbalance correction methods had median calibration intercepts of -4.5 or lower for scenarios with a 1% event fraction, -2.1 or lower for scenarios with a 10% event fraction, and -0.7 or lower for scenarios with a 30% event fraction. This was corrected by applying the recalibration procedure. Using the original (imbalanced) data: recalibration improved median calibration intercepts to values between -0.07 and 0.03 (Figures S12-17, Table S5). One exception involved the training of SLR models after RUS in the scenario with 1% event fraction, a training set size of 2500, and 24 predictors. Using RUS implied that the model with 24 predictors was trained on a dataset including only 25 events and 25 non-events on average, leading to lack of convergence of the SLR model, but not for the Ridge model.

The use of SMOTE, and to a lesser extent ROS, resulted probability estimates that were too extreme as evidenced by median calibration slopes below 1 (Figures 5 and S18-22, Table S6). This finding was more evident for lower event fraction, lower training set size, and a larger number of predictors. Median calibration slopes below 1 were also observed for SLR developed on unadjusted training data. The median slopes obtained for models developed on training data after ROS or SMOTE were still lower. When using SLR after RUS, overfitting was stronger than when the unadjusted training data were used, leading to very low calibration slopes in some scenarios.

Regarding classification, using a probability threshold of 0.5 for models trained on unadjusted data resulted in median sensitivities of 0% and median specificities of 100% when the true event fraction was 1% (Figures S23-40, Tables S7-8). More balanced results for sensitivity and specificity were obtained by either using imbalance correction methods or shifting the probability threshold (Figures S29-40).

**Discussion**

The key finding of our work is that training logistic regression models on imbalance corrected data did not lead to better AUROC compared to models trained on uncorrected data, but did result in strong and systematic overestimation of the probability for the minority class. This strong miscalibration reduces the clinical utility of the model: models yielding probability estimates that are clearly too high may lead to overtreatment. For example, if a model overestimates the risk of malignancy of a detected ovarian tumor, the decision to refer patients to advanced and specialized surgery may be taken too quickly.

Class imbalance is often framed as problematic in the context of prediction models that classify patients into low versus high risk groups.[1-3,28] Nevertheless, for clinical prediction models the accurate estimation of probabilities is essential to define such low risk and high risk groups. For instance, clinical staff using the model to support treatment decisions may choose probability thresholds that match the assumed misclassification costs. Hence, when probability estimation is important, calibration becomes a central performance criterion.[29,30]

The relation between correction for class imbalance and calibration of estimated probabilities is rarely made. For instance, it is not discussed in some key publications on class imbalance for prediction models.[1-3,29] A study from 2011 hinted at this link by stating that 'the predicted probability using a logistic regression model is closest to the true probability when the sample has the same class distribution as the original population', and that differences in class distribution between study sample and population should be avoided.[31] However, the authors did not systematically study typical imbalance correction methods, and the simulations were based on an unrealistic setting with only one predictor and a true AUROC around 0.99. Another study into imbalance corrections quantified calibration incorrectly by using Brier score and class-specific Brier scores.[32] Brier score is a statistically proper measure of overall measure of performance, that captures both discrimination and calibration. This study incorrectly claimed that using RUS improved probability estimates compared to using uncorrected data in the minority class based on observed lower values of the Brier score in the minority class. This, however, does not mean that the probability estimates are well calibrated, but simply means that the probabilities in the minority class are closer to 1. This is consistent with our findings: probability estimates under RUS are indeed miscalibrated toward too extreme values.

Another study did indicate that undersampling distorts probability estimates and increases the variance of the prediction model (which relates to the higher tendency of overfitting due to artificially reducing sample size).[33] However, the study focused on classification accuracy, by claiming that the effect of undersampling on accuracy depends on many factors such that it is difficult to know when it will lead to better accuracy. In contrast, our study suggests that, at least for logistic regression models, RUS (or ROS or SMOTE) is unlikely to lead to better discrimination or separability between the minority and minority classes.

It is well known that developing robust clinical prediction models, the sample size should be large enough to reduce overfitting.[13-15,17,34] Recent studies indicate that the most important factors to determine overall sample size are the event fraction, the number of considered parameters, and the expected performance of the model. From that perspective, undersampling is a very counterintuitive approach, because it deliberately decreases sample size available for model training, which may lead to artificially increased risk of overfitting. Our results are consistent with this expectation: RUS resulted in lower AUROC values on the test data.

Based on the results presented in this study, it is warranted to conduct follow-up studies that systematically study the impact of imbalance corrections on discrimination and calibration performance, in particular in the context of other algorithms. For instance, the calibration performance of increasingly popular approaches for prediction model development such as Random Forest, Support Vector Machines and Neural Networks remains to be investigated. Also, other imbalance correction methods exist, such as weighting, cost-sensitive learning, or variants of RUS, ROS and SMOTE.[3,28,35] We anticipate that risk miscalibration will remain present regardless of type of model or imbalance correction technique, unless the models are re-calibrated. However, class imbalance correction followed by recalibration is only worth the effort if imbalance correction leads to better discrimination of the resulting models.

In conclusion, our study shows that correcting class imbalance did not result in better prediction models based on standard or ridge logistic regression. The imbalance corrections resulted in inaccurate probability estimates without improving discrimination in terms of AUROC. We therefore warn researchers for the limitations of imbalance corrections when developing a prediction model.

Table 1. Test set performance metrics.

| Metric | Description |
|---|---|
| Concordance (c) statistic | The c statistic estimates the probability that a model gives a higher prediction (e.g. estimated probability) for a random individual with the event than for a random individual without the event. For binary outcomes, this equals the area under the receiver operating characteristic curve (AUROC). The c statistic is 1 when all patients with an event have a higher risk estimate than all patients without an event, and is 0.5 when risk estimates cannot differentiate at all between patients with and without the event. |
| Classification accuracy | The accuracy is the proportion of patients that are classified correctly, i.e. the proportion of patients that are either true positives (TP) or true negatives (TN): (TP + TN) / N. |
| Sensitivity | The sensitivity is the proportion of patients with the event that are classified as high risk: TP / (TP + FN). |
| Specificity | The specificity is the proportion of patients without the event that are classified as low risk: TN / (TN + FP). |
| Calibration intercept | The calibration intercept quantifies whether risk estimates are on average too high (overestimation, calibration intercept < 0) or too low (underestimation, calibration intercept >0). It is calculated as the intercept $a$ of the following logistic regression analysis: $logit(\pi_i) = a + LP$, where LP is the linear predictor (logit of the estimated risk from the model). The LP is added as an offset, meaning that its coefficient is fixed at 1. |
| Calibration slope | The calibration slope quantifies whether risk estimates are too extreme (too close to 0 or 1, calibration slope < 1) or too modest (too close to the event fraction, calibration slope > 1). It is calculated as the coefficient $b$ of the following logistic regression analysis: $logit(\pi_i) = a + b * LP$. |
| Flexible calibration curve | This curve estimates the accuracy of estimated risks conditional on the estimated risk (and hence not conditional on a patient's predictor values). It is based on the following flexible logistic regression model: $logit(\pi_i) = a + f(LP)$. The flexible function $f(.)$ was based on loess (locally estimated scatter plot smoothing). |
| Net Benefit | Assuming that we use a model to identify high risk patients, for which a given clinical intervention (treatment) is warranted, Net Benefit quantifies the utility of the model to make such treatment decisions. It exploits the link between the risk threshold and misclassification costs. Using a risk threshold t = 0.1 means that we accept to treat at most 10 patients per true positive. In other words, we tolerate 9 false positives (unnecessary treatments) per true positive (necessary treatment). This implies that the benefit of 1 true positive is 9 times higher than the harm of a false positive. So as long as you have <9 false positives per true positive, the benefits outweigh the harms. Net Benefit is therefore calculated as (TP – w*FP) / N, with w equal to odds(t) = 1/(1-t). Net Benefit is conditional on the adopted misclassification costs, and can therefore be calculated for several potential risk thresholds. A plot of Net Benefit for a range of thresholds is a decision curve. Net Benefit can also be calculated for two default strategies: treating everyone (all) or treating no one (none). Whatever the misclassification costs, treating no one has a Net Benefit of 0 by definition. Treating everyone has a positive Net Benefit when misclassification costs clearly favor true positives (t is low). If, for a given t, the Net Benefit of a model is not higher than the Net Benefit of the two default strategies, the model has no clinical utility for the misclassification costs associated with t. |

Figure 1. Flexible calibration curves on the test set for the Ridge models to diagnose ovarian cancer.

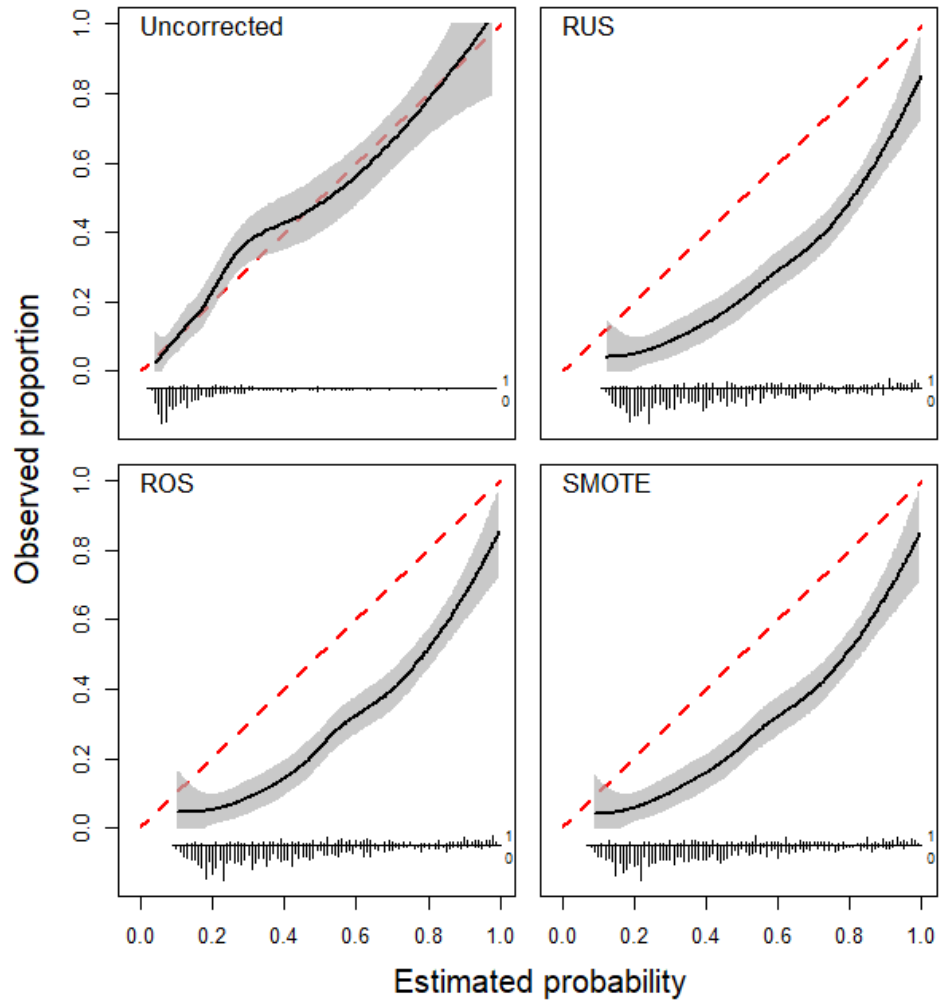

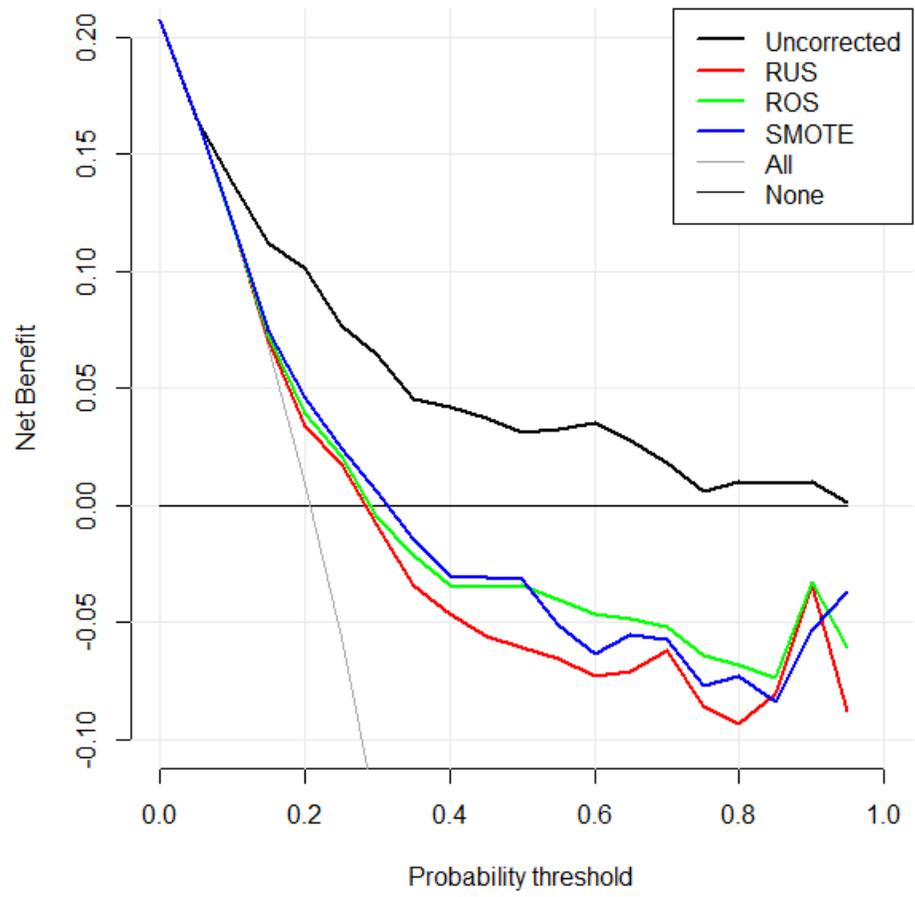

Figure 3. Test set AUROC for the Ridge models in the simulation scenarios with an event fraction of 1%.

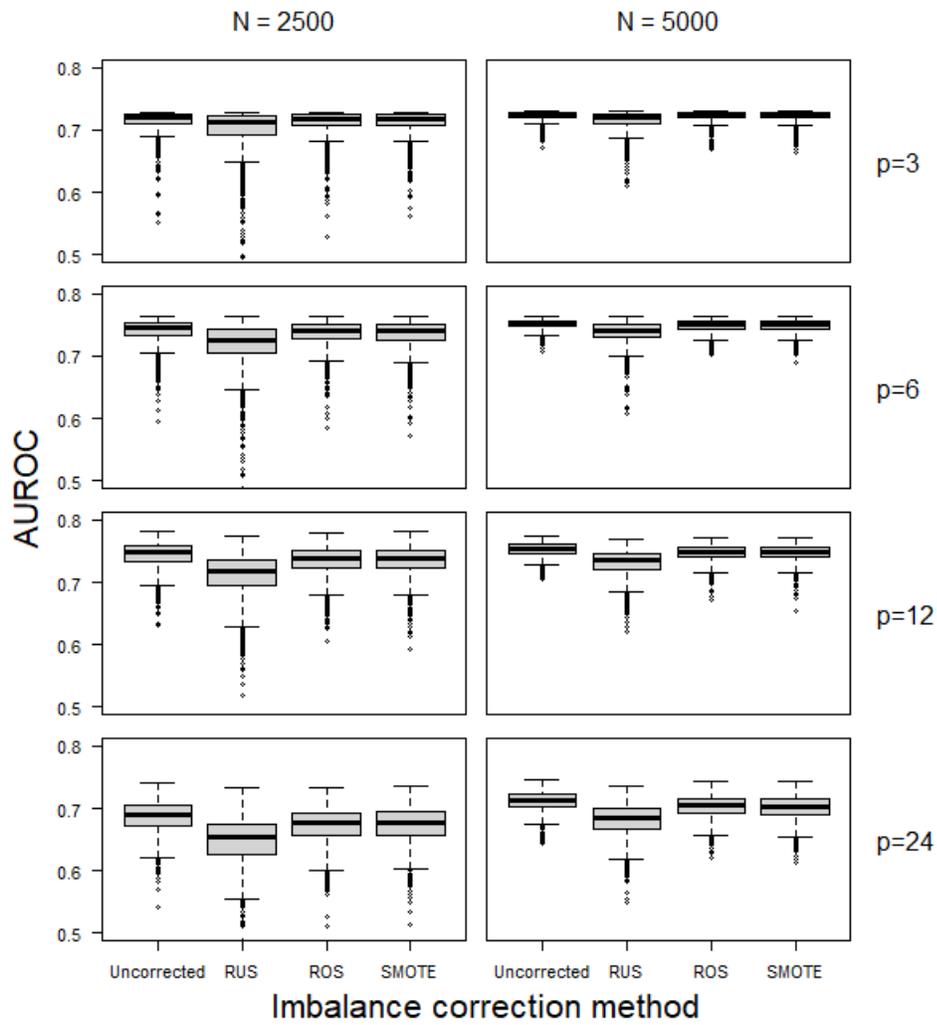

Figure 4. Test set calibration intercept for the Ridge models in the simulation scenarios with an event fraction of 1%.

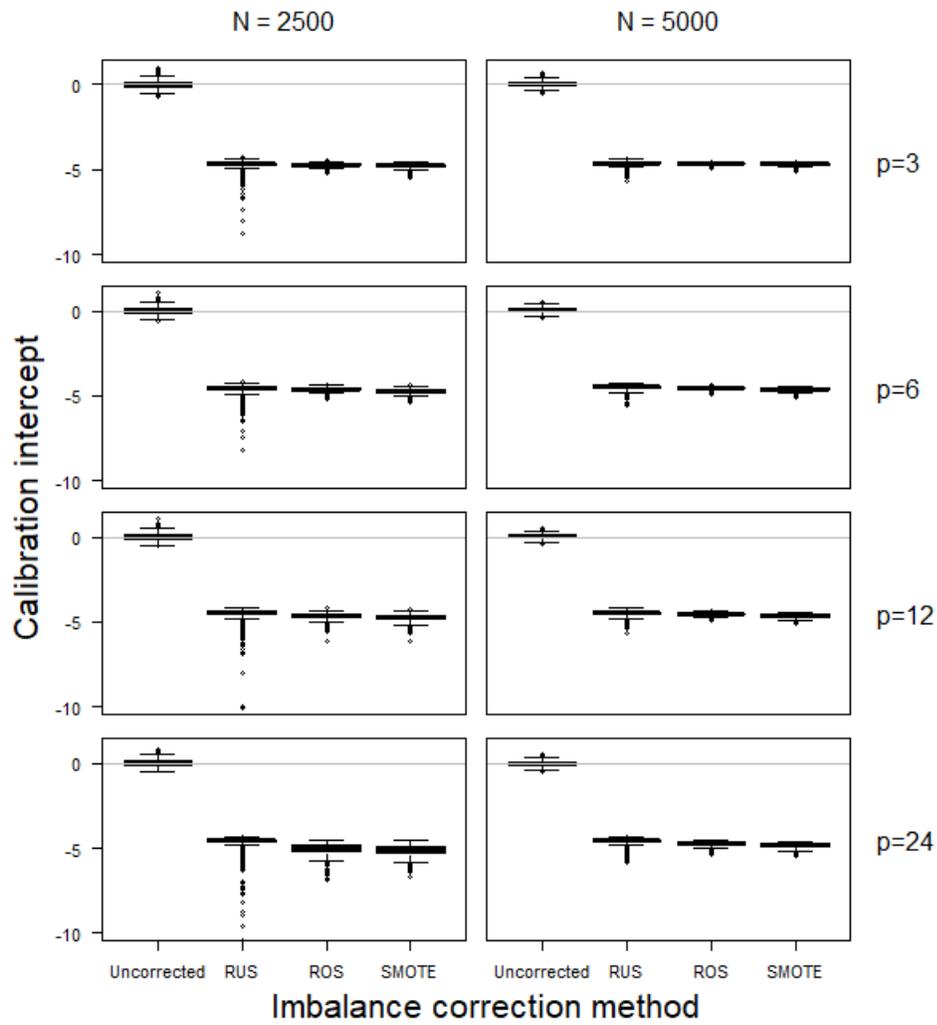

Figure 5. Test set calibration slope for the Ridge models in the simulation scenarios with an event fraction of 1%.

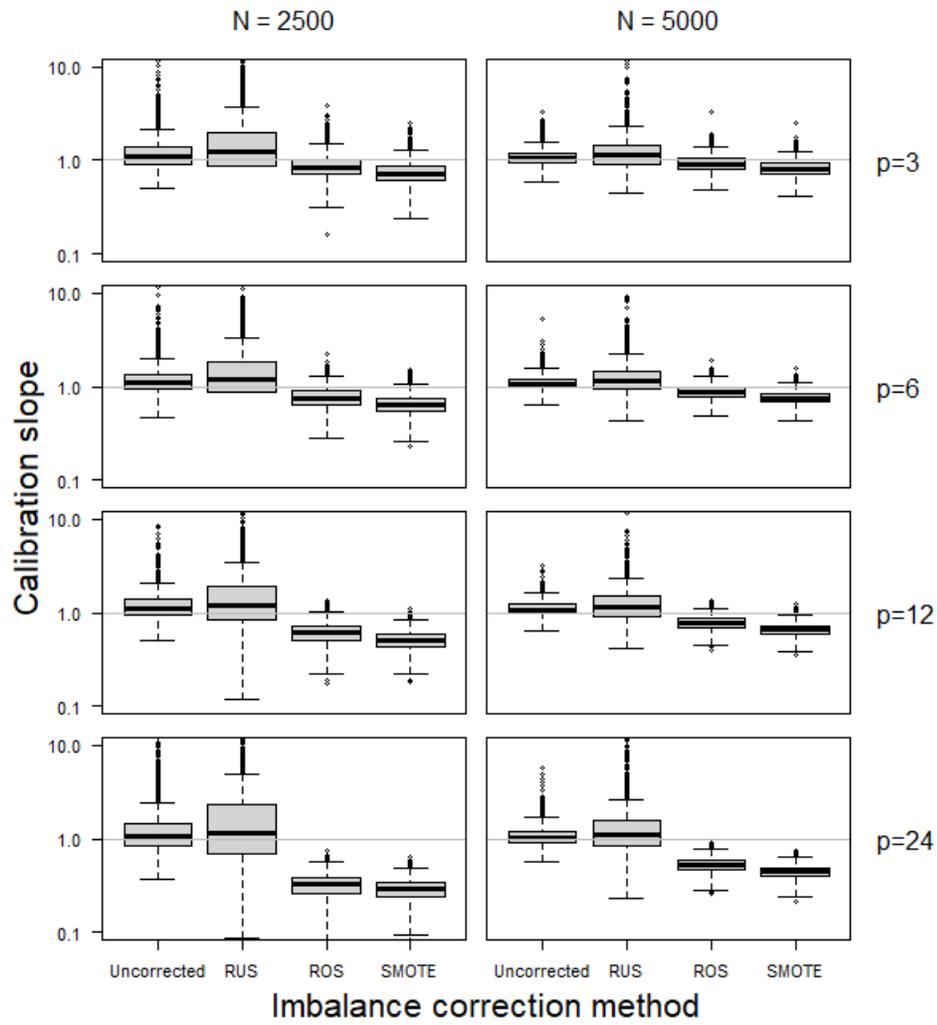

The harm of class imbalance corrections for risk prediction models:

illustration and simulation using logistic regression

van den Goorbergh R[1], van Smeden M[1], Timmerman D,[2,3] Van Calster B[2,4,5]

1 Julius Center for Health Sciences and Primary Care, UMC Utrecht, Utrecht University, Utrecht, Netherlands

2 Department of Development and Regeneration, KU Leuven, Leuven, Belgium

3 Department of Obstetrics and Gynecology, University Hospitals Leuven, Leuven, Belgium

4 Department of Biomedical Data Sciences, Leiden University Medical Center, Leiden, Netherlands

5 EPI-center, KU Leuven, Leuven, Belgium

SUPPLEMENTARY INFORMATION

**Supplementary Notes**

*Class imbalance methods*

When using RUS, the size of the majority class (i.e. the group of individuals with observed events or non-events, whichever is larger) is reduced by discarding a random set of cases until the majority class has the same size as the minority class. When using ROS, the size of the minority class is increased by resampling cases from the minority class, with replacement, until the minority class has the same size as the majority class. This results in an artificially balanced dataset containing duplicate cases for the minority class.

SMOTE is a form of oversampling that creates new, synthetic cases. Contrary to ROS, where the minority cases are simply duplicated and thus result in a data set with cases that are identical, SMOTE results in synthetic data points that are interpolations of the original minority class cases. The procedure is as follows: for every minority class case, the k nearest minority class neighbors in the predictor space are determined, based on the Euclidean distance. Then, the differences between the feature vector of the minority case and those of its k nearest neighbors are taken. These differences are then multiplied by a random number between 0 and 1 and added to the feature vector of the minority case. By creating synthetic data in this manner, there is more variation in the minority cases and hence, the models trained on this data set may be less prone to overfitting than when trained on ROS data. We used k=5 when implementing SMOTE.

Figure S1 illustrates these methods using two predictors. The original dataset has 100 cases (red dots) with the event and 1900 without the event (grey triangles) (upper left panel). The difference between ROS (lower left panel) and SMOTE (lower right panel) is obvious. ROS includes many duplicates of the original cases from the minority class. SMOTE creates synthetic cases that lie on the 'line' between two original minority cases.

SMOTE is designed to work with continuous variables. For use with ordinal or categorical variables, one may either use rounding or use an adaptation of SMOTE for mixed variable types.

*Note: Logistic regression models*

Let $Y$ denote the outcome taking on the value 1 for events and value 0 for non-events, $y_i$ the observed value of $Y$ for the for a given individual $i$, $\mathbf{X}$ a vector of $p$ predictor variables, $\mathbf{x_i}$ the observed values for the predictors for a given individual $i$, and $\pi_i = P(Y = 1|\mathbf{X} = \mathbf{x}_i)$ be the probability of an event for a given individual $i$. Assuming for simplicity only linear associations of the predictors and no interactions between predictors, logistic regression for $\pi_i$ can be defined as follows

$$log\left(\frac{\pi_i}{1-\pi_i}\right) = \alpha + \mathbf{x_i}\mathbf{\beta},$$

where $\alpha$ is the intercept and $\mathbf{\beta}$ a vector containing the regression coefficient for each predictor. Nonlinear effects of predictors (such as quadratic terms) and interaction terms between predictors can be incorporated by considering these terms as additional terms in $\mathbf{X}$. Using a dataset with $N$ individuals ($i = 1, \ldots, N$), the regression coefficients are estimated using maximum likelihood estimation, i.e. by maximizing the log-likelihood function

$$l(\alpha, \mathbf{\beta}) = \sum_{i=1}^{N}\{y_i log(\pi_i(\alpha, \mathbf{\beta})) + (1 - y_i)log(1 - \pi_i(\alpha, \mathbf{\beta}))\}.$$

We will refer to this model as standard logistic regression (SLR).

Alternatively, when using a ridge penalty, the following penalized version of the log-likelihood function is used for estimating $\mathbf{\beta}$, in order to penalize the coefficients towards zero:

$$l(\alpha, \mathbf{\beta})_{ridge} = l(\alpha, \mathbf{\beta}) - \lambda \sum_{j=1}^{p} \beta_j^2.$$

We refer to this model simply as Ridge. Here, $\lambda$ is a hyperparameter that controls the amount of penalization. We tuned $\lambda$ using 10-fold cross-validation on the deviance from a grid of 251 possible values between 0 (no penalization) and 64 (very strong penalization). The non-null values in this grid were equidistant on logarithmic scale.

*Coefficient estimation*

The intercepts and coefficients that result in the desired true AUROC and event fraction were estimated by numerical optimization using the optim() function from the `stats` R package. The method used for optimization was the Broyden-Fletcher-Goldfarb-Shanno (BFGS) algorithm. The function to be minimized was the sum of the difference between the observed AUROC and the desired AUROC squared and the difference between the observed prevalence and the desired prevalence squared. As this method can lead to slightly varying results, the minimization procedure was deployed 20 times after which the median coefficient values were chosen. This process was repeated for 20 generated data sets of size $N = 10^5$, after which again the median coefficient values of all 20 repetitions were chosen. The final coefficient values validated on independently generated data sets of size $N = 10^5$.

*Error handling*

Errors in the generation of the development datasets were closely monitored, a table summarizing the error occurrence per simulation cell is included in the article. Data separation in the development datasets was assumed when the apparent AUROC in the development data set was equal to 1, based on the maximum likelihood logistic regression model. Because, in practice, clinical prediction modelers should not develop prediction models on separated data, separated data sets were removed from the analysis. Very few cases of data separation, or no cases at all, were expected in the generated development data sets, given that the true AUROC was set to be approximately 0.75 and the minimum sample size was 2,500. However, data separation was likely to occur when random undersampling is used.

If a development data set contained cases of only one class, this data set was excluded from the analysis; the simulation results are based on complete case analysis. Development datasets with fewer than 8 events or non-events can cause severe problems in estimating tuning parameter $\lambda$ in ridge logistic regression using 10-fold cross-validation. In such cases, leave-one-out cross-validation was used to estimate $\lambda$. When there were less than 6 minority-class events in the development data set, the SMOTE-algorithm fails when using the default setting because it searches for the *k=5* nearest neighbors. In such cases *k* was set to the number of minority-class events minus 1. In the most extreme scenario with event fraction 0.01 and sample size 2,500, the probability of generating a data set with < 8 events was only 0.00002 (1 in 50,000).

Table S1. Number and percentage of the 2000 runs with data separation for simulation scenarios with event fraction of 0.01. No issues were encountered for event fractions of 0.1 or 0.3.

| N | p | Uncorrected | Imbalance correction method | | |
|---|---|---|---|---|---|
| | | | RUS | ROS | SMOTE |
| 2500 | 3 | 0 | 0 | 0 | 0 |
| 2500 | 6 | 0 | 3 (0.2%) | 0 | 0 |
| 2500 | 12 | 0 | 52 (2.6%) | 0 | 0 |
| 2500 | 24 | 0 | 1238 (61.9%) | 0 | 0 |
| 5000 | 3 | 0 | 0 | 0 | 0 |
| 5000 | 6 | 0 | 0 | 0 | 0 |
| 5000 | 12 | 0 | 0 | 0 | 0 |
| 5000 | 24 | 0 | 9 (0.5%) | 0 | 0 |

Table S2. Test set performance for the SLR and Ridge models to diagnose ovarian malignancy.

|  | Uncorrected | RUS | ROS | SMOTE |
|---|---|---|---|---|
| *SLR model* | | | | |
| AUC | 0.79 | 0.79 | 0.80 | 0.79 |
| (95% CI) | (0.75; 0.83) | (0.75; 0.83) | (0.75; 0.84) | (0.75; 0.83) |
| Accuracy, t=0.5 | 82 | 73 | 76 | 76 |
| Sensitivity, t=0.5 | 31 | 71 | 70 | 69 |
| Specificity, t=0.5 | 96 | 74 | 77 | 78 |
| Accuracy, t=0.192 | 75 | Nc | Nc | Nc |
| Sensitivity, t=0.192 | 71 | Nc | Nc | Nc |
| Specificity, t=0.192 | 76 | Nc | Nc | Nc |
| Calibration intercept | 0.06 | -1.50 | -1.37 | -1.32 |
| (95% CI) | (-0.16; 0.26) | (-1.72; -1.28) | (-1.59; -1.16) | (-1.54; -1.11) |
| Calibration slope | 0.99 | 0.85 | 0.93 | 0.92 |
| (95% CI) | (0.81; 1.20) | (0.68; 1.02) | (0.76; 1.12) | (0.75; 1.11) |
| *Ridge model* | | | | |
| AUC | 0.79 | 0.80 | 0.80 | 0.79 |
| (95% CI) | (0.75; 0.83) | (0.75; 0.83) | (0.75; 0.84) | (0.75; 0.83) |
| Accuracy, t=0.5 | 82 | 73 | 76 | 76 |
| Sensitivity, t=0.5 | 29 | 71 | 70 | 69 |
| Specificity, t=0.5 | 96 | 74 | 77 | 78 |
| Accuracy, t=0.192 | 75 | Nc | Nc | Nc |
| Sensitivity, t=0.192 | 70 | Nc | Nc | Nc |
| Specificity, t=0.192 | 76 | Nc | Nc | Nc |
| Calibration intercept | 0.05 | -1.48 | -1.38 | -1.33 |
| (95% CI) | (-0.16; 0.26) | (-1.70; -1.27) | (-1.60; -1.17) | (-1.55; -1.12) |
| Calibration slope | 1.03 | 0.93 | 0.95 | 0.94 |
| (95% CI) | (0.84; 1.25) | (0.76; 1.12) | (0.77; 1.14) | (0.77; 1.13) |

Table S3. Test set AUROC, reported as median (IQR) over the 2000 simulation runs for SLR and Ridge in the 24 simulation scenarios.

| Scenario | EF | N | p | SLR | | | | RIDGE | | | |
|---|---|---|---|---|---|---|---|---|---|---|---|
| | | | | Uncorrected | RUS | ROS | SMOTE | Uncorrected | RUS | ROS | SMOTE |
| 1 | 0.01 | 2500 | 3 | 0.72 (0.71;0.72) | 0.71 (0.69;0.72) | 0.72 (0.71;0.72) | 0.72 (0.71;0.72) | 0.72 (0.71;0.72) | 0.71 (0.69;0.72) | 0.72 (0.71;0.72) | 0.72 (0.71;0.72) |
| 2 | 0.01 | 2500 | 6 | 0.75 (0.73;0.75) | 0.72 (0.69;0.74) | 0.74 (0.73;0.75) | 0.74 (0.73;0.75) | 0.75 (0.73;0.75) | 0.73 (0.70;0.74) | 0.74 (0.73;0.75) | 0.74 (0.73;0.75) |
| 3 | 0.01 | 2500 | 12 | 0.75 (0.73;0.76) | 0.70 (0.67;0.72) | 0.74 (0.72;0.75) | 0.74 (0.72;0.75) | 0.75 (0.73;0.76) | 0.72 (0.69;0.74) | 0.74 (0.72;0.75) | 0.74 (0.72;0.75) |
| 4 | 0.01 | 2500 | 24 | 0.69 (0.67;0.70) | 0.59 (0.56;0.62) | 0.68 (0.66;0.69) | 0.68 (0.66;0.69) | 0.69 (0.67;0.71) | 0.65 (0.63;0.68) | 0.68 (0.66;0.69) | 0.68 (0.66;0.69) |
| 5 | 0.01 | 5000 | 3 | 0.73 (0.72;0.73) | 0.72 (0.71;0.73) | 0.72 (0.72;0.73) | 0.72 (0.72;0.73) | 0.73 (0.72;0.73) | 0.72 (0.71;0.73) | 0.72 (0.72;0.73) | 0.72 (0.72;0.73) |
| 6 | 0.01 | 5000 | 6 | 0.75 (0.75;0.76) | 0.74 (0.73;0.75) | 0.75 (0.74;0.76) | 0.75 (0.74;0.76) | 0.75 (0.75;0.76) | 0.74 (0.73;0.75) | 0.75 (0.74;0.76) | 0.75 (0.74;0.76) |
| 7 | 0.01 | 5000 | 12 | 0.75 (0.75;0.76) | 0.73 (0.71;0.74) | 0.75 (0.74;0.76) | 0.75 (0.74;0.76) | 0.75 (0.75;0.76) | 0.74 (0.72;0.75) | 0.75 (0.74;0.76) | 0.75 (0.74;0.76) |
| 8 | 0.01 | 5000 | 24 | 0.71 (0.70;0.72) | 0.67 (0.64;0.69) | 0.70 (0.69;0.72) | 0.70 (0.69;0.71) | 0.71 (0.70;0.72) | 0.68 (0.67;0.70) | 0.70 (0.69;0.72) | 0.70 (0.69;0.71) |
| 9 | 0.1 | 2500 | 3 | 0.73 (0.73;0.73) | 0.73 (0.73;0.73) | 0.73 (0.73;0.73) | 0.73 (0.73;0.73) | 0.73 (0.73;0.73) | 0.73 (0.73;0.73) | 0.73 (0.73;0.73) | 0.73 (0.73;0.73) |
| 10 | 0.1 | 2500 | 6 | 0.74 (0.74;0.74) | 0.74 (0.74;0.74) | 0.74 (0.74;0.74) | 0.74 (0.74;0.74) | 0.74 (0.74;0.74) | 0.74 (0.74;0.74) | 0.74 (0.74;0.74) | 0.74 (0.74;0.74) |
| 11 | 0.1 | 2500 | 12 | 0.75 (0.75;0.75) | 0.75 (0.74;0.75) | 0.75 (0.75;0.75) | 0.75 (0.75;0.75) | 0.75 (0.75;0.75) | 0.75 (0.74;0.75) | 0.75 (0.75;0.75) | 0.75 (0.75;0.75) |
| 12 | 0.1 | 2500 | 24 | 0.73 (0.73;0.73) | 0.72 (0.71;0.72) | 0.73 (0.72;0.73) | 0.73 (0.72;0.73) | 0.73 (0.73;0.73) | 0.72 (0.71;0.72) | 0.73 (0.72;0.73) | 0.73 (0.72;0.73) |
| 13 | 0.1 | 5000 | 3 | 0.73 (0.73;0.73) | 0.73 (0.73;0.73) | 0.73 (0.73;0.73) | 0.73 (0.73;0.73) | 0.73 (0.73;0.73) | 0.73 (0.73;0.73) | 0.73 (0.73;0.73) | 0.73 (0.73;0.73) |
| 14 | 0.1 | 5000 | 6 | 0.75 (0.75;0.75) | 0.74 (0.74;0.75) | 0.75 (0.74;0.75) | 0.75 (0.74;0.75) | 0.75 (0.75;0.75) | 0.74 (0.74;0.75) | 0.75 (0.74;0.75) | 0.75 (0.74;0.75) |
| 15 | 0.1 | 5000 | 12 | 0.75 (0.75;0.75) | 0.75 (0.75;0.75) | 0.75 (0.75;0.75) | 0.75 (0.75;0.75) | 0.75 (0.75;0.75) | 0.75 (0.75;0.75) | 0.75 (0.75;0.75) | 0.75 (0.75;0.75) |
| 16 | 0.1 | 5000 | 24 | 0.74 (0.74;0.74) | 0.73 (0.73;0.73) | 0.74 (0.74;0.74) | 0.74 (0.74;0.74) | 0.74 (0.74;0.74) | 0.73 (0.73;0.74) | 0.74 (0.74;0.74) | 0.74 (0.74;0.74) |
| 17 | 0.3 | 2500 | 3 | 0.74 (0.74;0.74) | 0.74 (0.74;0.74) | 0.74 (0.74;0.74) | 0.74 (0.74;0.74) | 0.74 (0.74;0.74) | 0.74 (0.74;0.74) | 0.74 (0.74;0.74) | 0.74 (0.74;0.74) |
| 18 | 0.3 | 2500 | 6 | 0.74 (0.74;0.74) | 0.74 (0.74;0.74) | 0.74 (0.74;0.74) | 0.74 (0.74;0.74) | 0.74 (0.74;0.74) | 0.74 (0.74;0.74) | 0.74 (0.74;0.74) | 0.74 (0.74;0.74) |
| 19 | 0.3 | 2500 | 12 | 0.75 (0.74;0.75) | 0.74 (0.74;0.75) | 0.75 (0.74;0.75) | 0.75 (0.74;0.75) | 0.75 (0.74;0.75) | 0.74 (0.74;0.75) | 0.75 (0.74;0.75) | 0.75 (0.74;0.75) |
| 20 | 0.3 | 2500 | 24 | 0.74 (0.73;0.74) | 0.73 (0.73;0.73) | 0.73 (0.73;0.74) | 0.73 (0.73;0.74) | 0.74 (0.73;0.74) | 0.73 (0.73;0.73) | 0.73 (0.73;0.74) | 0.73 (0.73;0.74) |
| 21 | 0.3 | 5000 | 3 | 0.74 (0.74;0.74) | 0.74 (0.74;0.74) | 0.74 (0.74;0.74) | 0.74 (0.74;0.74) | 0.74 (0.74;0.74) | 0.74 (0.74;0.74) | 0.74 (0.74;0.74) | 0.74 (0.74;0.74) |
| 22 | 0.3 | 5000 | 6 | 0.74 (0.74;0.74) | 0.74 (0.74;0.74) | 0.74 (0.74;0.74) | 0.74 (0.74;0.74) | 0.74 (0.74;0.74) | 0.74 (0.74;0.74) | 0.74 (0.74;0.74) | 0.74 (0.74;0.74) |
| 23 | 0.3 | 5000 | 12 | 0.75 (0.75;0.75) | 0.75 (0.75;0.75) | 0.75 (0.75;0.75) | 0.75 (0.75;0.75) | 0.75 (0.75;0.75) | 0.75 (0.75;0.75) | 0.75 (0.75;0.75) | 0.75 (0.75;0.75) |
| 24 | 0.3 | 5000 | 24 | 0.74 (0.74;0.74) | 0.74 (0.74;0.74) | 0.74 (0.74;0.74) | 0.74 (0.74;0.74) | 0.74 (0.74;0.74) | 0.74 (0.74;0.74) | 0.74 (0.74;0.74) | 0.74 (0.74;0.74) |

Table S4. Test set calibration intercept, reported as median (IQR) over the simulation 2000 runs for SLR and Ridge in the 24 simulation scenarios.

| | | | | SLR | | | | RIDGE | | | |
|---|---|---|---|---|---|---|---|---|---|---|---|
| Scen. | EF | N | p | Uncorrected | RUS | ROS | SMOTE | Uncorrected | RUS | ROS | SMOTE |
| 1 | 0.01 | 2500 | 3 | -0.04 (-0.17;0.11) | -4.84 (-5.10;-4.69) | -4.71 (-4.76;-4.67) | -4.78 (-4.85;-4.72) | -0.04 (-0.17;0.11) | -4.67 (-4.75;-4.60) | -4.70 (-4.75;-4.67) | -4.77 (-4.85;-4.71) |
| 2 | 0.01 | 2500 | 6 | -0.01 (-0.14;0.13) | -5.01 (-5.46;-4.74) | -4.61 (-4.68;-4.56) | -4.71 (-4.81;-4.64) | -0.01 (-0.14;0.13) | -4.54 (-4.65;-4.47) | -4.61 (-4.67;-4.55) | -4.70 (-4.80;-4.63) |
| 3 | 0.01 | 2500 | 12 | 0.02 (-0.11;0.15) | -5.92 (-7.08;-5.25) | -4.62 (-4.73;-4.54) | -4.76 (-4.89;-4.65) | 0.02 (-0.11;0.15) | -4.46 (-4.57;-4.39) | -4.60 (-4.70;-4.53) | -4.73 (-4.85;-4.63) |
| 4 | 0.01 | 2500 | 24 | -0.05 (-0.18;0.09) | <-100 (<-100;-14.4) | -5.04 (-5.25;-4.87) | -5.15 (-5.36;-4.99) | -0.02 (-0.15;0.11) | -4.56 (-4.66;-4.53) | -4.98 (-5.18;-4.82) | -5.08 (-5.27;-4.93) |
| 5 | 0.01 | 5000 | 3 | 0.02 (-0.08;0.12) | -4.71 (-4.84;-4.61) | -4.64 (-4.67;-4.62) | -4.68 (-4.73;-4.65) | 0.02 (-0.08;0.12) | -4.60 (-4.68;-4.55) | -4.64 (-4.67;-4.62) | -4.68 (-4.72;-4.65) |
| 6 | 0.01 | 5000 | 6 | 0.03 (-0.06;0.12) | -4.69 (-4.87;-4.56) | -4.54 (-4.58;-4.51) | -4.62 (-4.67;-4.57) | 0.03 (-0.06;0.12) | -4.49 (-4.57;-4.43) | -4.54 (-4.57;-4.51) | -4.61 (-4.67;-4.57) |
| 7 | 0.01 | 5000 | 12 | 0.01 (-0.08;0.11) | -4.93 (-5.20;-4.73) | -4.54 (-4.58;-4.49) | -4.65 (-4.71;-4.58) | 0.01 (-0.08;0.11) | -4.45 (-4.55;-4.40) | -4.53 (-4.58;-4.49) | -4.64 (-4.70;-4.58) |
| 8 | 0.01 | 5000 | 24 | -0.03 (-0.13;0.07) | -6.10 (-6.89;-5.58) | -4.72 (-4.80;-4.66) | -4.85 (-4.95;-4.78) | -0.02 (-0.12;0.07) | -4.56 (-4.63;-4.51) | -4.71 (-4.78;-4.65) | -4.83 (-4.92;-4.76) |
| 9 | 0.1 | 2500 | 3 | 0.01 (-0.04;0.05) | -2.20 (-2.23;-2.16) | -2.19 (-2.21;-2.18) | -2.15 (-2.18;-2.12) | 0.01 (-0.04;0.05) | -2.18 (-2.21;-2.15) | -2.19 (-2.20;-2.17) | -2.14 (-2.17;-2.11) |
| 10 | 0.1 | 2500 | 6 | 0.00 (-0.05;0.04) | -2.17 (-2.21;-2.14) | -2.15 (-2.17;-2.14) | -2.12 (-2.15;-2.08) | 0.00 (-0.05;0.04) | -2.14 (-2.18;-2.11) | -2.15 (-2.16;-2.13) | -2.11 (-2.14;-2.08) |
| 11 | 0.1 | 2500 | 12 | 0.00 (-0.04;0.05) | -2.16 (-2.20;-2.12) | -2.12 (-2.14;-2.10) | -2.09 (-2.13;-2.06) | 0.00 (-0.04;0.05) | -2.10 (-2.14;-2.07) | -2.11 (-2.13;-2.10) | -2.09 (-2.12;-2.05) |
| 12 | 0.1 | 2500 | 24 | -0.02 (-0.06;0.03) | -2.26 (-2.31;-2.22) | -2.17 (-2.19;-2.15) | -2.15 (-2.19;-2.12) | -0.02 (-0.06;0.03) | -2.15 (-2.18;-2.12) | -2.16 (-2.17;-2.14) | -2.14 (-2.17;-2.10) |
| 13 | 0.1 | 5000 | 3 | 0.00 (-0.03;0.03) | -2.20 (-2.23;-2.18) | -2.20 (-2.21;-2.19) | -2.15 (-2.18;-2.12) | 0.00 (-0.03;0.03) | -2.19 (-2.22;-2.17) | -2.20 (-2.21;-2.19) | -2.15 (-2.18;-2.11) |
| 14 | 0.1 | 5000 | 6 | 0.01 (-0.02;0.05) | -2.15 (-2.18;-2.12) | -2.14 (-2.15;-2.13) | -2.10 (-2.13;-2.07) | 0.01 (-0.02;0.05) | -2.13 (-2.16;-2.11) | -2.14 (-2.15;-2.12) | -2.10 (-2.13;-2.07) |
| 15 | 0.1 | 5000 | 12 | 0.00 (-0.03;0.03) | -2.14 (-2.17;-2.11) | -2.12 (-2.13;-2.11) | -2.09 (-2.13;-2.06) | 0.00 (-0.03;0.03) | -2.11 (-2.14;-2.09) | -2.12 (-2.13;-2.11) | -2.09 (-2.12;-2.05) |
| 16 | 0.1 | 5000 | 24 | 0.00 (-0.03;0.03) | -2.20 (-2.23;-2.17) | -2.15 (-2.17;-2.14) | -2.13 (-2.16;-2.10) | 0.00 (-0.03;0.03) | -2.14 (-2.17;-2.12) | -2.15 (-2.16;-2.14) | -2.12 (-2.15;-2.09) |
| 17 | 0.3 | 2500 | 3 | 0.01 (-0.03;0.04) | -0.84 (-0.85;-0.82) | -0.84 (-0.85;-0.82) | -0.69 (-0.72;-0.66) | 0.01 (-0.03;0.04) | -0.84 (-0.85;-0.82) | -0.84 (-0.85;-0.82) | -0.69 (-0.72;-0.66) |
| 18 | 0.3 | 2500 | 6 | 0.00 (-0.03;0.03) | -0.84 (-0.86;-0.83) | -0.84 (-0.85;-0.82) | -0.70 (-0.73;-0.67) | 0.00 (-0.03;0.03) | -0.84 (-0.85;-0.82) | -0.84 (-0.85;-0.82) | -0.70 (-0.73;-0.66) |
| 19 | 0.3 | 2500 | 12 | 0.00 (-0.04;0.03) | -0.83 (-0.85;-0.81) | -0.83 (-0.84;-0.81) | -0.70 (-0.73;-0.67) | 0.00 (-0.04;0.03) | -0.82 (-0.84;-0.81) | -0.82 (-0.84;-0.81) | -0.70 (-0.73;-0.66) |
| 20 | 0.3 | 2500 | 24 | 0.01 (-0.03;0.04) | -0.84 (-0.86;-0.83) | -0.83 (-0.84;-0.81) | -0.69 (-0.73;-0.66) | 0.01 (-0.03;0.04) | -0.83 (-0.84;-0.81) | -0.82 (-0.83;-0.81) | -0.69 (-0.72;-0.65) |
| 21 | 0.3 | 5000 | 3 | 0.00 (-0.02;0.02) | -0.84 (-0.85;-0.83) | -0.84 (-0.85;-0.83) | -0.70 (-0.72;-0.67) | 0.00 (-0.02;0.02) | -0.84 (-0.85;-0.83) | -0.84 (-0.85;-0.83) | -0.69 (-0.72;-0.67) |
| 22 | 0.3 | 5000 | 6 | 0.00 (-0.02;0.03) | -0.83 (-0.84;-0.82) | -0.83 (-0.84;-0.82) | -0.69 (-0.71;-0.67) | 0.00 (-0.02;0.03) | -0.83 (-0.84;-0.82) | -0.83 (-0.84;-0.82) | -0.69 (-0.71;-0.67) |
| 23 | 0.3 | 5000 | 12 | 0.00 (-0.02;0.02) | -0.83 (-0.84;-0.81) | -0.82 (-0.83;-0.81) | -0.70 (-0.72;-0.67) | 0.00 (-0.02;0.02) | -0.82 (-0.83;-0.81) | -0.82 (-0.83;-0.81) | -0.69 (-0.72;-0.67) |
| 24 | 0.3 | 5000 | 24 | -0.01 (-0.03;0.01) | -0.85 (-0.86;-0.84) | -0.84 (-0.85;-0.83) | -0.71 (-0.73;-0.68) | -0.01 (-0.03;0.01) | -0.84 (-0.85;-0.83) | -0.84 (-0.85;-0.83) | -0.70 (-0.73;-0.68) |

Table S5. Test set calibration intercept after recalibration for RUS/ROS/SMOTE, reported as median (IQR) over the 2000 runs for SLR and Ridge in the 24 simulation scenarios.

| Scenario | EF | N | p | SLR | | | RIDGE | | |
|---|---|---|---|---|---|---|---|---|---|
| | | | | RUS | ROS | SMOTE | RUS | ROS | SMOTE |
| 1 | 0.01 | 2500 | 3 | -0.04 (-0.17;0.11) | -0.04 (-0.17;0.11) | -0.04 (-0.18;0.11) | -0.04 (-0.17;0.11) | -0.04 (-0.17;0.11) | -0.04 (-0.18;0.11) |
| 2 | 0.01 | 2500 | 6 | 0.00 (-0.14;0.13) | -0.01 (-0.14;0.13) | -0.02 (-0.15;0.14) | 0.00 (-0.14;0.13) | -0.01 (-0.14;0.13) | -0.02 (-0.15;0.14) |
| 3 | 0.01 | 2500 | 12 | 0.02 (-0.11;0.16) | 0.01 (-0.12;0.16) | 0.01 (-0.13;0.16) | 0.02 (-0.11;0.16) | 0.01 (-0.12;0.16) | 0.01 (-0.13;0.16) |
| 4 | 0.01 | 2500 | 24 | -0.02 (-0.15;0.11) | -0.06 (-0.20;0.09) | -0.07 (-0.21;0.09) | -0.02 (-0.15;0.11) | -0.06 (-0.20;0.09) | -0.07 (-0.21;0.09) |
| 5 | 0.01 | 5000 | 3 | 0.02 (-0.08;0.12) | 0.02 (-0.08;0.12) | 0.02 (-0.08;0.12) | 0.02 (-0.08;0.12) | 0.02 (-0.08;0.12) | 0.02 (-0.08;0.12) |
| 6 | 0.01 | 5000 | 6 | 0.02 (-0.07;0.12) | 0.03 (-0.06;0.12) | 0.03 (-0.06;0.13) | 0.02 (-0.07;0.12) | 0.03 (-0.06;0.12) | 0.03 (-0.06;0.13) |
| 7 | 0.01 | 5000 | 12 | 0.01 (-0.08;0.1) | 0.01 (-0.08;0.11) | 0.01 (-0.08;0.11) | 0.01 (-0.08;0.10) | 0.01 (-0.08;0.11) | 0.01 (-0.08;0.11) |
| 8 | 0.01 | 5000 | 24 | -0.02 (-0.12;0.07) | -0.03 (-0.13;0.06) | -0.04 (-0.14;0.06) | -0.02 (-0.12;0.07) | -0.03 (-0.13;0.06) | -0.04 (-0.14;0.06) |
| 9 | 0.1 | 2500 | 3 | 0.01 (-0.04;0.05) | 0.01 (-0.04;0.05) | 0.01 (-0.04;0.05) | 0.01 (-0.04;0.05) | 0.01 (-0.04;0.05) | 0.01 (-0.04;0.05) |
| 10 | 0.1 | 2500 | 6 | 0.00 (-0.05;0.04) | 0.00 (-0.05;0.04) | 0.00 (-0.05;0.04) | 0.00 (-0.05;0.04) | 0.00 (-0.05;0.04) | 0.00 (-0.05;0.04) |
| 11 | 0.1 | 2500 | 12 | 0.00 (-0.04;0.05) | 0.00 (-0.04;0.05) | 0.00 (-0.04;0.05) | 0.00 (-0.04;0.05) | 0.00 (-0.04;0.05) | 0.00 (-0.04;0.05) |
| 12 | 0.1 | 2500 | 24 | -0.02 (-0.06;0.03) | -0.02 (-0.06;0.03) | -0.02 (-0.06;0.03) | -0.02 (-0.06;0.03) | -0.02 (-0.06;0.03) | -0.02 (-0.06;0.03) |
| 13 | 0.1 | 5000 | 3 | 0.00 (-0.03;0.03) | 0.00 (-0.03;0.03) | 0.00 (-0.03;0.03) | 0.00 (-0.03;0.03) | 0.00 (-0.03;0.03) | 0.00 (-0.03;0.03) |
| 14 | 0.1 | 5000 | 6 | 0.01 (-0.02;0.05) | 0.01 (-0.02;0.05) | 0.01 (-0.02;0.05) | 0.01 (-0.02;0.05) | 0.01 (-0.02;0.05) | 0.01 (-0.02;0.05) |
| 15 | 0.1 | 5000 | 12 | 0.00 (-0.04;0.03) | 0.00 (-0.04;0.03) | 0.00 (-0.04;0.03) | 0.00 (-0.04;0.03) | 0.00 (-0.04;0.03) | 0.00 (-0.04;0.03) |
| 16 | 0.1 | 5000 | 24 | 0.00 (-0.03;0.03) | 0.00 (-0.03;0.03) | 0.00 (-0.03;0.03) | 0.00 (-0.03;0.03) | 0.00 (-0.03;0.03) | 0.00 (-0.03;0.03) |
| 17 | 0.3 | 2500 | 3 | 0.01 (-0.03;0.04) | 0.01 (-0.03;0.04) | 0.01 (-0.03;0.04) | 0.01 (-0.03;0.04) | 0.01 (-0.03;0.04) | 0.01 (-0.03;0.04) |
| 18 | 0.3 | 2500 | 6 | 0.00 (-0.03;0.03) | 0.00 (-0.03;0.03) | 0.00 (-0.03;0.03) | 0.00 (-0.03;0.03) | 0.00 (-0.03;0.03) | 0.00 (-0.03;0.03) |
| 19 | 0.3 | 2500 | 12 | 0.00 (-0.04;0.03) | 0.00 (-0.04;0.03) | 0.00 (-0.04;0.03) | 0.00 (-0.04;0.03) | 0.00 (-0.04;0.03) | 0.00 (-0.04;0.03) |
| 20 | 0.3 | 2500 | 24 | 0.01 (-0.03;0.04) | 0.01 (-0.03;0.04) | 0.01 (-0.03;0.04) | 0.01 (-0.03;0.04) | 0.01 (-0.03;0.04) | 0.01 (-0.03;0.04) |
| 21 | 0.3 | 5000 | 3 | 0.00 (-0.02;0.02) | 0.00 (-0.02;0.02) | 0.00 (-0.02;0.02) | 0.00 (-0.02;0.02) | 0.00 (-0.02;0.02) | 0.00 (-0.02;0.02) |
| 22 | 0.3 | 5000 | 6 | 0.00 (-0.02;0.03) | 0.00 (-0.02;0.03) | 0.00 (-0.02;0.03) | 0.00 (-0.02;0.03) | 0.00 (-0.02;0.03) | 0.00 (-0.02;0.03) |
| 23 | 0.3 | 5000 | 12 | 0.00 (-0.02;0.02) | 0.00 (-0.02;0.02) | 0.00 (-0.02;0.02) | 0.00 (-0.02;0.02) | 0.00 (-0.02;0.02) | 0.00 (-0.02;0.02) |
| 24 | 0.3 | 5000 | 24 | -0.01 (-0.03;0.01) | -0.01 (-0.03;0.01) | -0.01 (-0.03;0.01) | -0.01 (-0.03;0.01) | -0.01 (-0.03;0.01) | -0.01 (-0.03;0.01) |

Table S6. Test set calibration slope, reported as median (IQR) over the 2000 runs for SLR and Ridge in the 24 simulation scenarios.

| | | | | SLR | | | | RIDGE | | | |
|---|---|---|---|---|---|---|---|---|---|---|---|
| Scenario | EF | N | p | Uncorrected | RUS | ROS | SMOTE | Uncorrected | RUS | ROS | SMOTE |
| 1 | 0.01 | 2500 | 3 | 0.89 (0.77;1.05) | 0.71 (0.54;0.90) | 0.83 (0.70;1.01) | 0.71 (0.60;0.87) | 1.09 (0.90;1.39) | 1.23 (0.88;2.00) | 0.84 (0.71;1.02) | 0.72 (0.60;0.88) |
| 2 | 0.01 | 2500 | 6 | 0.85 (0.74;0.96) | 0.54 (0.42;0.69) | 0.75 (0.64;0.89) | 0.63 (0.54;0.74) | 1.11 (0.93;1.37) | 1.22 (0.86;1.86) | 0.76 (0.64;0.90) | 0.64 (0.55;0.75) |
| 3 | 0.01 | 2500 | 12 | 0.74 (0.66;0.83) | 0.32 (0.22;0.43) | 0.60 (0.50;0.70) | 0.50 (0.43;0.59) | 1.11 (0.94;1.38) | 1.21 (0.85;1.90) | 0.61 (0.51;0.72) | 0.51 (0.44;0.60) |
| 4 | 0.01 | 2500 | 24 | 0.48 (0.41;0.54) | 0.01 (0.01;0.06) | 0.31 (0.26;0.38) | 0.28 (0.23;0.33) | 1.07 (0.85;1.48) | 1.13 (0.69;2.36) | 0.32 (0.26;0.39) | 0.29 (0.24;0.34) |
| 5 | 0.01 | 5000 | 3 | 0.95 (0.85;1.06) | 0.83 (0.69;1.00) | 0.91 (0.81;1.04) | 0.81 (0.71;0.93) | 1.06 (0.94;1.20) | 1.12 (0.91;1.47) | 0.91 (0.81;1.04) | 0.81 (0.71;0.93) |
| 6 | 0.01 | 5000 | 6 | 0.93 (0.85;1.02) | 0.75 (0.63;0.89) | 0.88 (0.78;0.99) | 0.76 (0.67;0.85) | 1.07 (0.97;1.20) | 1.16 (0.93;1.47) | 0.88 (0.79;0.99) | 0.76 (0.68;0.86) |
| 7 | 0.01 | 5000 | 12 | 0.86 (0.79;0.94) | 0.57 (0.48;0.67) | 0.77 (0.70;0.86) | 0.65 (0.59;0.73) | 1.07 (0.97;1.23) | 1.15 (0.92;1.49) | 0.78 (0.70;0.87) | 0.66 (0.59;0.73) |
| 8 | 0.01 | 5000 | 24 | 0.65 (0.60;0.71) | 0.26 (0.20;0.33) | 0.53 (0.46;0.59) | 0.44 (0.39;0.49) | 1.04 (0.91;1.22) | 1.10 (0.83;1.57) | 0.53 (0.47;0.60) | 0.45 (0.39;0.50) |
| 9 | 0.1 | 2500 | 3 | 1.00 (0.94;1.05) | 0.98 (0.91;1.06) | 0.99 (0.93;1.06) | 0.95 (0.89;1.00) | 1.03 (0.97;1.09) | 1.05 (0.96;1.14) | 1.00 (0.94;1.07) | 0.95 (0.89;1.01) |
| 10 | 0.1 | 2500 | 6 | 0.96 (0.92;1.02) | 0.93 (0.86;1.00) | 0.95 (0.90;1.01) | 0.87 (0.82;0.93) | 1.01 (0.96;1.07) | 1.03 (0.95;1.12) | 0.97 (0.91;1.03) | 0.88 (0.83;0.94) |
| 11 | 0.1 | 2500 | 12 | 0.93 (0.89;0.98) | 0.87 (0.81;0.93) | 0.92 (0.87;0.97) | 0.81 (0.77;0.86) | 1.01 (0.96;1.06) | 1.04 (0.95;1.12) | 0.94 (0.88;0.99) | 0.83 (0.78;0.88) |
| 12 | 0.1 | 2500 | 24 | 0.85 (0.81;0.89) | 0.73 (0.68;0.79) | 0.82 (0.77;0.86) | 0.71 (0.67;0.75) | 0.99 (0.94;1.06) | 1.03 (0.93;1.13) | 0.85 (0.80;0.90) | 0.74 (0.70;0.78) |
| 13 | 0.1 | 5000 | 3 | 0.99 (0.96;1.03) | 0.98 (0.93;1.04) | 0.99 (0.95;1.04) | 0.95 (0.92;1.00) | 1.01 (0.97;1.05) | 1.02 (0.96;1.08) | 1.00 (0.95;1.04) | 0.96 (0.92;1.00) |
| 14 | 0.1 | 5000 | 6 | 0.99 (0.96;1.03) | 0.97 (0.92;1.02) | 0.99 (0.95;1.03) | 0.91 (0.87;0.95) | 1.02 (0.98;1.05) | 1.02 (0.96;1.08) | 0.99 (0.95;1.04) | 0.92 (0.88;0.96) |
| 15 | 0.1 | 5000 | 12 | 0.97 (0.94;1.00) | 0.93 (0.89;0.98) | 0.96 (0.92;1.00) | 0.86 (0.82;0.89) | 1.01 (0.97;1.05) | 1.02 (0.97;1.08) | 0.97 (0.93;1.01) | 0.87 (0.83;0.90) |
| 16 | 0.1 | 5000 | 24 | 0.93 (0.90;0.96) | 0.86 (0.82;0.90) | 0.91 (0.88;0.94) | 0.79 (0.76;0.82) | 1.00 (0.97;1.04) | 1.03 (0.97;1.09) | 0.93 (0.89;0.96) | 0.81 (0.78;0.84) |
| 17 | 0.3 | 2500 | 3 | 0.99 (0.96;1.03) | 0.99 (0.95;1.04) | 0.99 (0.95;1.04) | 0.98 (0.94;1.01) | 1.01 (0.97;1.05) | 1.01 (0.97;1.06) | 1.00 (0.96;1.05) | 0.99 (0.95;1.02) |
| 18 | 0.3 | 2500 | 6 | 0.99 (0.96;1.03) | 0.99 (0.94;1.03) | 0.99 (0.95;1.03) | 0.95 (0.92;0.99) | 1.02 (0.98;1.06) | 1.02 (0.97;1.07) | 1.00 (0.97;1.05) | 0.97 (0.93;1.01) |
| 19 | 0.3 | 2500 | 12 | 0.97 (0.93;1.00) | 0.95 (0.91;0.99) | 0.96 (0.92;0.99) | 0.90 (0.87;0.94) | 1.01 (0.97;1.04) | 1.01 (0.97;1.06) | 0.98 (0.95;1.02) | 0.93 (0.90;0.97) |
| 20 | 0.3 | 2500 | 24 | 0.94 (0.91;0.98) | 0.91 (0.88;0.96) | 0.93 (0.89;0.97) | 0.87 (0.83;0.90) | 1.03 (0.99;1.07) | 1.04 (0.98;1.09) | 0.98 (0.94;1.02) | 0.92 (0.88;0.96) |
| 21 | 0.3 | 5000 | 3 | 1.00 (0.98;1.03) | 1.00 (0.97;1.03) | 1.00 (0.97;1.03) | 0.99 (0.96;1.02) | 1.01 (0.98;1.04) | 1.01 (0.98;1.04) | 1.01 (0.98;1.04) | 1.00 (0.97;1.02) |
| 22 | 0.3 | 5000 | 6 | 0.98 (0.95;1.01) | 0.98 (0.94;1.01) | 0.98 (0.95;1.01) | 0.94 (0.92;0.97) | 0.99 (0.96;1.02) | 0.99 (0.96;1.02) | 0.99 (0.96;1.02) | 0.95 (0.93;0.98) |
| 23 | 0.3 | 5000 | 12 | 0.99 (0.96;1.01) | 0.98 (0.95;1.01) | 0.98 (0.96;1.01) | 0.93 (0.90;0.95) | 1.01 (0.98;1.03) | 1.01 (0.98;1.04) | 1.00 (0.97;1.02) | 0.94 (0.92;0.97) |
| 24 | 0.3 | 5000 | 24 | 0.97 (0.95;1.00) | 0.96 (0.93;0.99) | 0.97 (0.94;0.99) | 0.90 (0.88;0.93) | 1.02 (0.99;1.05) | 1.02 (0.98;1.06) | 0.99 (0.96;1.02) | 0.93 (0.90;0.96) |

Table S7. Test set sensitivity reported as median (IQR) over the 2000 runs for SLR and Ridge in the 24 simulation scenarios. For uncorrected training sets, we used either the default threshold of 0.5 or a threshold based on the true event fraction. For RUS/ROS/SMOTE, the default threshold of 0.5 was used.

| Scenario | EF | N | p | Imbalance correction method (threshold) | | | | |
|---|---|---|---|---|---|---|---|---|
| | | | | Uncorrected (0.5) | Uncorrected (EF) | RUS (0.5) | ROS (0.5) | SMOTE (0.5) |
| | | | | SLR | | | | |
| 1 | 0.01 | 2500 | 3 | 0 (0;0) | 0.60 (0.53;0.67) | 0.64 (0.60;0.68) | 0.63 (0.59;0.66) | 0.62 (0.59;0.65) |
| 2 | 0.01 | 2500 | 6 | 0 (0;0) | 0.66 (0.60;0.73) | 0.66 (0.62;0.71) | 0.65 (0.61;0.67) | 0.63 (0.60;0.66) |
| 3 | 0.01 | 2500 | 12 | 0 (0;0) | 0.67 (0.61;0.72) | 0.65 (0.61;0.69) | 0.62 (0.58;0.65) | 0.60 (0.56;0.64) |
| 4 | 0.01 | 2500 | 24 | 0 (0;0) | 0.52 (0.46;0.59) | 0.58 (0.53;0.63) | 0.48 (0.44;0.52) | 0.46 (0.41;0.50) |
| 5 | 0.01 | 5000 | 3 | 0 (0;0) | 0.62 (0.57;0.67) | 0.65 (0.62;0.68) | 0.65 (0.62;0.67) | 0.64 (0.62;0.66) |
| 6 | 0.01 | 5000 | 6 | 0 (0;0) | 0.69 (0.65;0.73) | 0.68 (0.65;0.71) | 0.67 (0.65;0.69) | 0.66 (0.64;0.68) |
| 7 | 0.01 | 5000 | 12 | 0 (0;0) | 0.69 (0.65;0.72) | 0.67 (0.64;0.69) | 0.65 (0.63;0.67) | 0.64 (0.61;0.66) |
| 8 | 0.01 | 5000 | 24 | 0 (0;0) | 0.60 (0.56;0.65) | 0.62 (0.59;0.65) | 0.57 (0.55;0.60) | 0.55 (0.52;0.58) |
| 9 | 0.1 | 2500 | 3 | 0.02 (0.01;0.02) | 0.67 (0.65;0.69) | 0.67 (0.66;0.68) | 0.67 (0.66;0.67) | 0.64 (0.63;0.66) |
| 10 | 0.1 | 2500 | 6 | 0.03 (0.02;0.04) | 0.70 (0.68;0.71) | 0.68 (0.67;0.69) | 0.68 (0.67;0.68) | 0.65 (0.64;0.67) |
| 11 | 0.1 | 2500 | 12 | 0.05 (0.04;0.06) | 0.71 (0.69;0.72) | 0.68 (0.67;0.69) | 0.68 (0.67;0.69) | 0.66 (0.64;0.67) |
| 12 | 0.1 | 2500 | 24 | 0.04 (0.03;0.05) | 0.67 (0.65;0.69) | 0.66 (0.65;0.67) | 0.65 (0.64;0.66) | 0.62 (0.61;0.63) |
| 13 | 0.1 | 5000 | 3 | 0.02 (0.01;0.02) | 0.67 (0.65;0.68) | 0.67 (0.66;0.68) | 0.67 (0.66;0.67) | 0.64 (0.63;0.66) |
| 14 | 0.1 | 5000 | 6 | 0.03 (0.03;0.04) | 0.70 (0.69;0.72) | 0.68 (0.68;0.69) | 0.68 (0.68;0.69) | 0.66 (0.65;0.67) |
| 15 | 0.1 | 5000 | 12 | 0.05 (0.04;0.05) | 0.71 (0.70;0.73) | 0.69 (0.68;0.69) | 0.69 (0.68;0.69) | 0.66 (0.65;0.67) |
| 16 | 0.1 | 5000 | 24 | 0.03 (0.03;0.04) | 0.68 (0.67;0.70) | 0.67 (0.66;0.68) | 0.67 (0.66;0.67) | 0.64 (0.63;0.65) |
| 17 | 0.3 | 2500 | 3 | 0.31 (0.30;0.33) | 0.68 (0.66;0.69) | 0.67 (0.67;0.68) | 0.67 (0.67;0.68) | 0.61 (0.60;0.63) |
| 18 | 0.3 | 2500 | 6 | 0.32 (0.31;0.34) | 0.68 (0.67;0.69) | 0.67 (0.67;0.68) | 0.67 (0.67;0.68) | 0.62 (0.61;0.63) |
| 19 | 0.3 | 2500 | 12 | 0.35 (0.33;0.36) | 0.69 (0.68;0.70) | 0.68 (0.67;0.69) | 0.68 (0.67;0.69) | 0.63 (0.62;0.64) |
| 20 | 0.3 | 2500 | 24 | 0.33 (0.31;0.34) | 0.68 (0.66;0.69) | 0.67 (0.67;0.68) | 0.67 (0.66;0.67) | 0.61 (0.60;0.63) |
| 21 | 0.3 | 5000 | 3 | 0.31 (0.30;0.32) | 0.68 (0.67;0.69) | 0.68 (0.67;0.68) | 0.68 (0.67;0.68) | 0.61 (0.60;0.62) |
| 22 | 0.3 | 5000 | 6 | 0.32 (0.31;0.33) | 0.68 (0.67;0.69) | 0.67 (0.67;0.68) | 0.67 (0.67;0.68) | 0.61 (0.61;0.62) |
| 23 | 0.3 | 5000 | 12 | 0.35 (0.34;0.36) | 0.69 (0.68;0.70) | 0.68 (0.68;0.69) | 0.68 (0.68;0.69) | 0.63 (0.63;0.64) |
| 24 | 0.3 | 5000 | 24 | 0.33 (0.31;0.34) | 0.68 (0.67;0.69) | 0.68 (0.67;0.68) | 0.68 (0.67;0.68) | 0.62 (0.61;0.63) |
| | | | | RIDGE | | | | |
| 1 | 0.01 | 2500 | 3 | 0 (0;0) | 0.62 (0.54;0.70) | 0.64 (0.60;0.68) | 0.63 (0.59;0.66) | 0.62 (0.59;0.65) |
| 2 | 0.01 | 2500 | 6 | 0 (0;0) | 0.71 (0.64;0.78) | 0.67 (0.63;0.71) | 0.65 (0.61;0.67) | 0.63 (0.60;0.66) |
| 3 | 0.01 | 2500 | 12 | 0 (0;0) | 0.75 (0.68;0.81) | 0.67 (0.63;0.71) | 0.62 (0.58;0.65) | 0.60 (0.56;0.64) |
| 4 | 0.01 | 2500 | 24 | 0 (0;0) | 0.67 (0.57;0.76) | 0.61 (0.57;0.65) | 0.48 (0.44;0.52) | 0.46 (0.41;0.50) |
| 5 | 0.01 | 5000 | 3 | 0 (0;0) | 0.63 (0.58;0.69) | 0.65 (0.62;0.68) | 0.65 (0.62;0.67) | 0.64 (0.62;0.66) |
| 6 | 0.01 | 5000 | 6 | 0 (0;0) | 0.72 (0.68;0.75) | 0.68 (0.65;0.71) | 0.67 (0.65;0.69) | 0.66 (0.64;0.68) |
| 7 | 0.01 | 5000 | 12 | 0 (0;0) | 0.73 (0.69;0.77) | 0.67 (0.65;0.70) | 0.65 (0.63;0.67) | 0.64 (0.61;0.66) |
| 8 | 0.01 | 5000 | 24 | 0 (0;0) | 0.68 (0.63;0.74) | 0.63 (0.60;0.66) | 0.57 (0.55;0.60) | 0.55 (0.52;0.58) |
| 9 | 0.1 | 2500 | 3 | 0.01 (0.01;0.02) | 0.67 (0.65;0.69) | 0.67 (0.66;0.68) | 0.67 (0.66;0.67) | 0.64 (0.63;0.66) |
| 10 | 0.1 | 2500 | 6 | 0.03 (0.02;0.03) | 0.70 (0.68;0.72) | 0.68 (0.67;0.69) | 0.68 (0.67;0.68) | 0.65 (0.64;0.67) |
| 11 | 0.1 | 2500 | 12 | 0.03 (0.03;0.04) | 0.72 (0.70;0.74) | 0.68 (0.67;0.69) | 0.68 (0.67;0.69) | 0.65 (0.64;0.67) |
| 12 | 0.1 | 2500 | 24 | 0.02 (0.01;0.02) | 0.69 (0.67;0.71) | 0.66 (0.65;0.67) | 0.65 (0.64;0.66) | 0.62 (0.61;0.63) |
| 13 | 0.1 | 5000 | 3 | 0.02 (0.01;0.02) | 0.67 (0.66;0.68) | 0.67 (0.66;0.68) | 0.67 (0.66;0.67) | 0.64 (0.63;0.66) |
| 14 | 0.1 | 5000 | 6 | 0.03 (0.02;0.04) | 0.70 (0.69;0.72) | 0.68 (0.68;0.69) | 0.68 (0.68;0.69) | 0.66 (0.65;0.67) |
| 15 | 0.1 | 5000 | 12 | 0.04 (0.03;0.05) | 0.72 (0.71;0.73) | 0.69 (0.68;0.69) | 0.69 (0.68;0.69) | 0.66 (0.65;0.67) |
| 16 | 0.1 | 5000 | 24 | 0.02 (0.02;0.03) | 0.69 (0.68;0.71) | 0.67 (0.66;0.68) | 0.67 (0.66;0.67) | 0.64 (0.63;0.65) |
| 17 | 0.3 | 2500 | 3 | 0.31 (0.29;0.33) | 0.68 (0.66;0.69) | 0.67 (0.67;0.68) | 0.67 (0.67;0.68) | 0.61 (0.60;0.63) |
| 18 | 0.3 | 2500 | 6 | 0.32 (0.30;0.33) | 0.68 (0.67;0.69) | 0.67 (0.67;0.68) | 0.67 (0.67;0.68) | 0.62 (0.60;0.63) |
| 19 | 0.3 | 2500 | 12 | 0.34 (0.32;0.35) | 0.69 (0.68;0.71) | 0.68 (0.67;0.69) | 0.68 (0.67;0.69) | 0.63 (0.62;0.64) |
| 20 | 0.3 | 2500 | 24 | 0.30 (0.28;0.32) | 0.68 (0.67;0.69) | 0.67 (0.67;0.68) | 0.67 (0.66;0.67) | 0.61 (0.60;0.63) |
| 21 | 0.3 | 5000 | 3 | 0.31 (0.30;0.32) | 0.68 (0.67;0.69) | 0.68 (0.67;0.68) | 0.68 (0.67;0.68) | 0.61 (0.60;0.62) |
| 22 | 0.3 | 5000 | 6 | 0.31 (0.30;0.32) | 0.68 (0.67;0.69) | 0.67 (0.67;0.68) | 0.67 (0.67;0.68) | 0.61 (0.61;0.62) |
| 23 | 0.3 | 5000 | 12 | 0.34 (0.33;0.35) | 0.69 (0.69;0.70) | 0.68 (0.68;0.69) | 0.68 (0.68;0.69) | 0.63 (0.63;0.64) |
| 24 | 0.3 | 5000 | 24 | 0.31 (0.30;0.33) | 0.68 (0.68;0.69) | 0.68 (0.67;0.68) | 0.68 (0.67;0.68) | 0.62 (0.61;0.63) |

Table S8. Test set specificity reported as median (IQR) over the 2000 runs for SLR and Ridge in the 24 simulation scenarios. For uncorrected training sets, we used either the default threshold of 0.5 or a threshold based on the true event fraction. For RUS/ROS/SMOTE, the default threshold of 0.5 was used.

| Scenario | EF | N | p | Imbalance correction method (threshold) | | | | |
|---|---|---|---|---|---|---|---|---|
| | | | | Uncorrected (0.5) | Uncorrected (EF) | RUS (0.5) | ROS (0.5) | SMOTE (0.5) |
| | | | | SLR | | | | |
| 1 | 0.01 | 2500 | 3 | 1 (1;1) | 0.70 (0.64;0.76) | 0.65 (0.60;0.68) | 0.67 (0.65;0.70) | 0.68 (0.65;0.71) |
| 2 | 0.01 | 2500 | 6 | 1 (1;1) | 0.69 (0.63;0.74) | 0.65 (0.60;0.69) | 0.70 (0.68;0.72) | 0.71 (0.68;0.74) |
| 3 | 0.01 | 2500 | 12 | 1 (1;1) | 0.70 (0.65;0.74) | 0.63 (0.59;0.67) | 0.73 (0.70;0.75) | 0.74 (0.72;0.76) |
| 4 | 0.01 | 2500 | 24 | 1 (1;1) | 0.74 (0.69;0.78) | 0.57 (0.52;0.62) | 0.75 (0.73;0.77) | 0.77 (0.75;0.80) |
| 5 | 0.01 | 5000 | 3 | 1 (1;1) | 0.70 (0.65;0.74) | 0.66 (0.63;0.68) | 0.67 (0.66;0.69) | 0.68 (0.66;0.69) |
| 6 | 0.01 | 5000 | 6 | 1 (1;1) | 0.68 (0.64;0.72) | 0.67 (0.64;0.69) | 0.69 (0.68;0.71) | 0.70 (0.68;0.72) |
| 7 | 0.01 | 5000 | 12 | 1 (1;1) | 0.68 (0.64;0.72) | 0.66 (0.63;0.68) | 0.71 (0.69;0.73) | 0.72 (0.70;0.74) |
| 8 | 0.01 | 5000 | 24 | 1 (1;1) | 0.71 (0.67;0.74) | 0.62 (0.59;0.65) | 0.72 (0.70;0.74) | 0.74 (0.72;0.75) |
| 9 | 0.1 | 2500 | 3 | 1 (1;1) | 0.67 (0.64;0.68) | 0.66 (0.65;0.67) | 0.66 (0.66;0.67) | 0.69 (0.68;0.70) |
| 10 | 0.1 | 2500 | 6 | 1 (1;1) | 0.66 (0.64;0.68) | 0.67 (0.66;0.69) | 0.68 (0.67;0.69) | 0.70 (0.69;0.71) |
| 11 | 0.1 | 2500 | 12 | 1 (0.99;1) | 0.66 (0.64;0.68) | 0.68 (0.67;0.69) | 0.69 (0.68;0.70) | 0.71 (0.70;0.72) |
| 12 | 0.1 | 2500 | 24 | 1 (0.99;1) | 0.66 (0.65;0.68) | 0.66 (0.65;0.67) | 0.68 (0.67;0.69) | 0.70 (0.69;0.71) |
| 13 | 0.1 | 5000 | 3 | 1 (1;1) | 0.66 (0.65;0.68) | 0.66 (0.65;0.67) | 0.66 (0.66;0.67) | 0.69 (0.68;0.70) |
| 14 | 0.1 | 5000 | 6 | 1 (1;1) | 0.66 (0.65;0.68) | 0.68 (0.67;0.68) | 0.68 (0.67;0.68) | 0.70 (0.69;0.71) |
| 15 | 0.1 | 5000 | 12 | 1 (0.99;1) | 0.66 (0.64;0.67) | 0.68 (0.67;0.69) | 0.69 (0.68;0.69) | 0.71 (0.70;0.72) |
| 16 | 0.1 | 5000 | 24 | 1 (1;1) | 0.66 (0.65;0.68) | 0.67 (0.66;0.68) | 0.68 (0.67;0.69) | 0.70 (0.69;0.71) |
| 17 | 0.3 | 2500 | 3 | 0.92 (0.91;0.92) | 0.67 (0.66;0.68) | 0.67 (0.67;0.68) | 0.67 (0.67;0.68) | 0.73 (0.72;0.74) |
| 18 | 0.3 | 2500 | 6 | 0.92 (0.91;0.92) | 0.67 (0.66;0.68) | 0.67 (0.67;0.68) | 0.67 (0.67;0.68) | 0.73 (0.72;0.74) |
| 19 | 0.3 | 2500 | 12 | 0.91 (0.90;0.91) | 0.67 (0.66;0.68) | 0.67 (0.67;0.68) | 0.68 (0.67;0.68) | 0.72 (0.71;0.73) |
| 20 | 0.3 | 2500 | 24 | 0.91 (0.90;0.92) | 0.67 (0.65;0.68) | 0.66 (0.66;0.67) | 0.67 (0.66;0.68) | 0.72 (0.71;0.73) |
| 21 | 0.3 | 5000 | 3 | 0.92 (0.91;0.92) | 0.67 (0.66;0.68) | 0.67 (0.67;0.68) | 0.67 (0.67;0.68) | 0.73 (0.72;0.74) |
| 22 | 0.3 | 5000 | 6 | 0.91 (0.91;0.92) | 0.67 (0.66;0.67) | 0.67 (0.67;0.67) | 0.67 (0.67;0.67) | 0.72 (0.72;0.73) |
| 23 | 0.3 | 5000 | 12 | 0.91 (0.90;0.91) | 0.67 (0.66;0.68) | 0.68 (0.67;0.68) | 0.68 (0.67;0.68) | 0.72 (0.72;0.73) |
| 24 | 0.3 | 5000 | 24 | 0.91 (0.91;0.92) | 0.67 (0.66;0.68) | 0.67 (0.66;0.67) | 0.67 (0.67;0.68) | 0.72 (0.72;0.73) |
| | | | | RIDGE | | | | |
| 1 | 0.01 | 2500 | 3 | 1 (1;1) | 0.68 (0.61;0.75) | 0.65 (0.60;0.68) | 0.67 (0.65;0.70) | 0.68 (0.65;0.71) |
| 2 | 0.01 | 2500 | 6 | 1 (1;1) | 0.65 (0.57;0.71) | 0.65 (0.61;0.69) | 0.70 (0.68;0.72) | 0.71 (0.68;0.74) |
| 3 | 0.01 | 2500 | 12 | 1 (1;1) | 0.62 (0.54;0.69) | 0.64 (0.60;0.68) | 0.73 (0.70;0.75) | 0.74 (0.72;0.76) |
| 4 | 0.01 | 2500 | 24 | 1 (1;1) | 0.61 (0.52;0.69) | 0.61 (0.57;0.64) | 0.75 (0.73;0.77) | 0.77 (0.75;0.80) |
| 5 | 0.01 | 5000 | 3 | 1 (1;1) | 0.68 (0.63;0.73) | 0.66 (0.63;0.68) | 0.67 (0.66;0.69) | 0.68 (0.66;0.69) |
| 6 | 0.01 | 5000 | 6 | 1 (1;1) | 0.65 (0.61;0.69) | 0.67 (0.64;0.69) | 0.69 (0.68;0.71) | 0.70 (0.68;0.72) |
| 7 | 0.01 | 5000 | 12 | 1 (1;1) | 0.63 (0.59;0.68) | 0.66 (0.64;0.69) | 0.71 (0.69;0.73) | 0.72 (0.70;0.74) |
| 8 | 0.01 | 5000 | 24 | 1 (1;1) | 0.63 (0.57;0.68) | 0.63 (0.60;0.66) | 0.72 (0.70;0.74) | 0.74 (0.72;0.75) |
| 9 | 0.1 | 2500 | 3 | 1 (1;1) | 0.66 (0.64;0.68) | 0.66 (0.65;0.67) | 0.66 (0.66;0.67) | 0.69 (0.68;0.70) |
| 10 | 0.1 | 2500 | 6 | 1 (1;1) | 0.66 (0.64;0.68) | 0.67 (0.66;0.69) | 0.68 (0.67;0.69) | 0.70 (0.69;0.71) |
| 11 | 0.1 | 2500 | 12 | 1 (1;1) | 0.65 (0.63;0.67) | 0.68 (0.67;0.69) | 0.69 (0.68;0.70) | 0.71 (0.70;0.72) |
| 12 | 0.1 | 2500 | 24 | 1 (1;1) | 0.64 (0.62;0.66) | 0.66 (0.65;0.67) | 0.68 (0.67;0.69) | 0.70 (0.69;0.72) |
| 13 | 0.1 | 5000 | 3 | 1 (1;1) | 0.66 (0.65;0.68) | 0.66 (0.65;0.67) | 0.66 (0.66;0.67) | 0.69 (0.68;0.70) |
| 14 | 0.1 | 5000 | 6 | 1 (1;1) | 0.66 (0.64;0.67) | 0.68 (0.67;0.68) | 0.68 (0.67;0.68) | 0.70 (0.69;0.71) |
| 15 | 0.1 | 5000 | 12 | 1 (1;1) | 0.65 (0.64;0.66) | 0.68 (0.67;0.69) | 0.69 (0.68;0.69) | 0.71 (0.70;0.72) |
| 16 | 0.1 | 5000 | 24 | 1 (1;1) | 0.65 (0.64;0.67) | 0.67 (0.66;0.68) | 0.68 (0.67;0.69) | 0.70 (0.69;0.71) |
| 17 | 0.3 | 2500 | 3 | 0.92 (0.91;0.93) | 0.67 (0.66;0.68) | 0.67 (0.67;0.68) | 0.67 (0.67;0.68) | 0.73 (0.72;0.74) |
| 18 | 0.3 | 2500 | 6 | 0.92 (0.91;0.93) | 0.67 (0.66;0.68) | 0.67 (0.67;0.68) | 0.67 (0.67;0.68) | 0.73 (0.72;0.74) |
| 19 | 0.3 | 2500 | 12 | 0.91 (0.90;0.92) | 0.67 (0.65;0.68) | 0.67 (0.67;0.68) | 0.68 (0.67;0.68) | 0.72 (0.71;0.74) |
| 20 | 0.3 | 2500 | 24 | 0.92 (0.91;0.93) | 0.66 (0.65;0.68) | 0.66 (0.66;0.67) | 0.67 (0.66;0.68) | 0.72 (0.71;0.74) |
| 21 | 0.3 | 5000 | 3 | 0.92 (0.91;0.92) | 0.67 (0.66;0.68) | 0.67 (0.67;0.68) | 0.67 (0.67;0.68) | 0.73 (0.72;0.74) |
| 22 | 0.3 | 5000 | 6 | 0.92 (0.91;0.92) | 0.67 (0.66;0.67) | 0.67 (0.67;0.67) | 0.67 (0.67;0.67) | 0.72 (0.72;0.73) |
| 23 | 0.3 | 5000 | 12 | 0.91 (0.91;0.92) | 0.67 (0.66;0.68) | 0.68 (0.67;0.68) | 0.68 (0.67;0.68) | 0.72 (0.72;0.73) |
| 24 | 0.3 | 5000 | 24 | 0.92 (0.91;0.92) | 0.66 (0.65;0.67) | 0.67 (0.66;0.67) | 0.67 (0.67;0.68) | 0.72 (0.72;0.73) |

Figure S1. Visualization of imbalance correction methods.

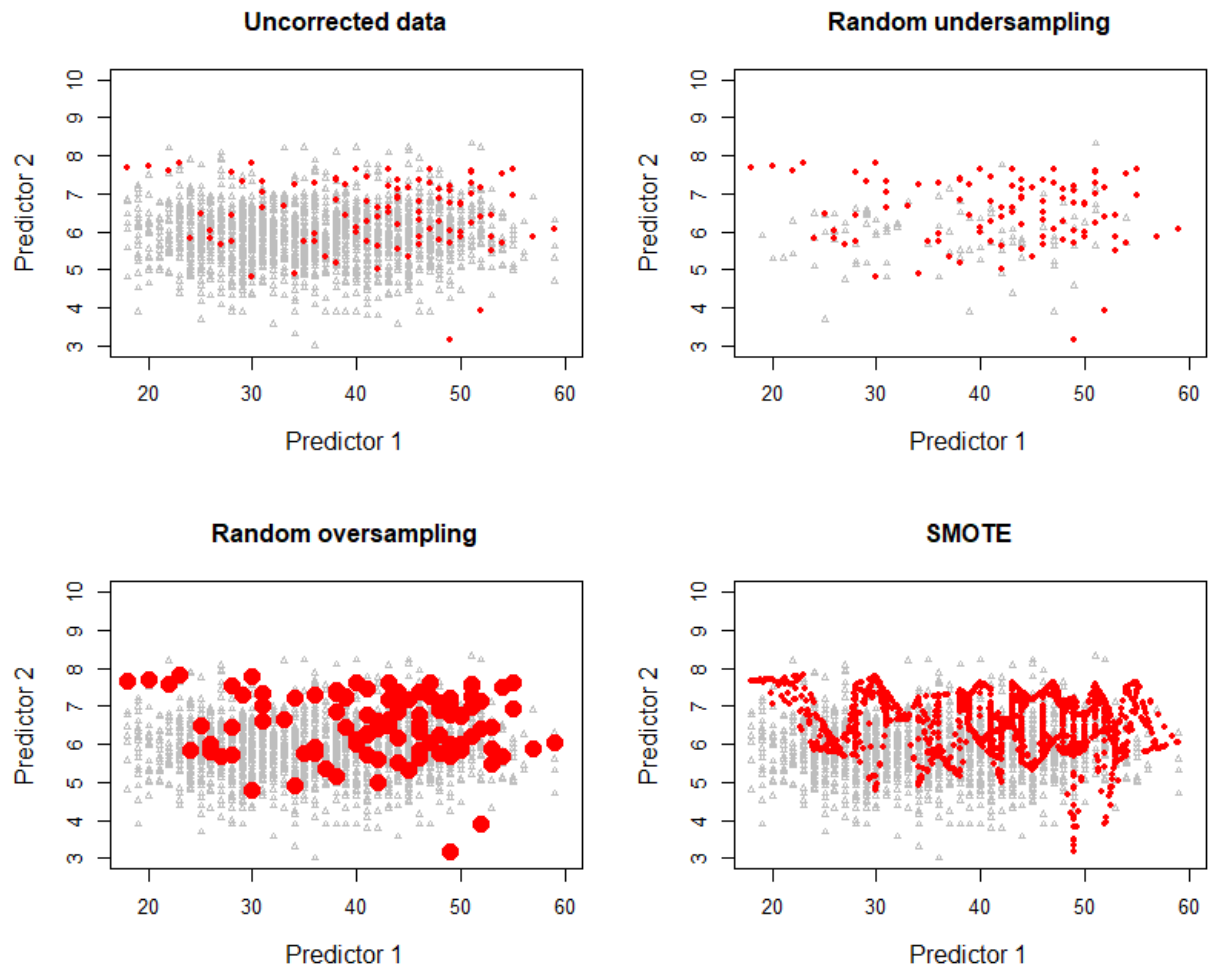

Figure S2. Test set AUROC for the Ridge models in the simulation scenarios with an event fraction of 10%.

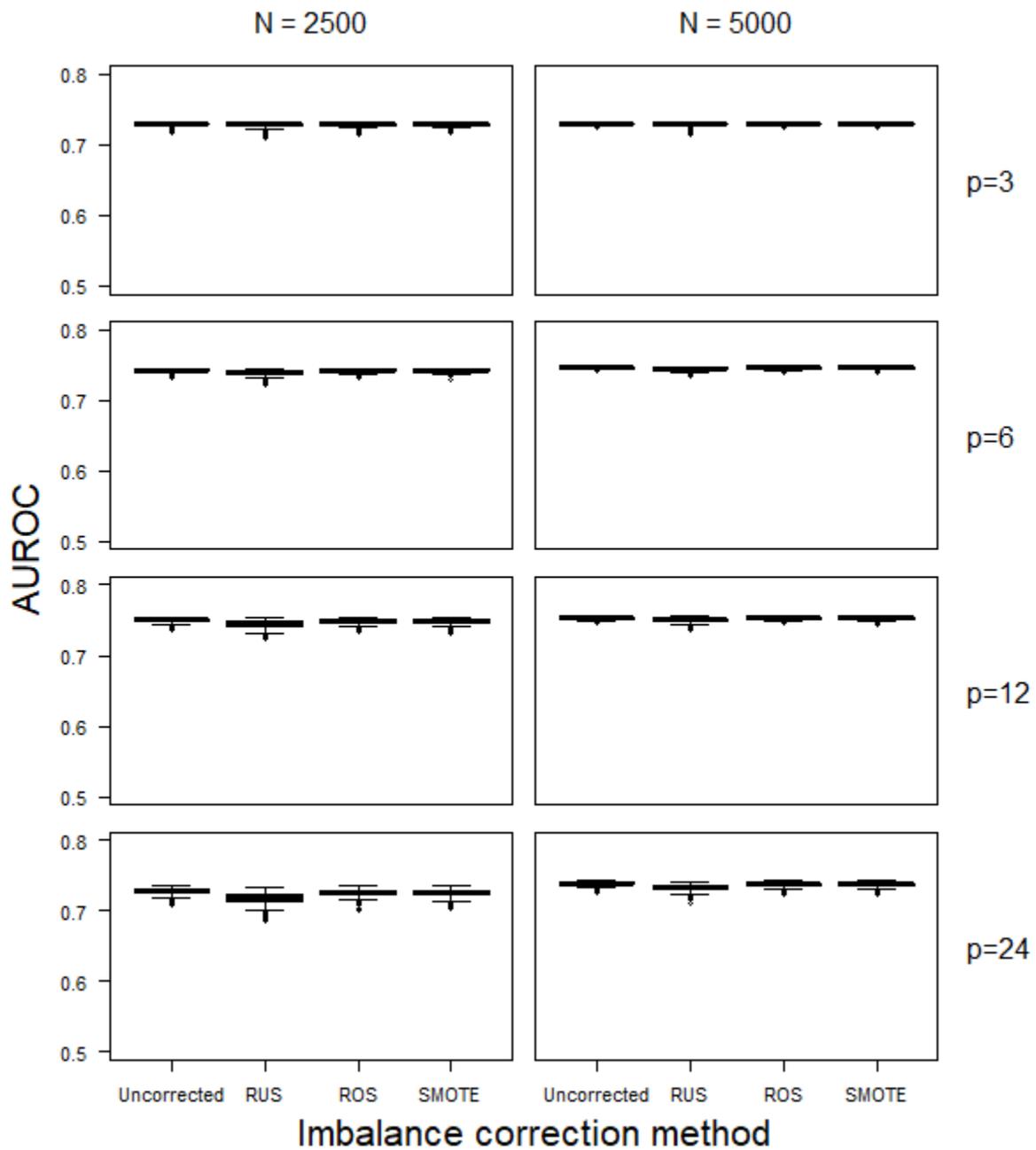

Figure S3. Test set AUROC for the Ridge models in the simulation scenarios with an event fraction of 30%.

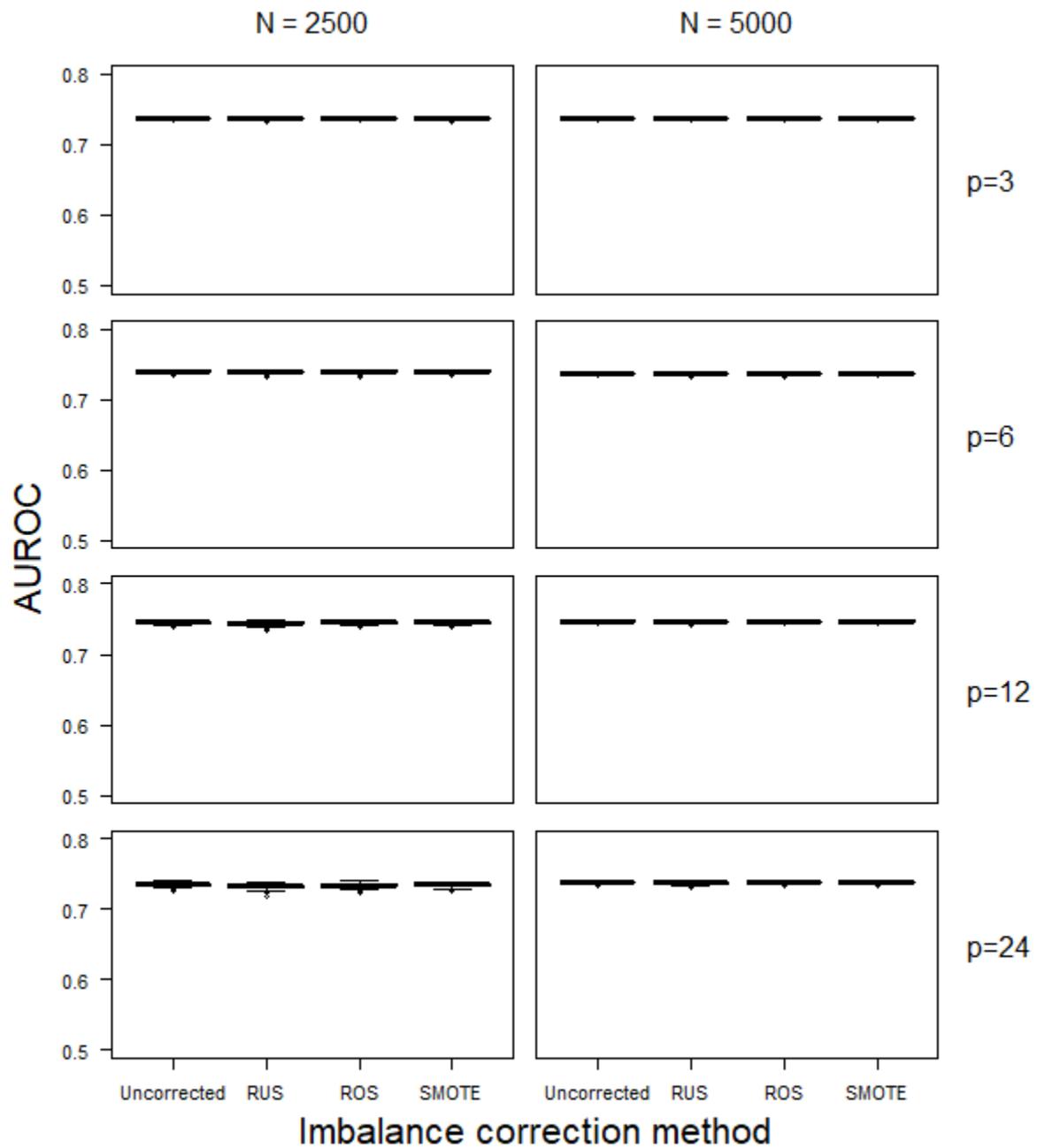

Figure S4. Test set AUROC for the SLR models in the simulation scenarios with an event fraction of 1%.

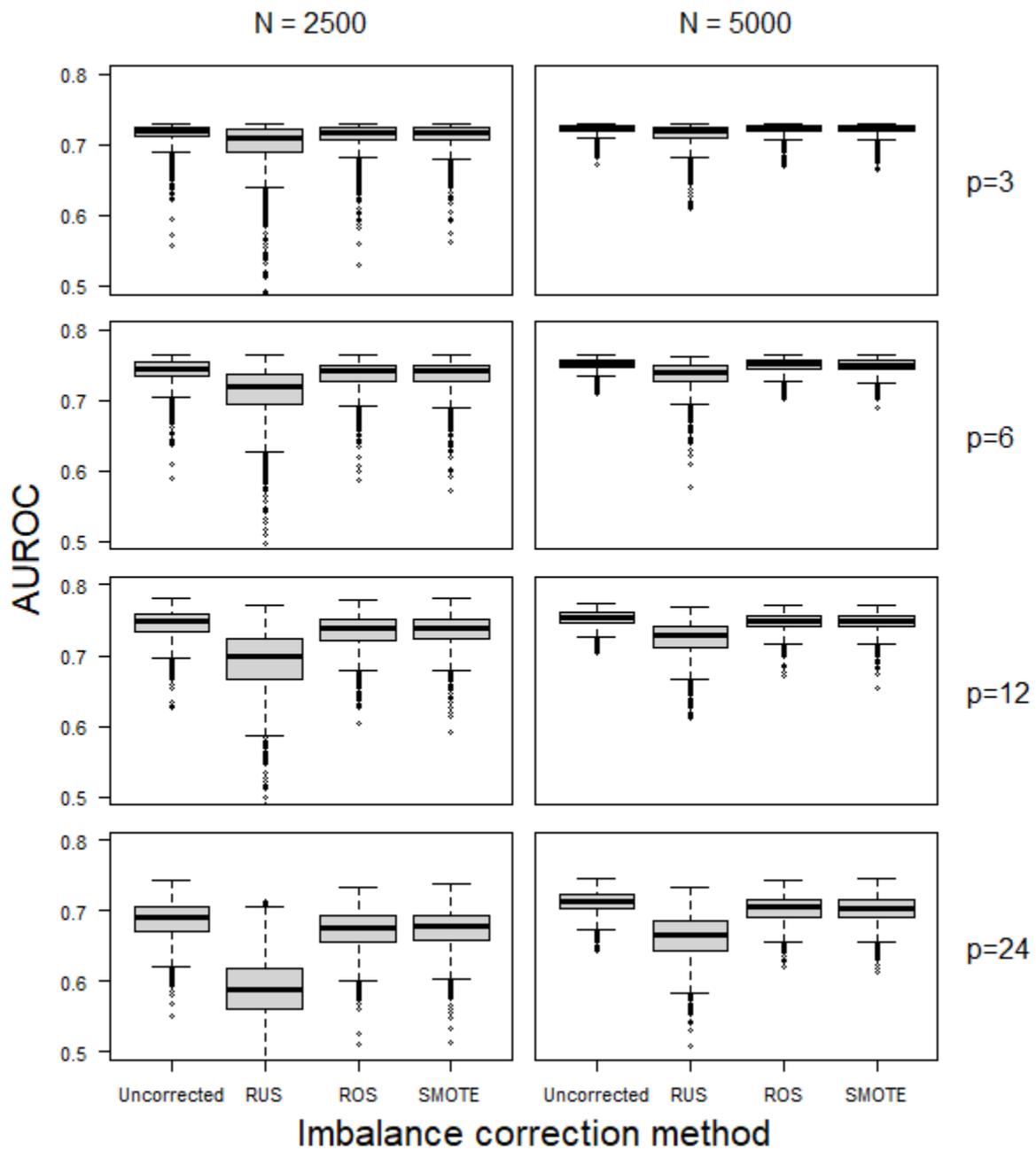

Figure S5. Test set AUROC for the SLR models in the simulation scenarios with an event fraction of 10%.

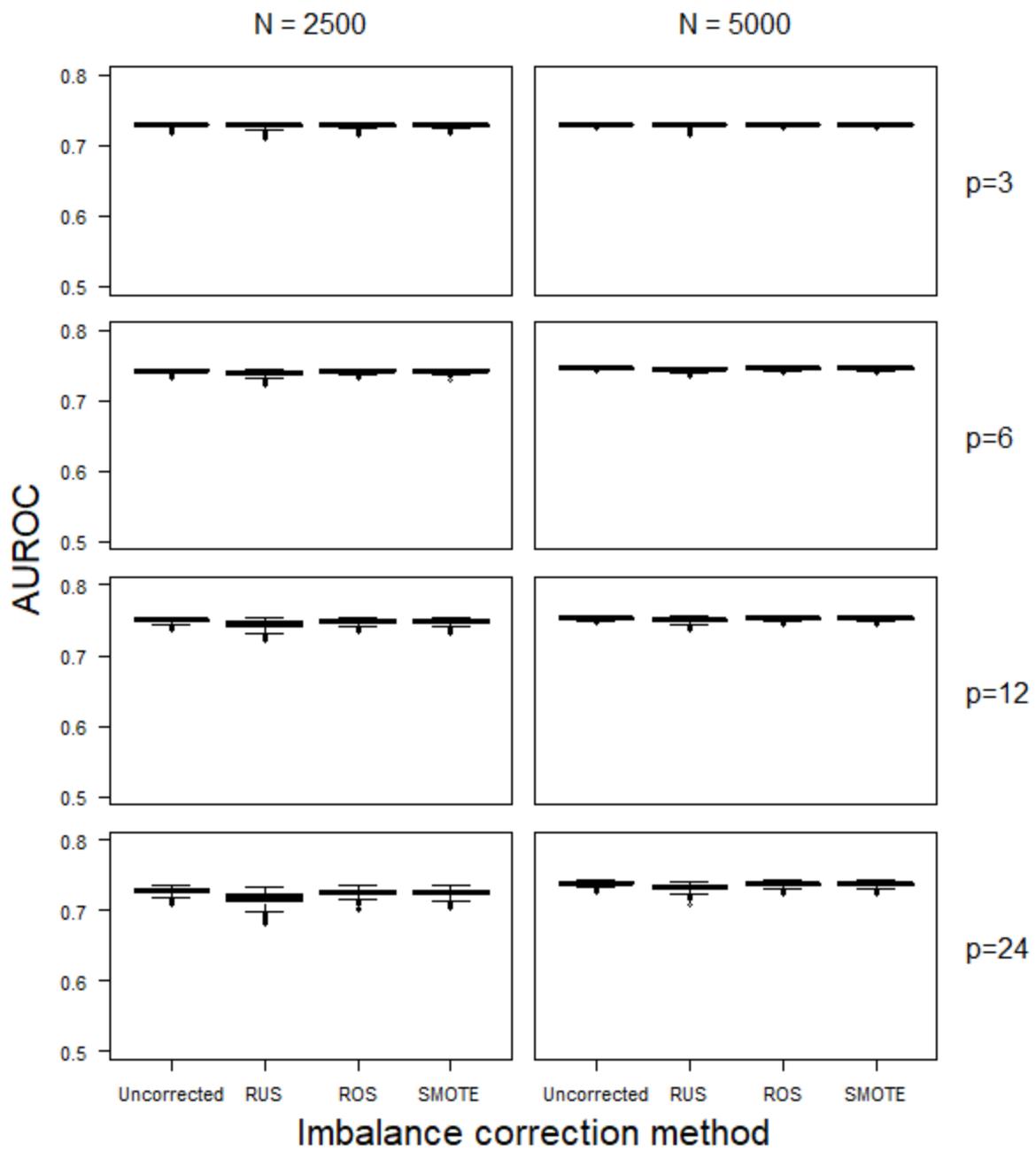

Figure S6. Test set AUROC for the SLR models in the simulation scenarios with an event fraction of 30%.

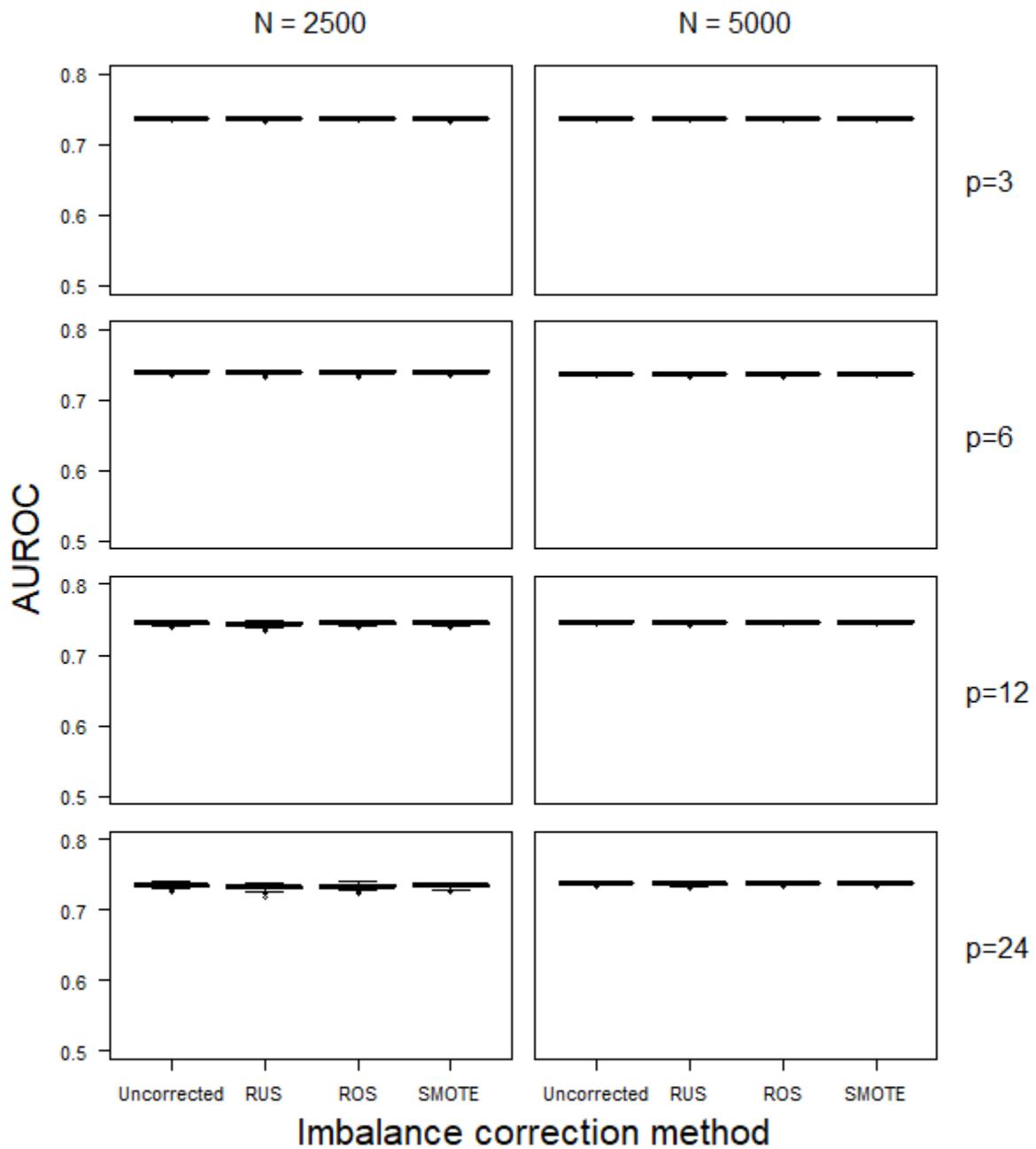

Figure S7. Test set calibration intercept for the Ridge models in the simulation scenarios with an event fraction of 10%.

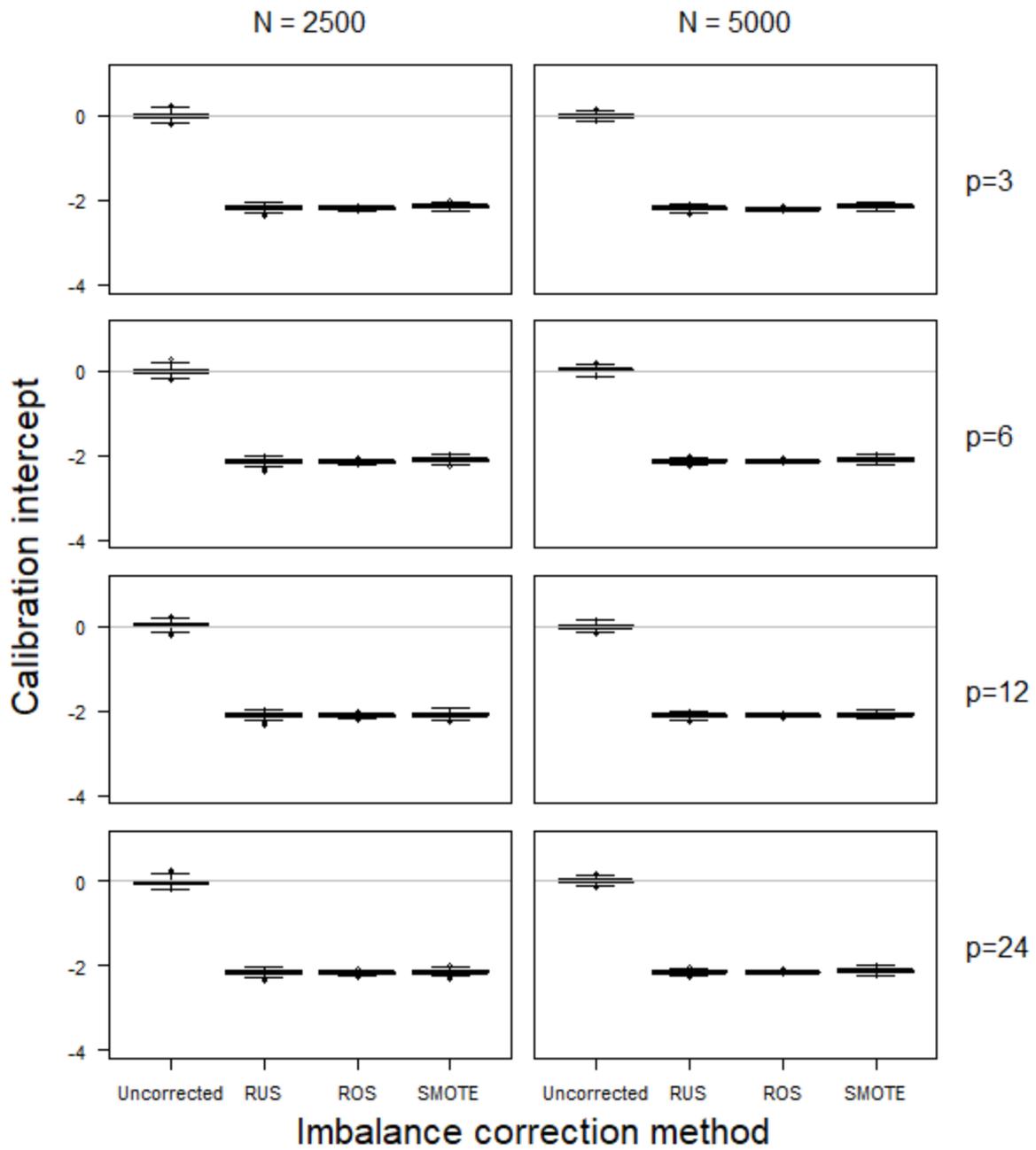

Figure S8. Test set calibration intercept for the Ridge models in the simulation scenarios with an event fraction of 30%.

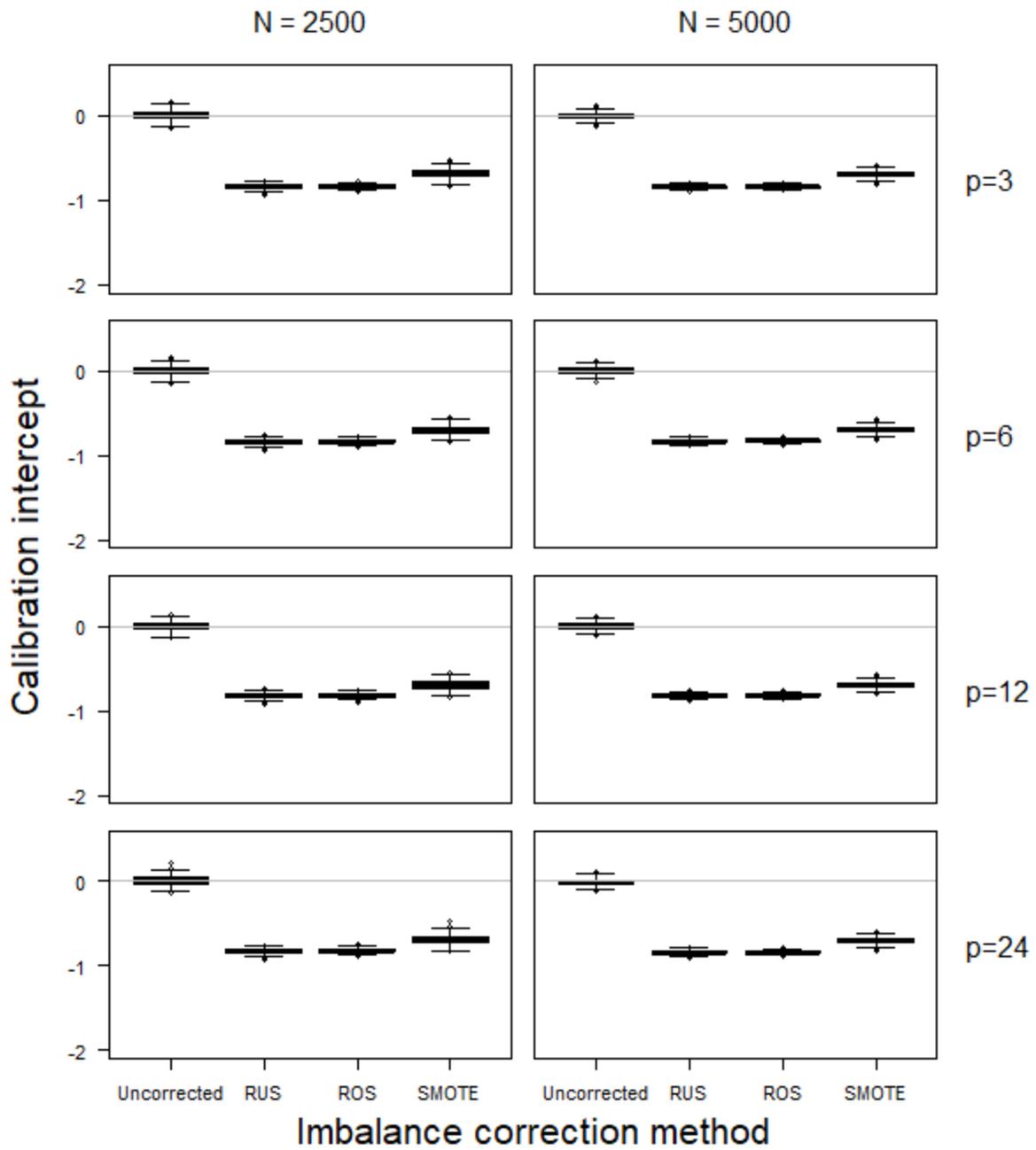

Figure S9. Test set calibration intercept for the SLR models in the simulation scenarios with an event fraction of 1%.

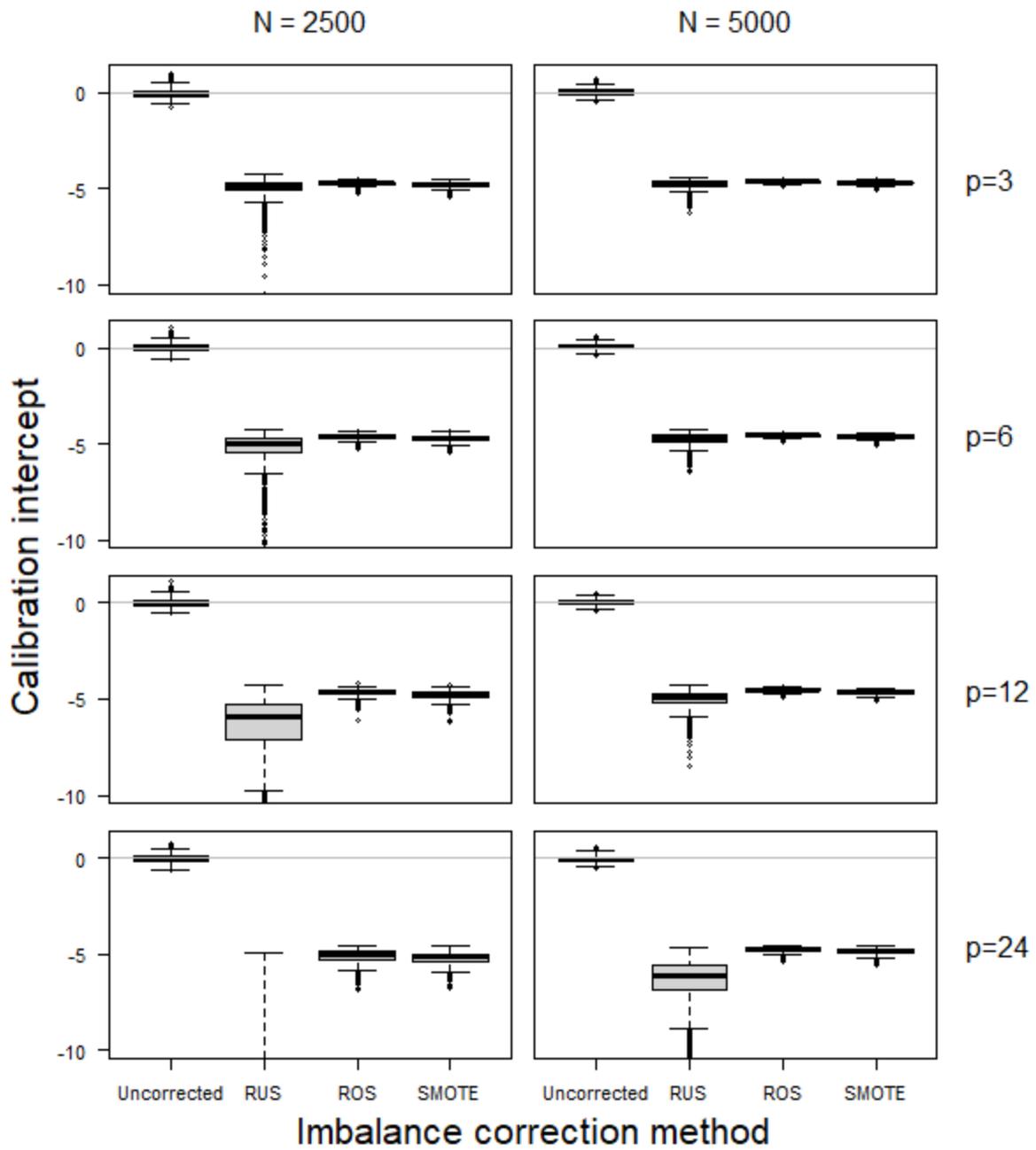

Figure S10. Test set calibration intercept for the SLR models in the simulation scenarios with an event fraction of 10%.

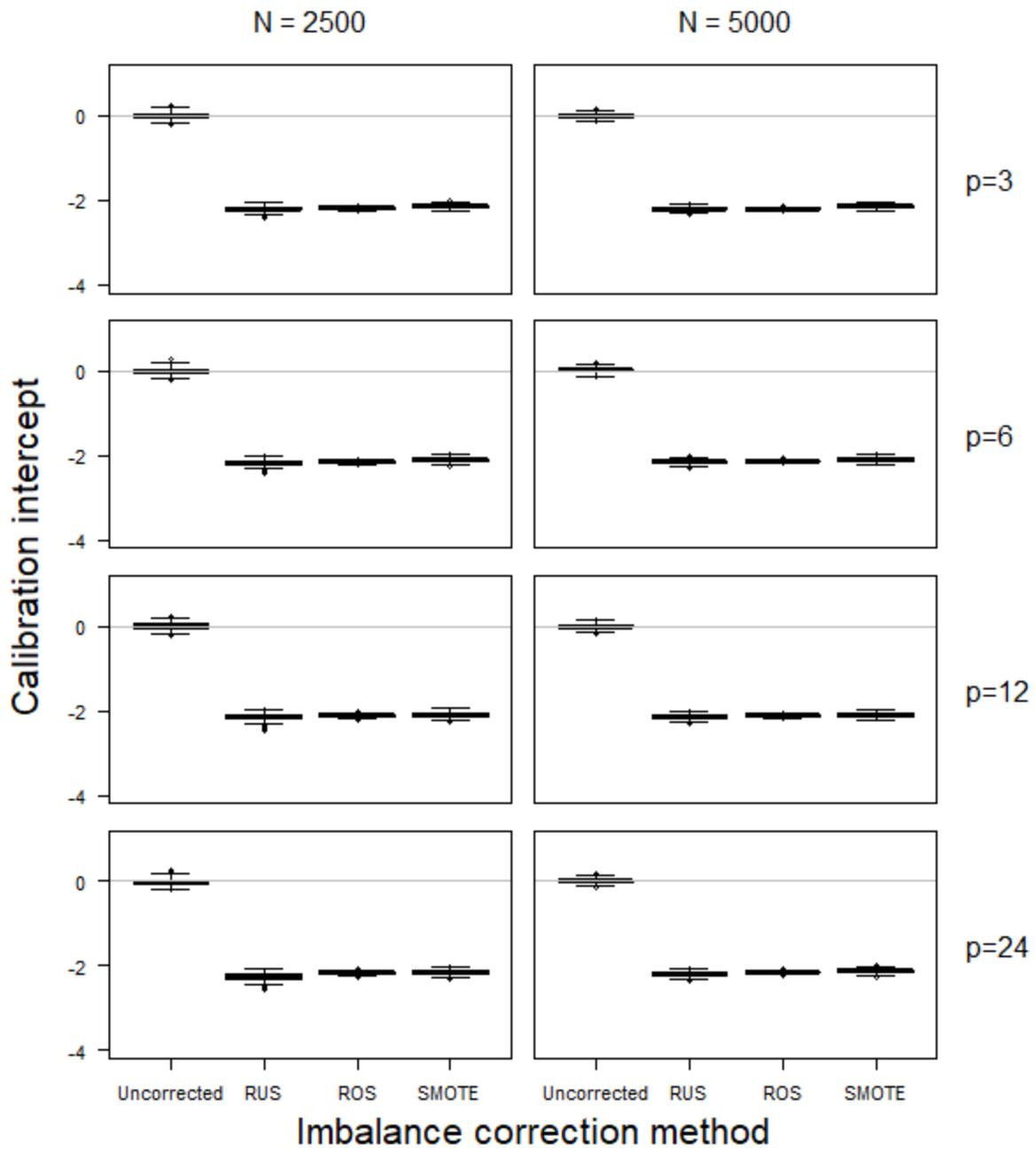

Figure S11. Test set calibration intercept for the SLR models in the simulation scenarios with an event fraction of 30%.

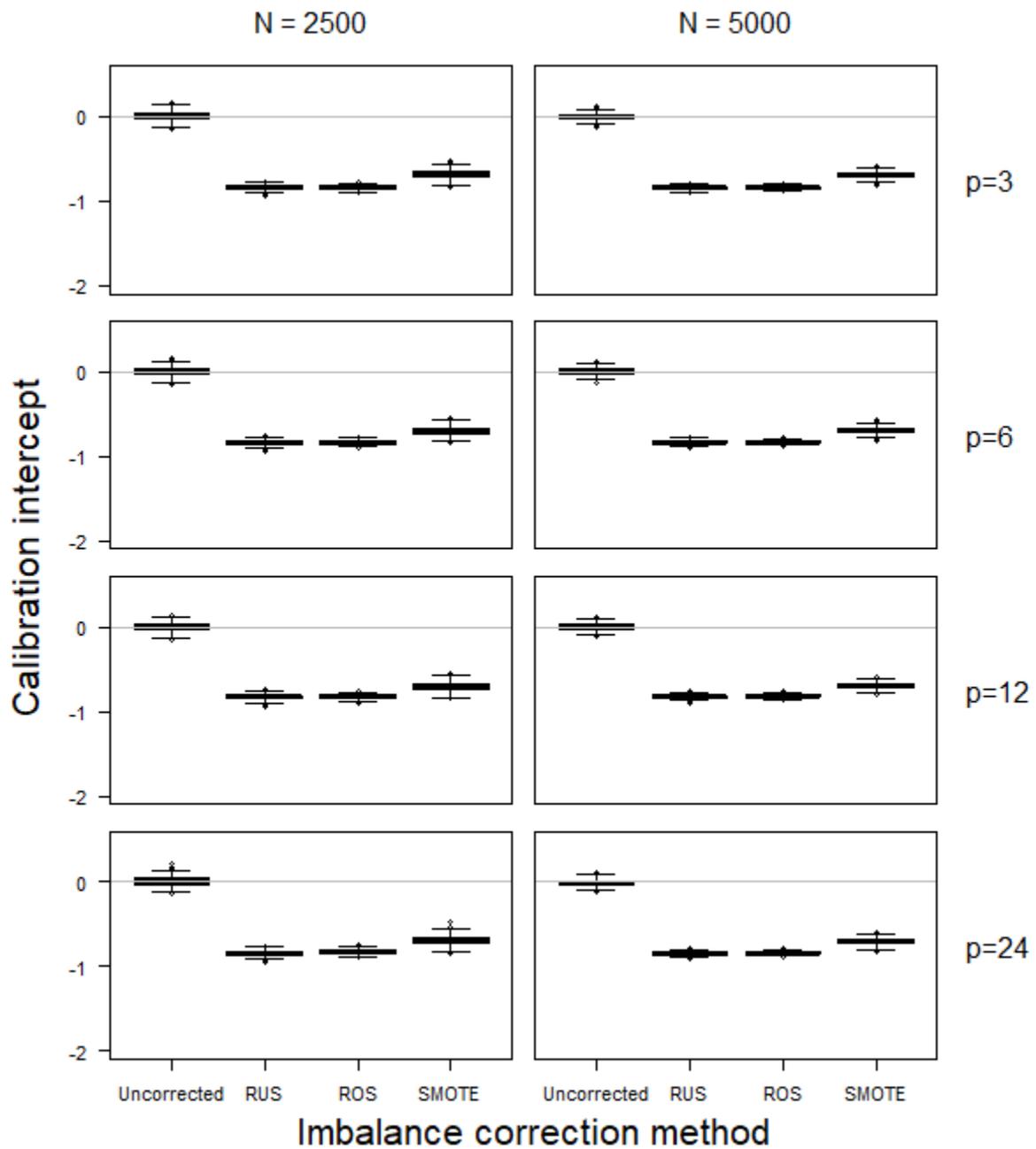

Figure S12. Test set calibration intercept for the Ridge models in the simulation scenarios with an event fraction of 1%, after recalibration of the models based on RUS, ROS, or SMOTE.

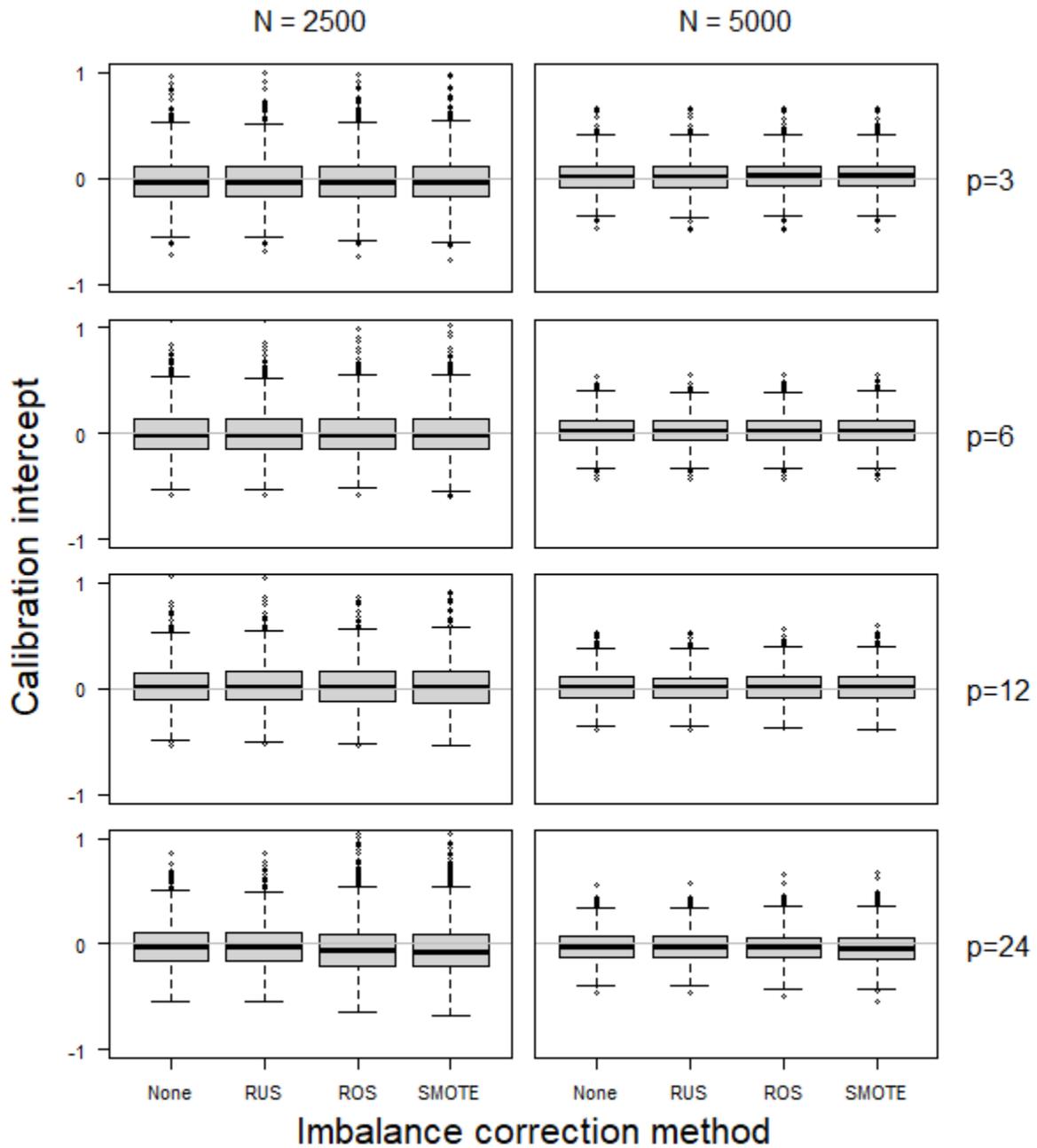

Figure S13. Test set calibration intercept for the Ridge models in the simulation scenarios with an event fraction of 10%, after recalibration of the models based on RUS, ROS, or SMOTE.

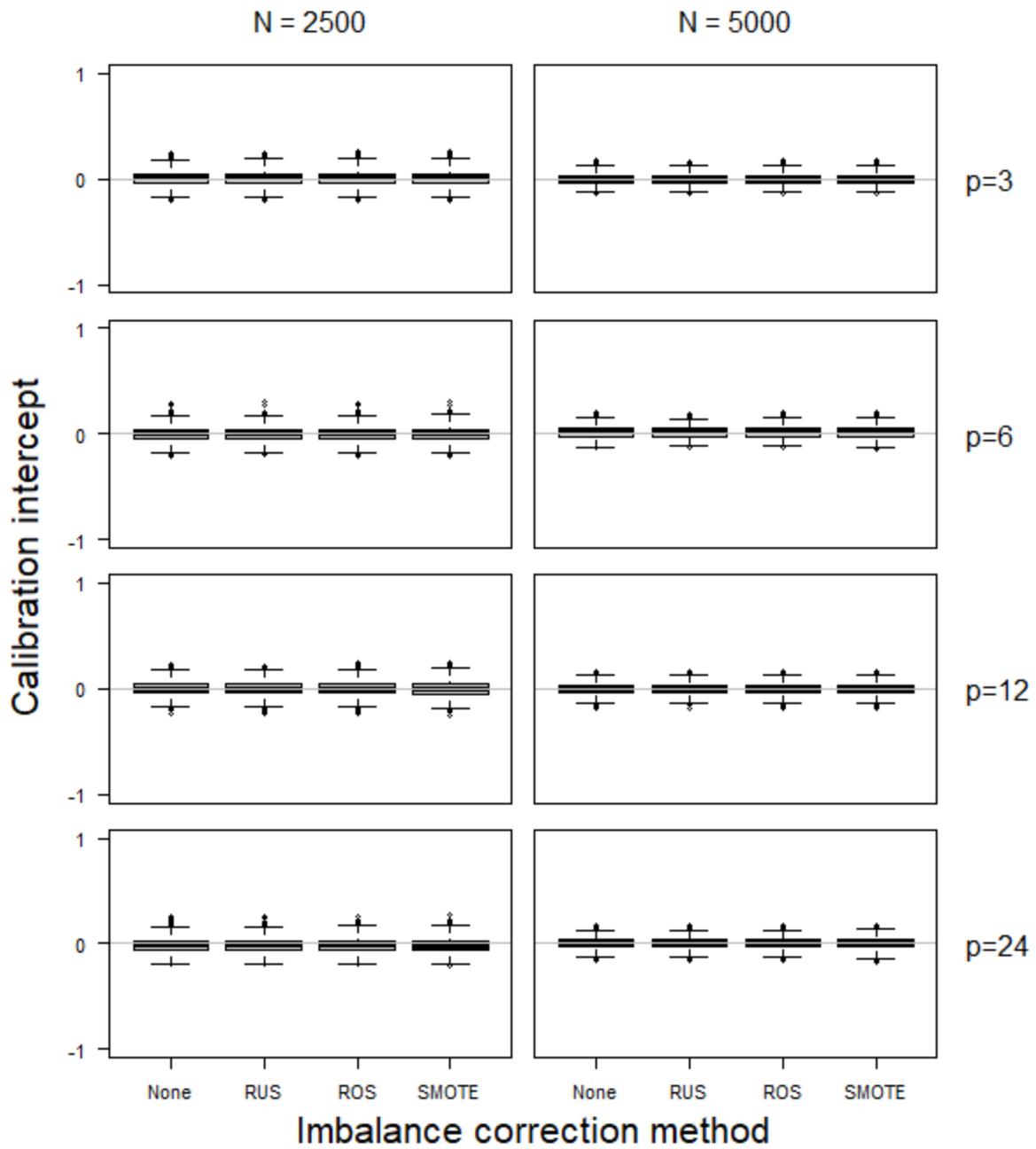

Figure S14. Test set calibration intercept for the Ridge models in the simulation scenarios with an event fraction of 30%, after recalibration of the models based on RUS, ROS, or SMOTE.

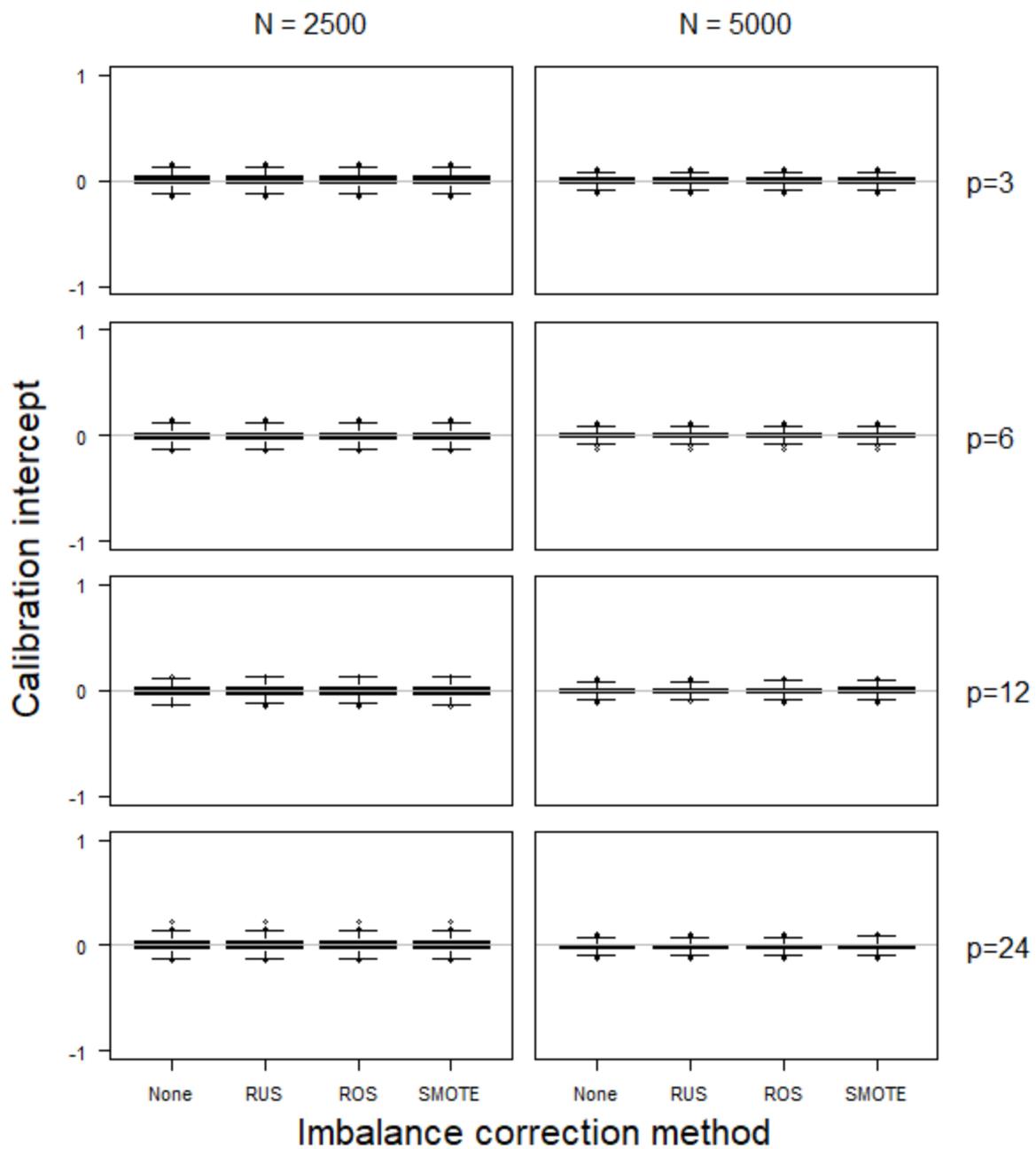

Figure S15. Test set calibration intercept for the SLR models in the simulation scenarios with an event fraction of 1%, after recalibration of the models based on RUS, ROS, or SMOTE.

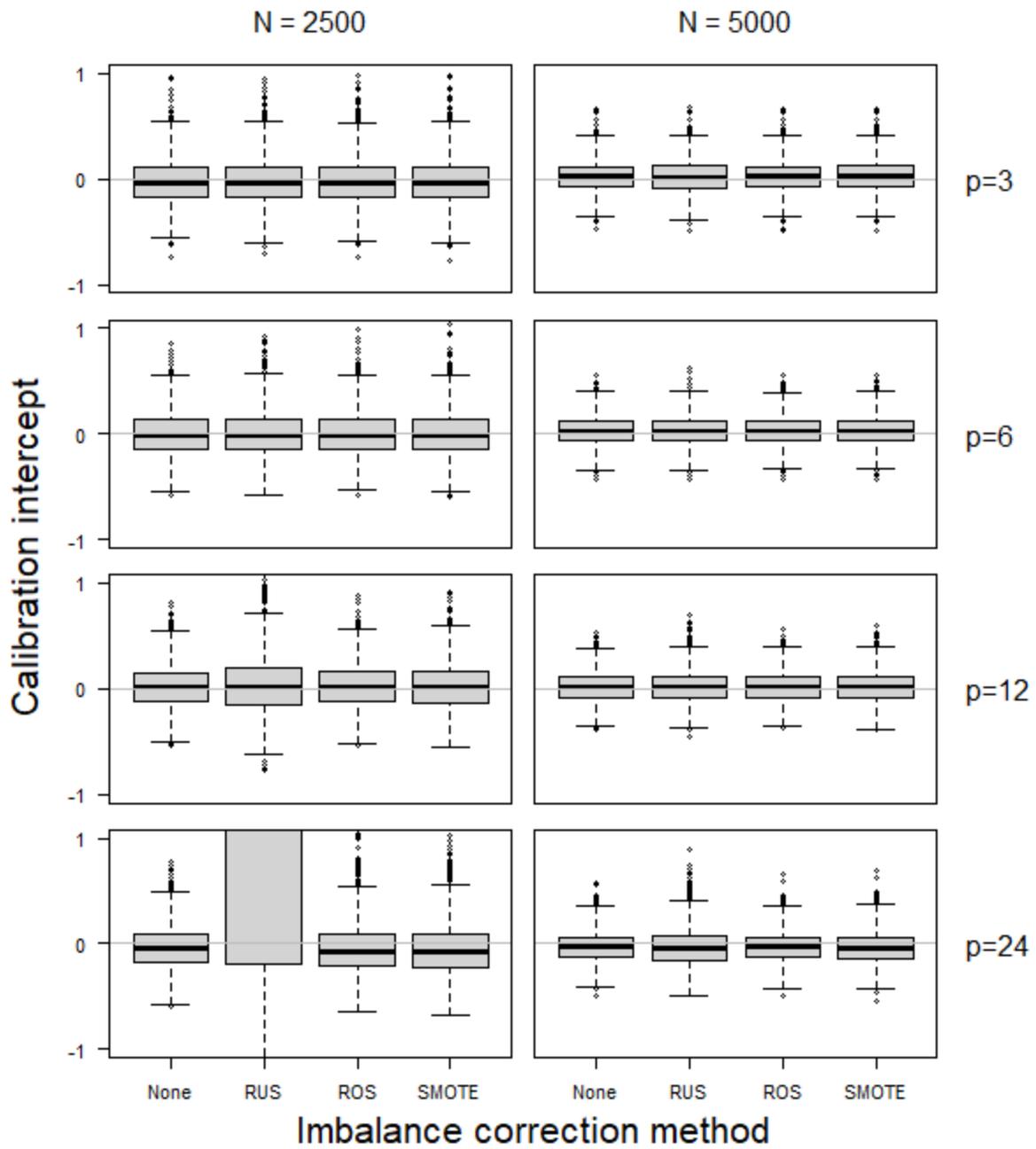

Figure S16. Test set calibration intercept for the SLR models in the simulation scenarios with an event fraction of 10%, after recalibration of the models based on RUS, ROS, or SMOTE.

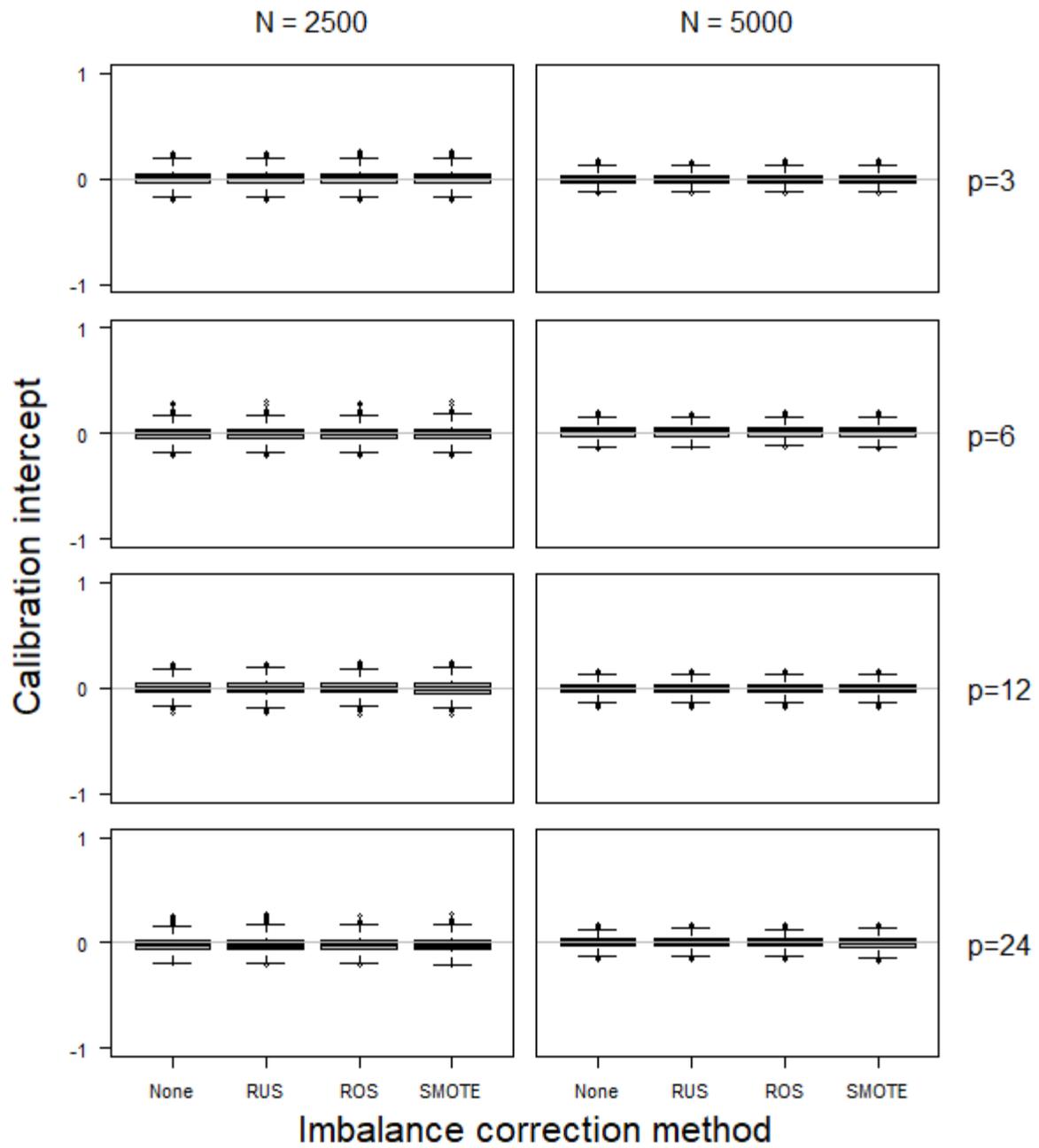

Figure S17. Test set calibration intercept for the SLR models in the simulation scenarios with an event fraction of 30%, after recalibration of the models based on RUS, ROS, or SMOTE.

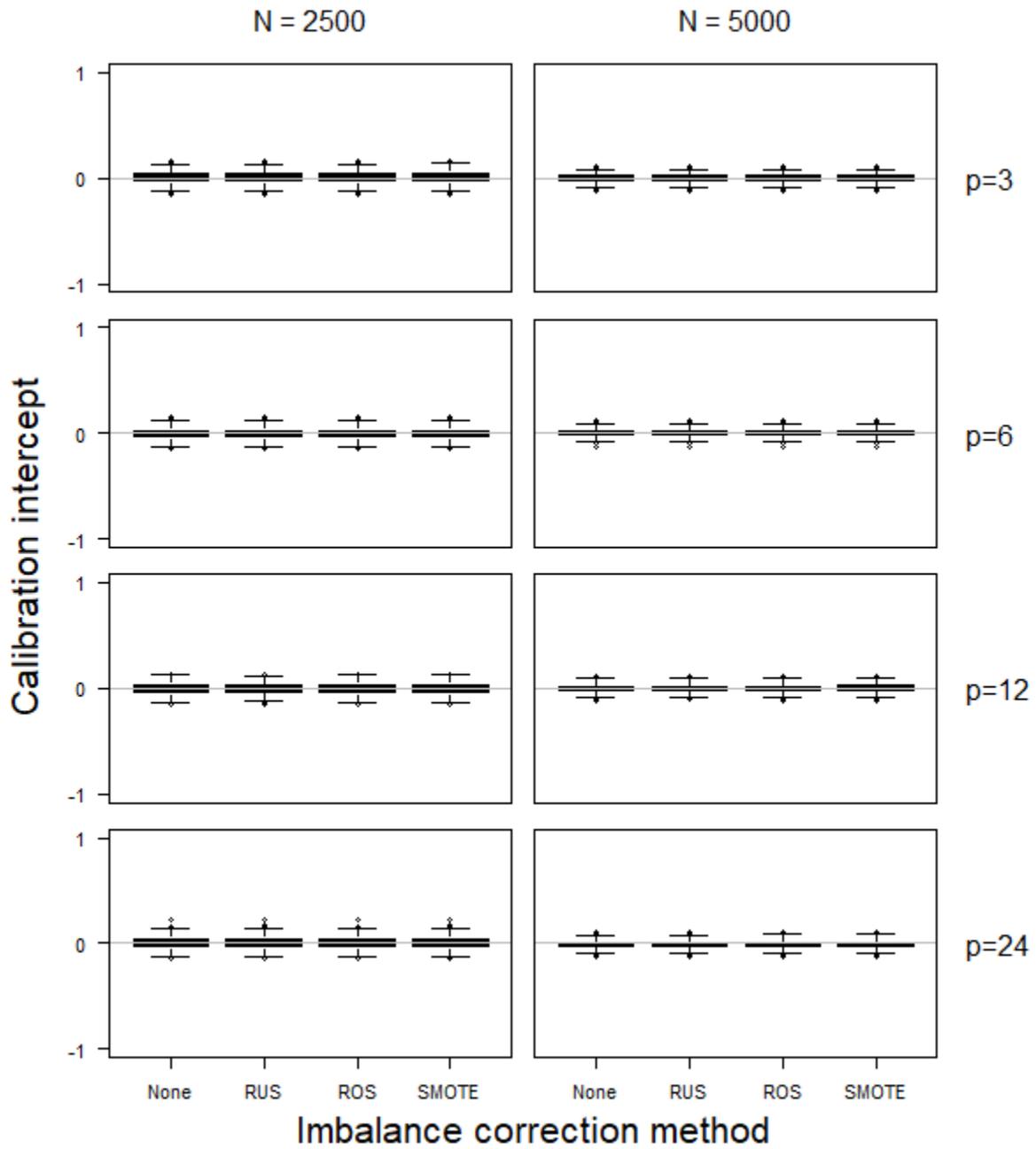

Figure S18. Test set calibration slope for the Ridge models in the simulation scenarios with an event fraction of 10%.

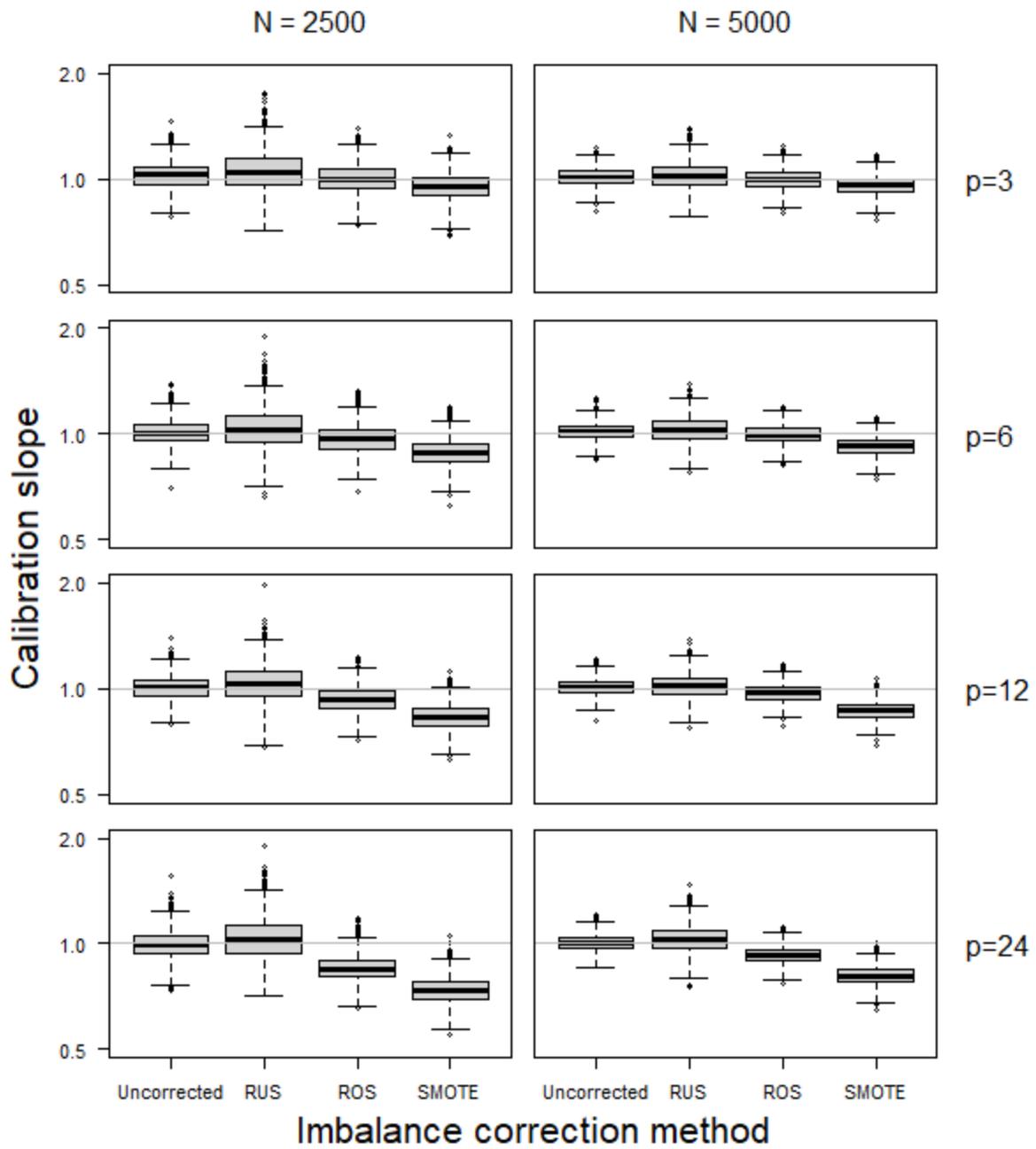

Figure S19. Test set calibration slope for the Ridge models in the simulation scenarios with an event fraction of 30%.

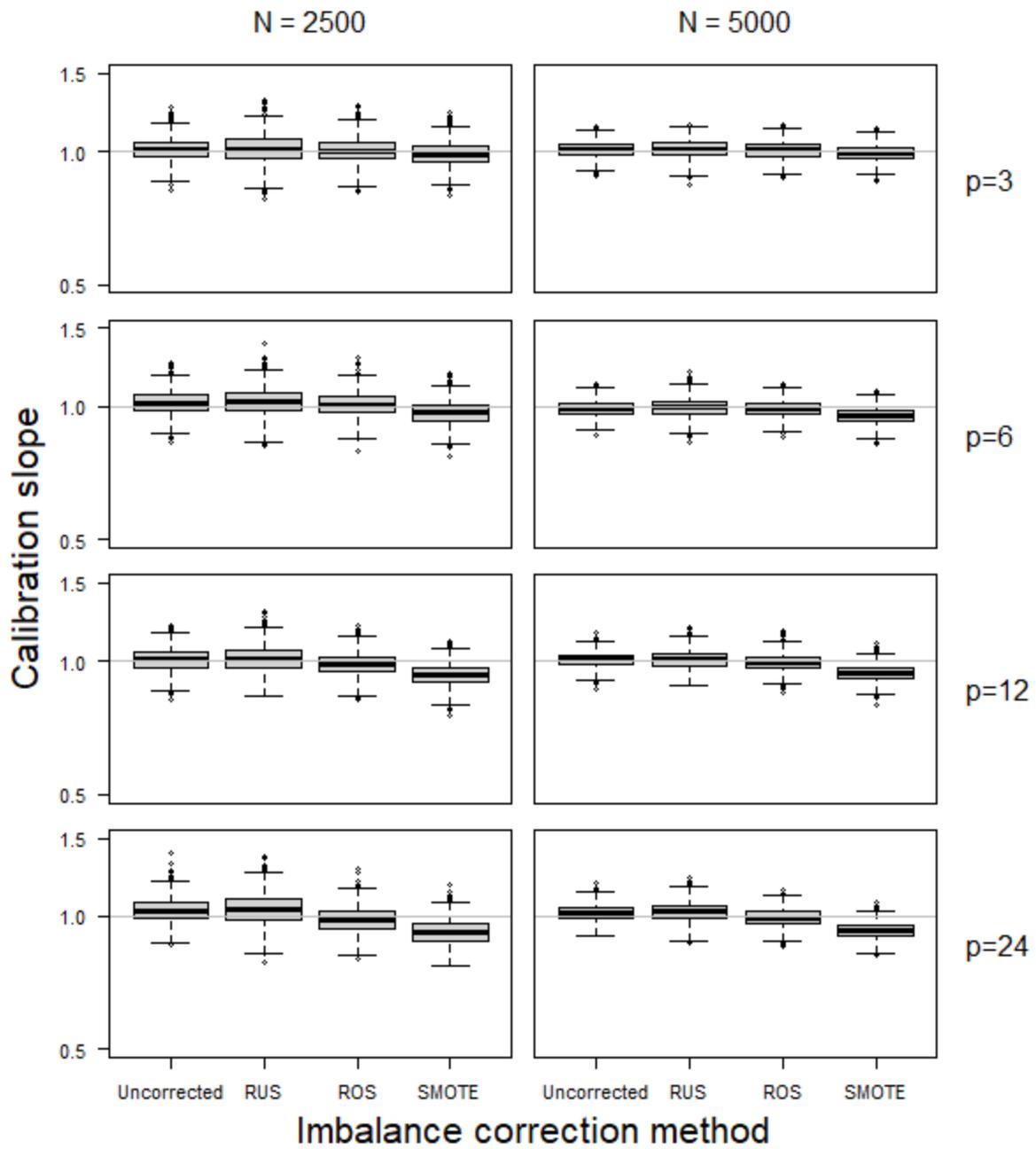

Figure S20. Test set calibration slope for the SLR models in the simulation scenarios with an event fraction of 1%.

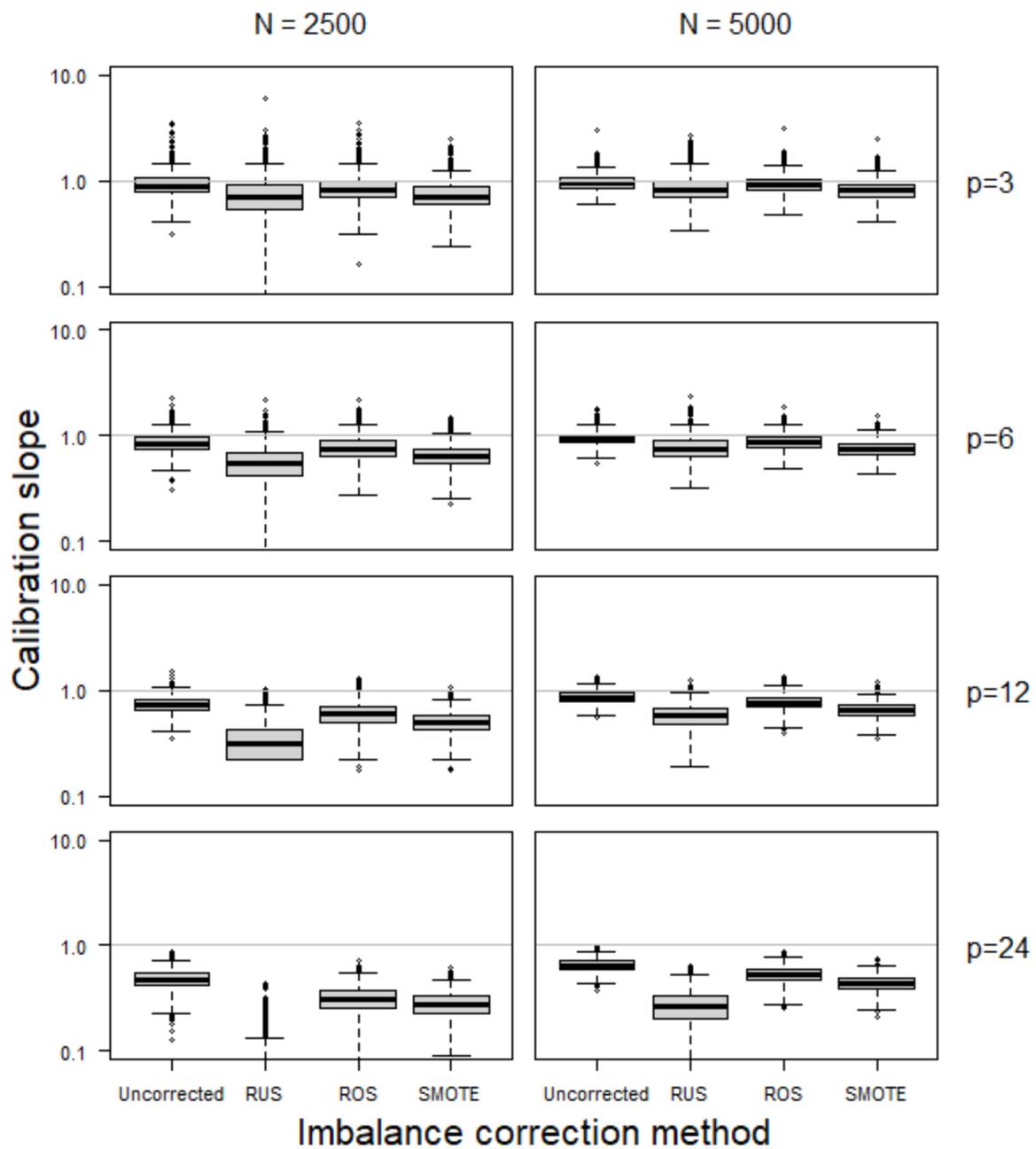

Figure S21. Test set calibration slope for the SLR models in the simulation scenarios with an event fraction of 10%.

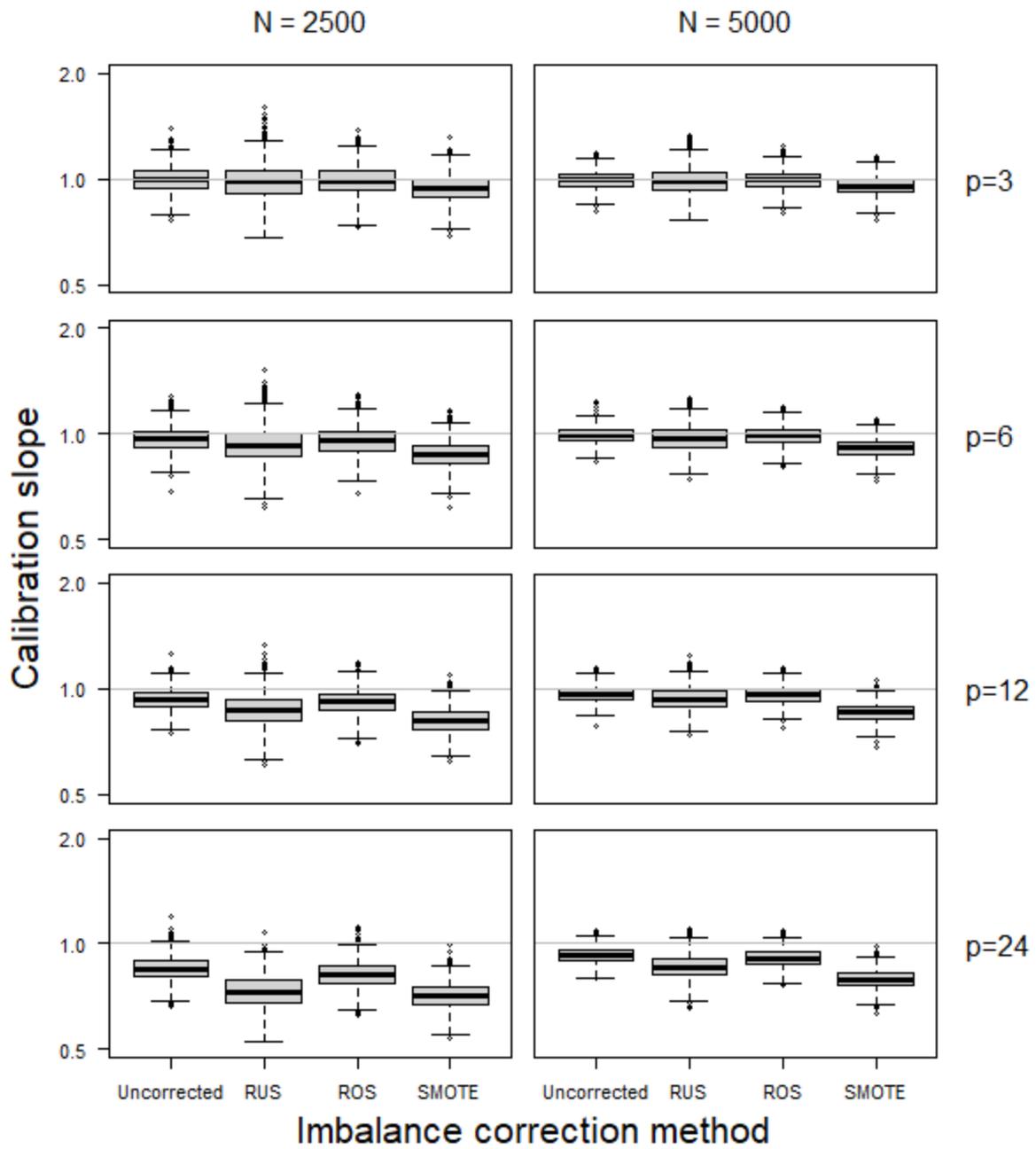

Figure S22. Test set calibration slope for the SLR models in the simulation scenarios with an event fraction of 30%.

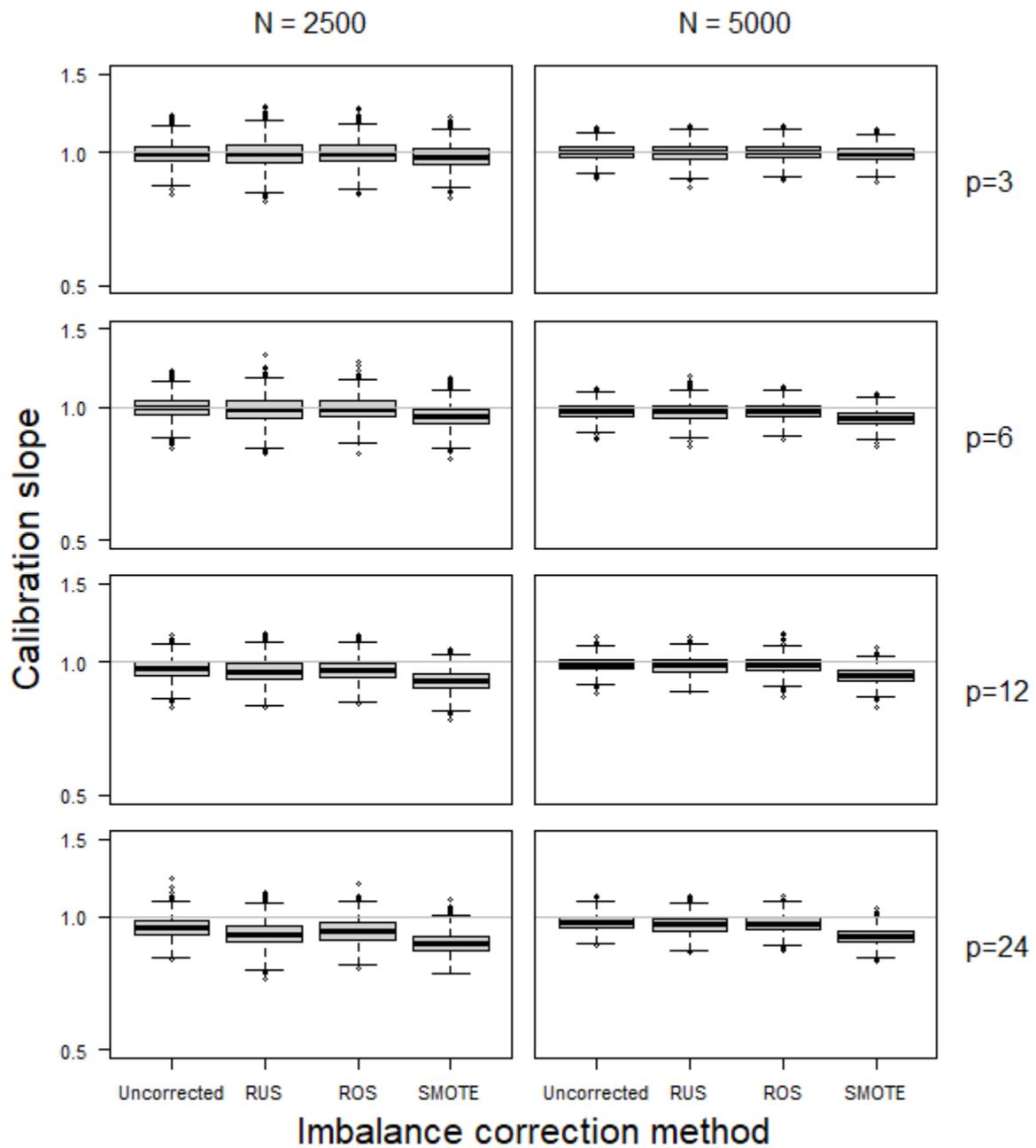

Figure S23. Test set classification accuracy for the Ridge models in the simulation scenarios with an event fraction of 1%. For uncorrected training sets, we used either the default threshold of 0.5 ("Unc .5") or a threshold based on the true event fraction ("Unc EF"). For RUS/ROS/SMOTE, the default threshold of 0.5 was used.

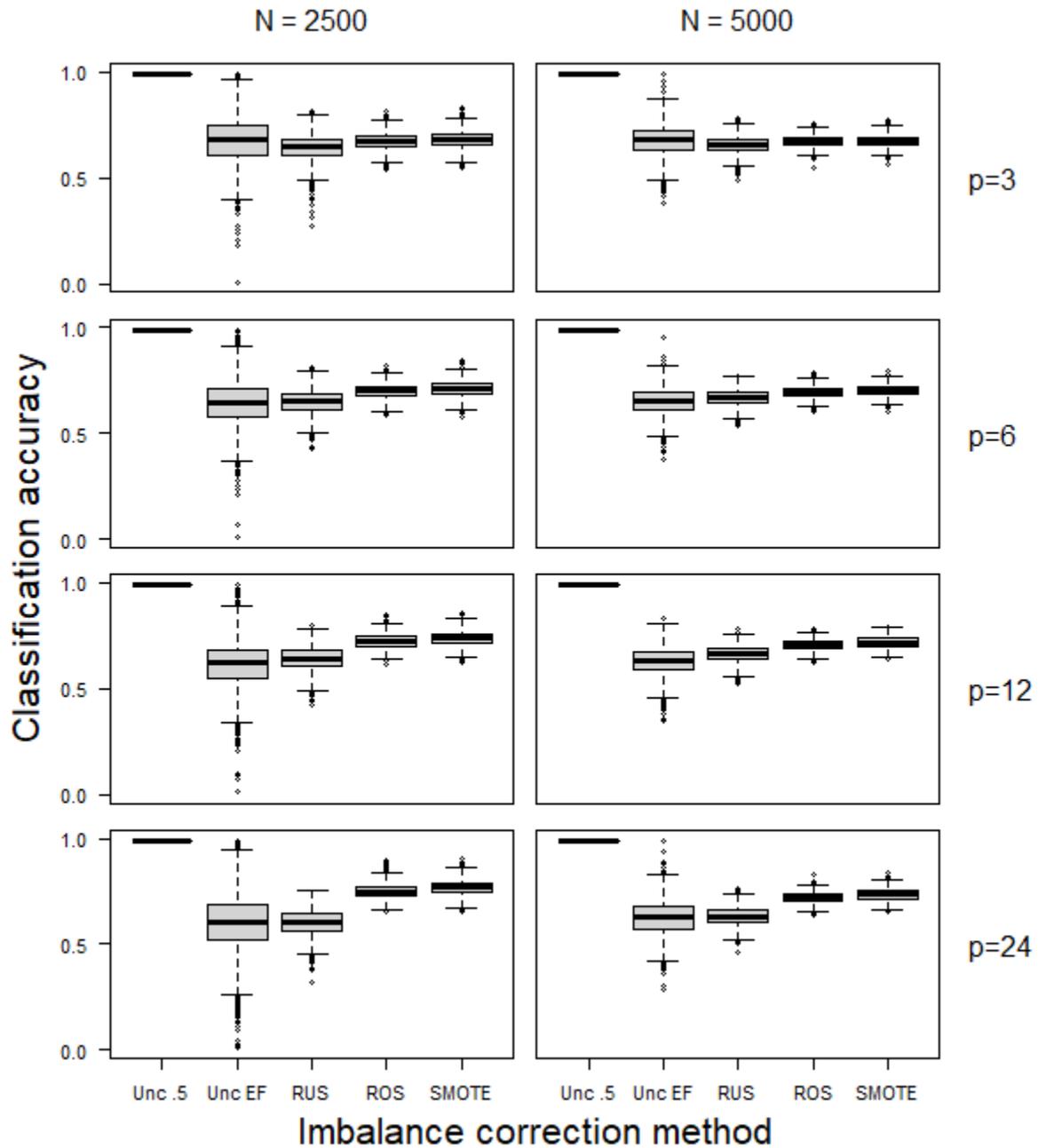

Figure S24. Test set classification accuracy for the Ridge models in the simulation scenarios with an event fraction of 10%. For uncorrected training sets, we used either the default threshold of 0.5 ("Unc .5") or a threshold based on the true event fraction ("Unc EF"). For RUS/ROS/SMOTE, the default threshold of 0.5 was used.

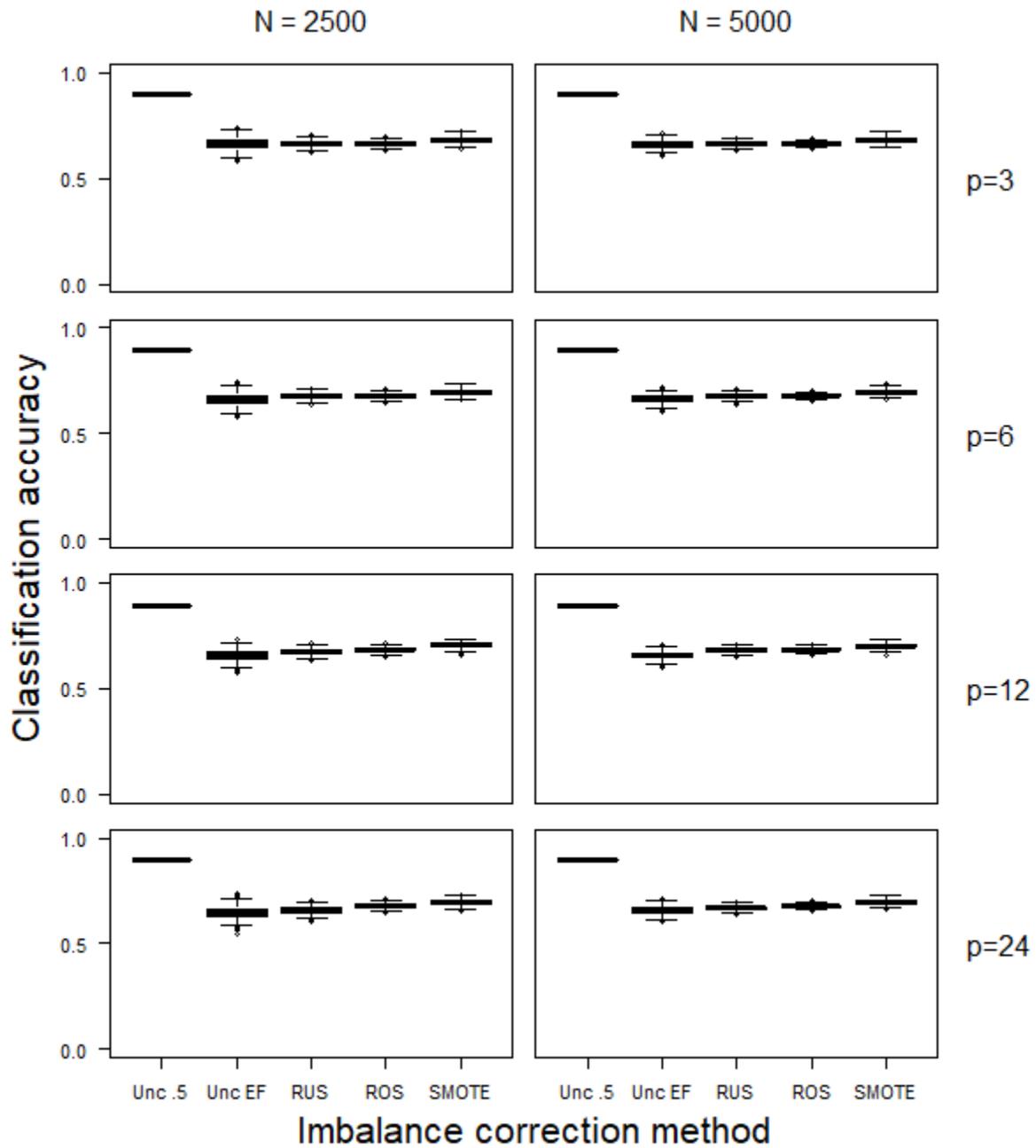

Figure S25. Test set classification accuracy for the Ridge models in the simulation scenarios with an event fraction of 30%. For uncorrected training sets, we used either the default threshold of 0.5 ("Unc .5") or a threshold based on the true event fraction ("Unc EF"). For RUS/ROS/SMOTE, the default threshold of 0.5 was used.

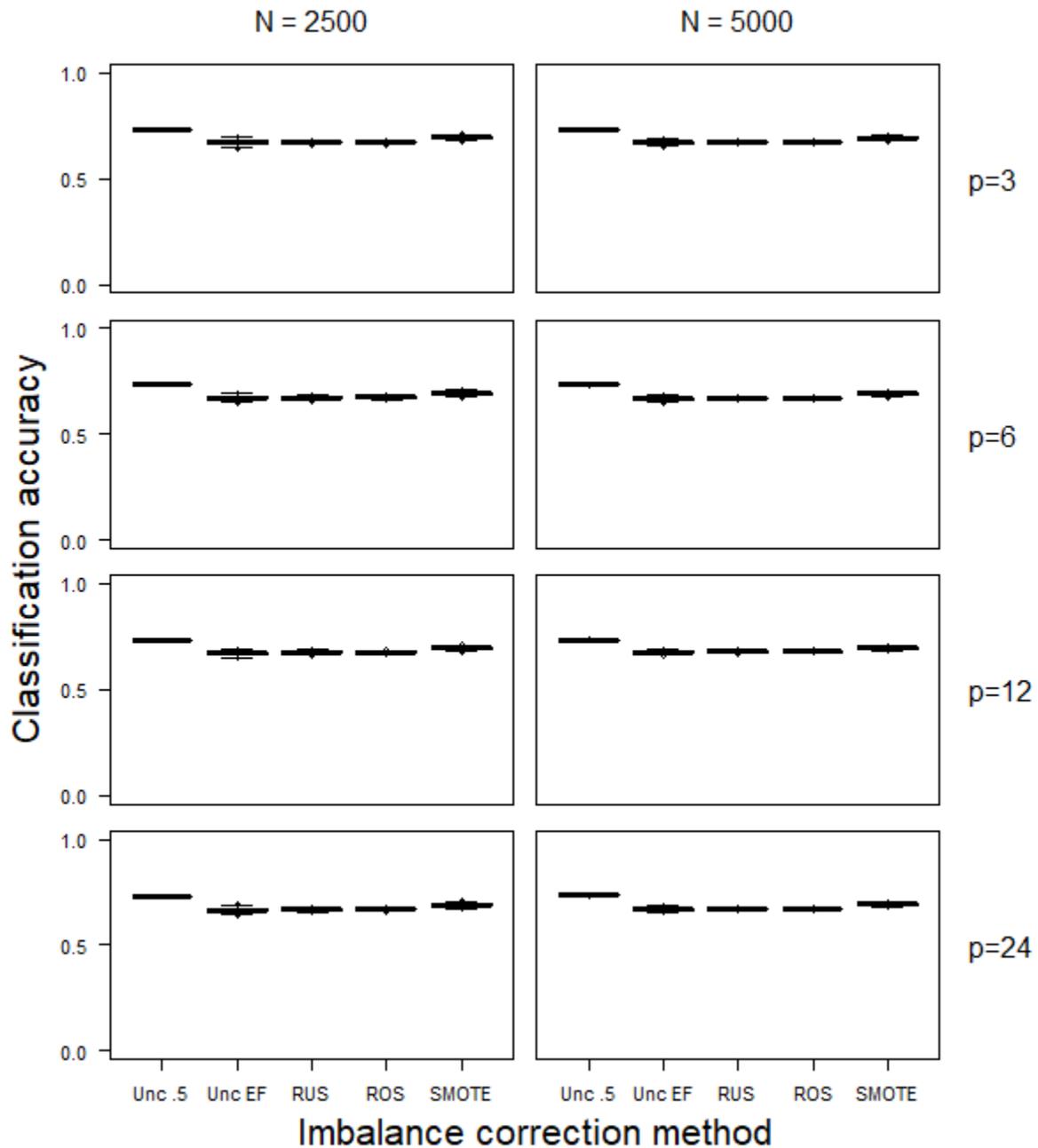

Figure S26. Test set classification accuracy for the SLR models in the simulation scenarios with an event fraction of 1%. For uncorrected training sets, we used either the default threshold of 0.5 ("Unc .5") or a threshold based on the true event fraction ("Unc EF"). For RUS/ROS/SMOTE, the default threshold of 0.5 was used.

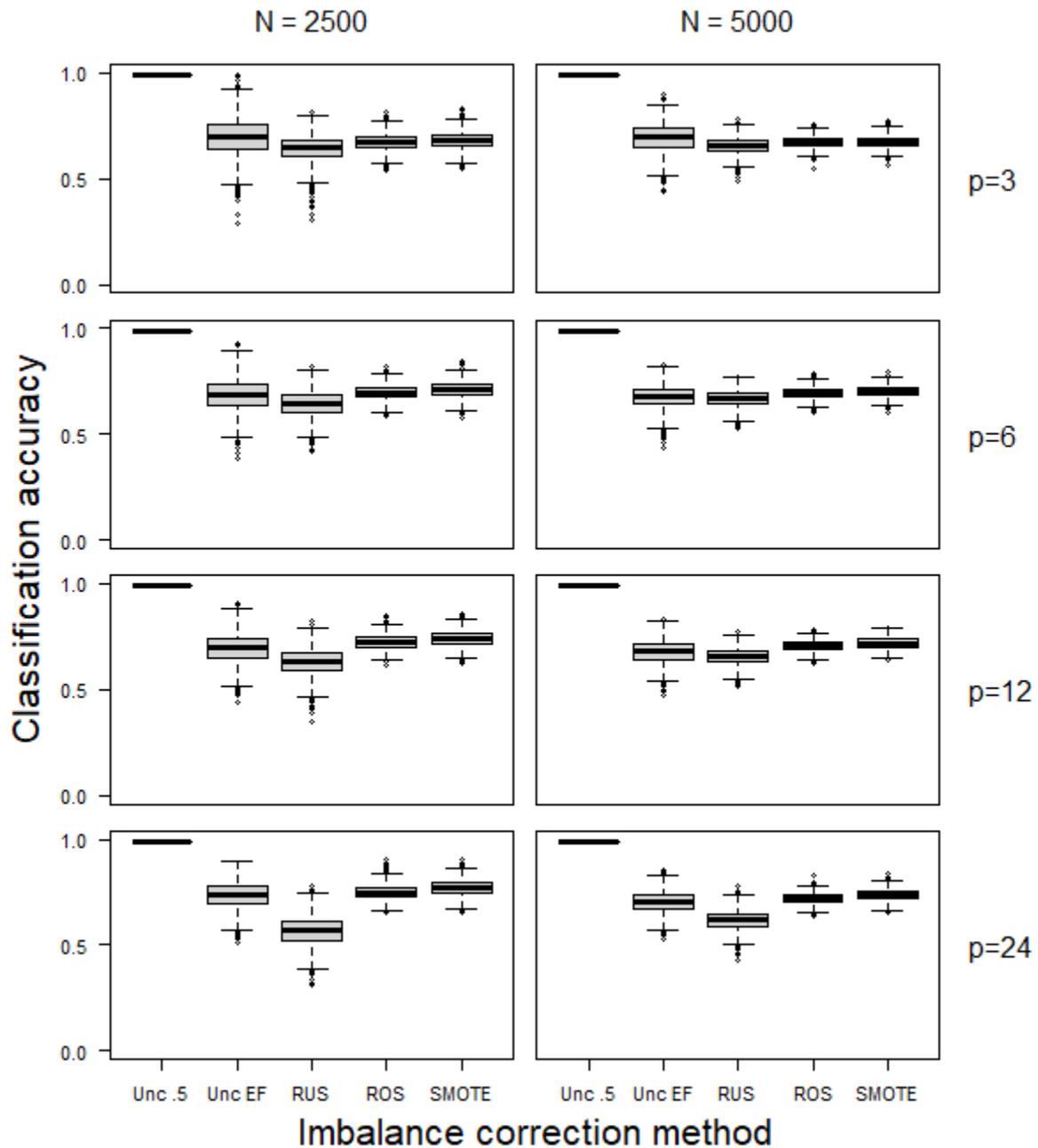

Figure S27. Test set classification accuracy for the SLR models in the simulation scenarios with an event fraction of 10%. For uncorrected training sets, we used either the default threshold of 0.5 ("Unc .5") or a threshold based on the true event fraction ("Unc EF"). For RUS/ROS/SMOTE, the default threshold of 0.5 was used.

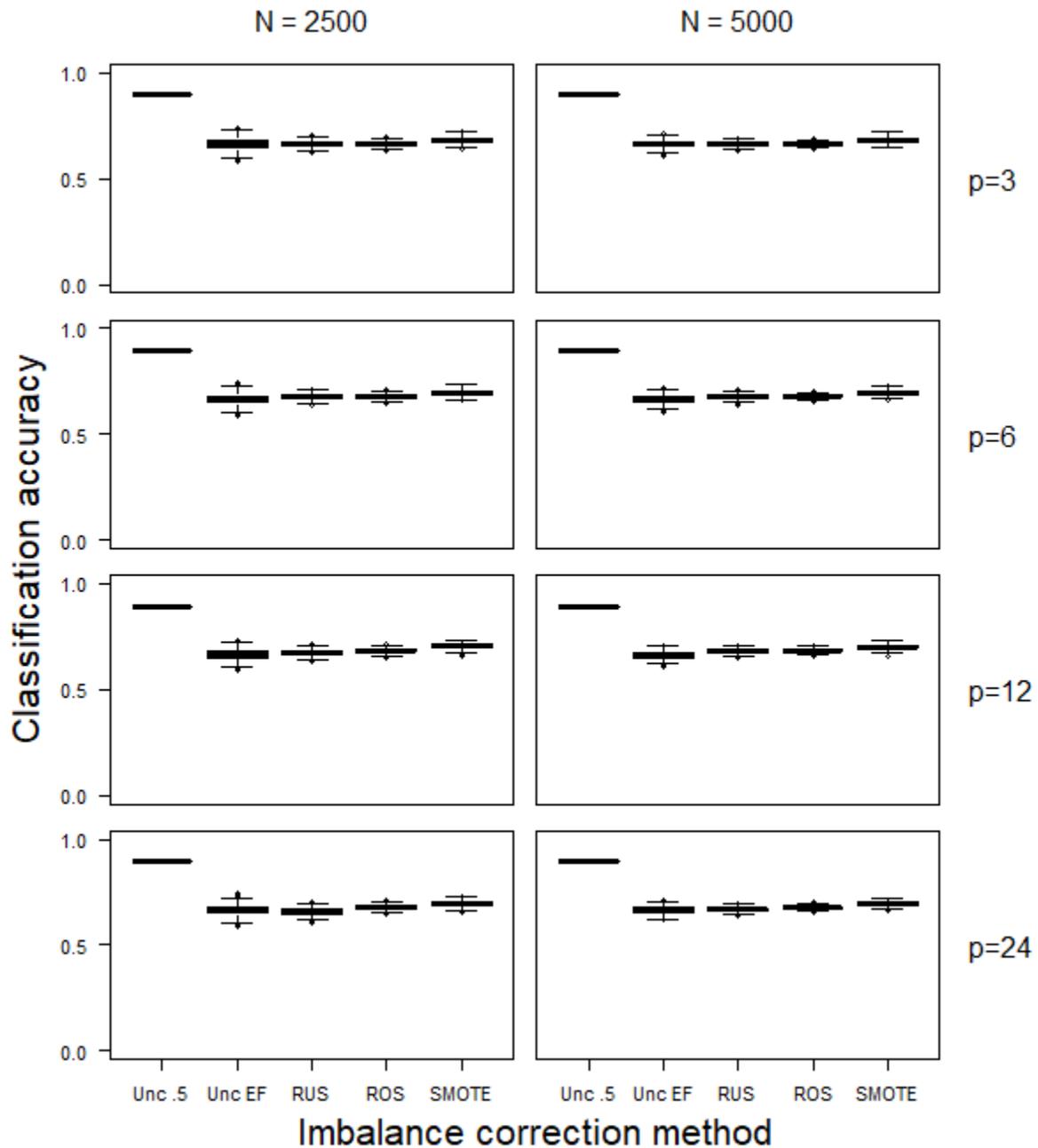

Figure S28. Test set classification accuracy for the SLR models in the simulation scenarios with an event fraction of 30%. For uncorrected training sets, we used either the default threshold of 0.5 ("Unc .5") or a threshold based on the true event fraction ("Unc EF"). For RUS/ROS/SMOTE, the default threshold of 0.5 was used.

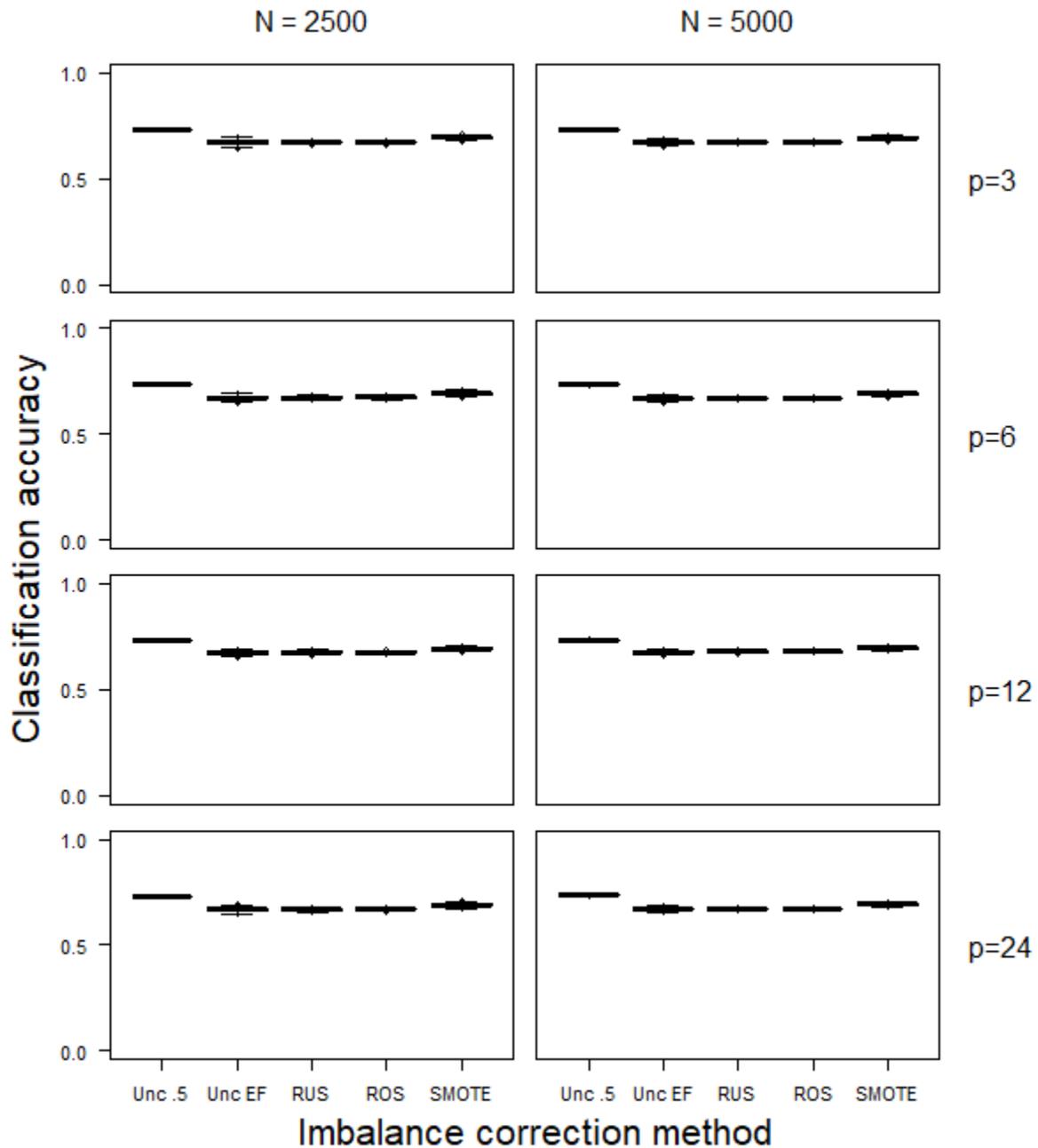

Figure S29. Test set sensitivity for the Ridge models in the simulation scenarios with an event fraction of 1%. For uncorrected training sets, we used either the default threshold of 0.5 ("Unc .5") or a threshold based on the true event fraction ("Unc EF"). For RUS/ROS/SMOTE, the default threshold of 0.5 was used.

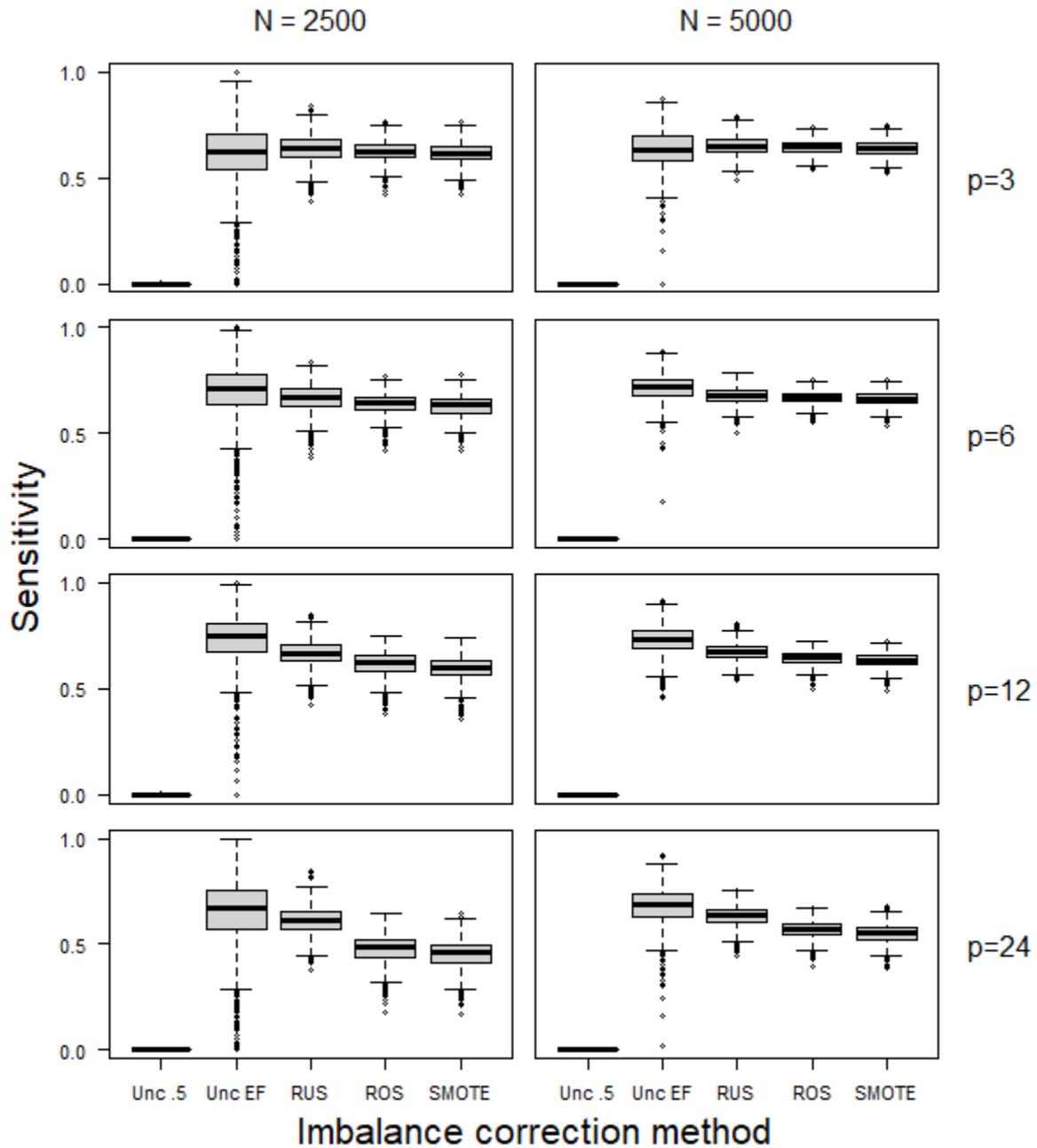

Figure S30. Test set sensitivity for the Ridge models in the simulation scenarios with an event fraction of 10%. For uncorrected training sets, we used either the default threshold of 0.5 ("Unc .5") or a threshold based on the true event fraction ("Unc EF"). For RUS/ROS/SMOTE, the default threshold of 0.5 was used.

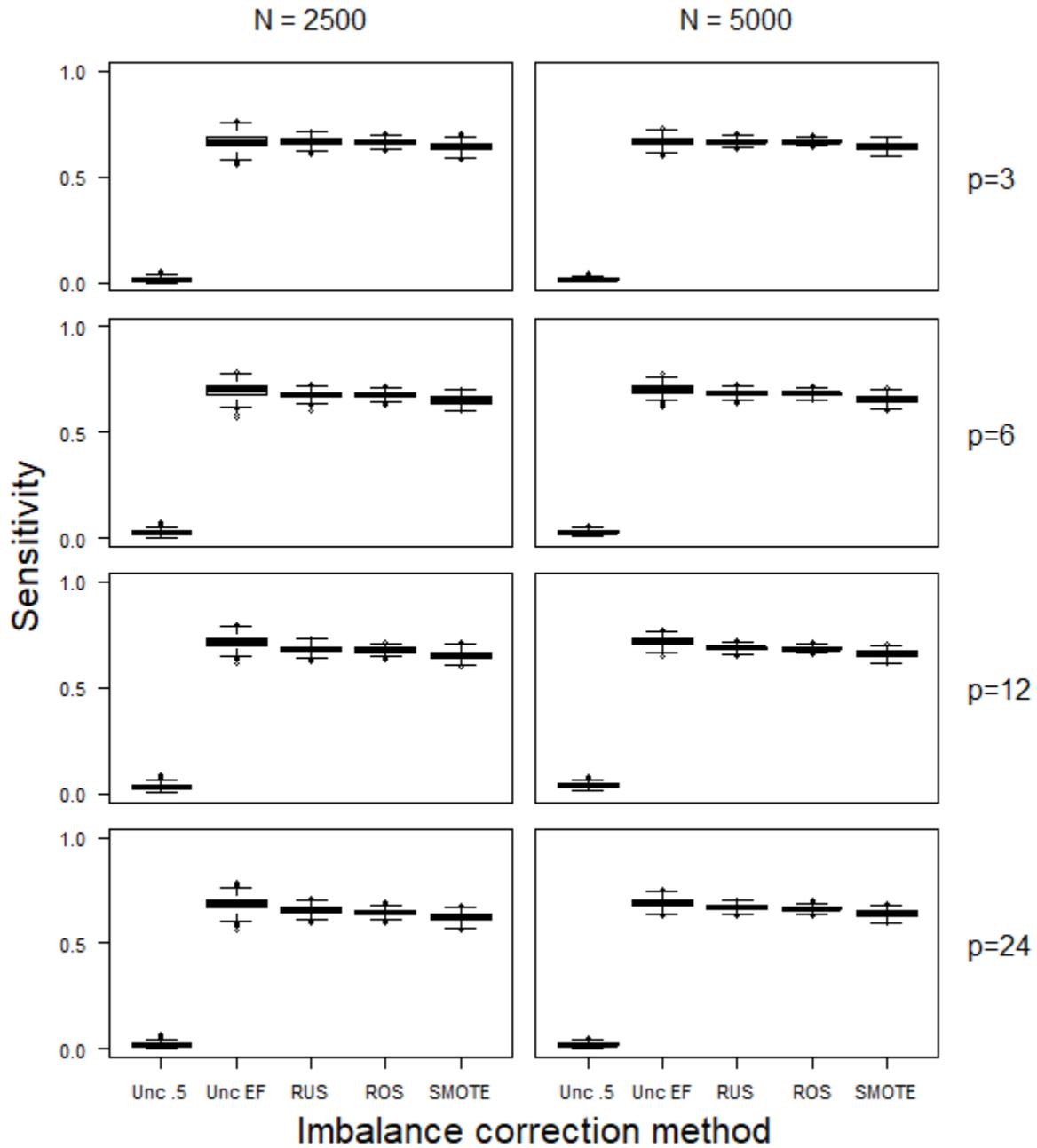

Figure S31. Test set sensitivity for the Ridge models in the simulation scenarios with an event fraction of 30%. For uncorrected training sets, we used either the default threshold of 0.5 ("Unc .5") or a threshold based on the true event fraction ("Unc EF"). For RUS/ROS/SMOTE, the default threshold of 0.5 was used.

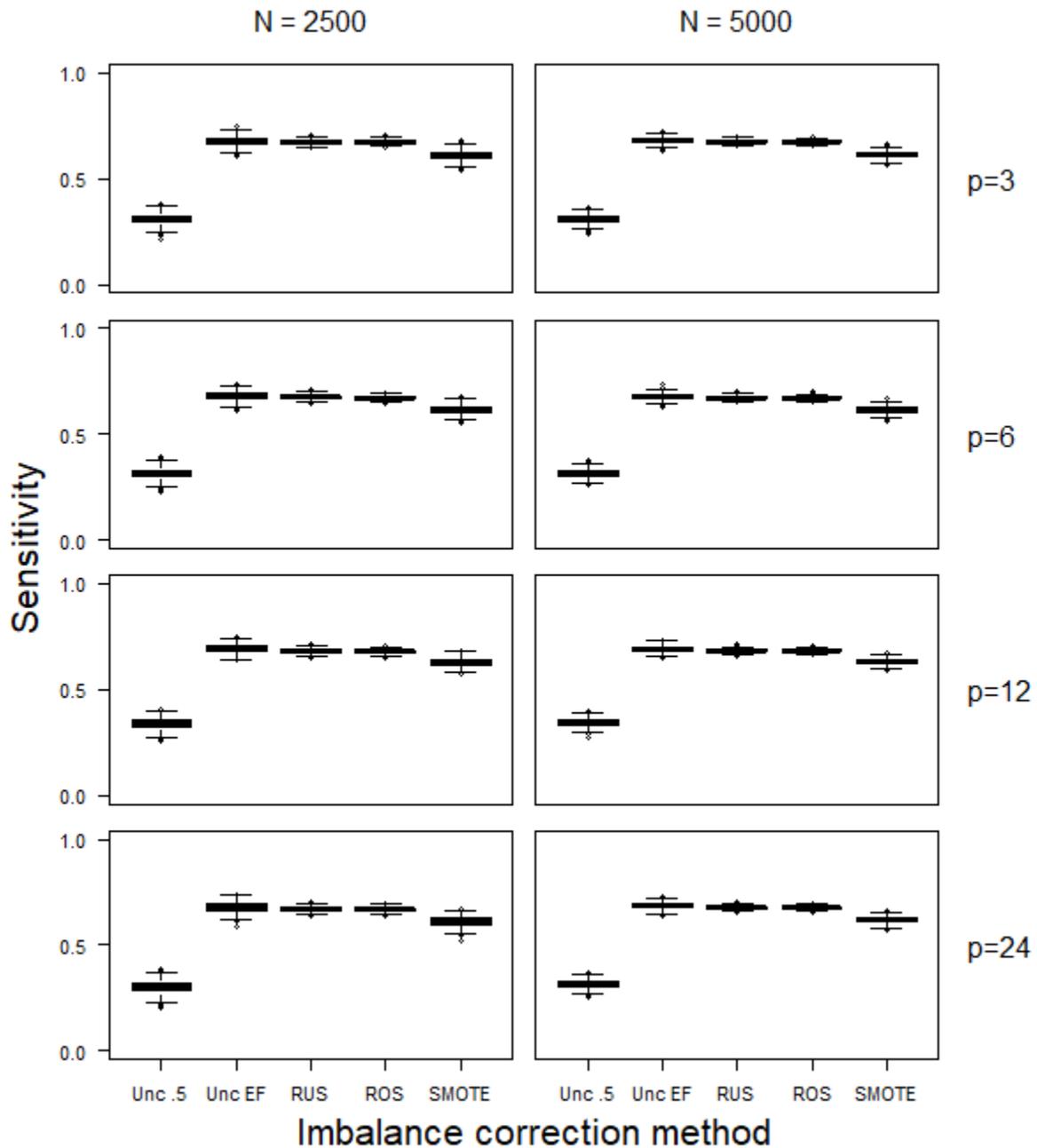

Figure S32. Test set sensitivity for the SLR models in the simulation scenarios with an event fraction of 1%. For uncorrected training sets, we used either the default threshold of 0.5 ("Unc .5") or a threshold based on the true event fraction ("Unc EF"). For RUS/ROS/SMOTE, the default threshold of 0.5 was used.

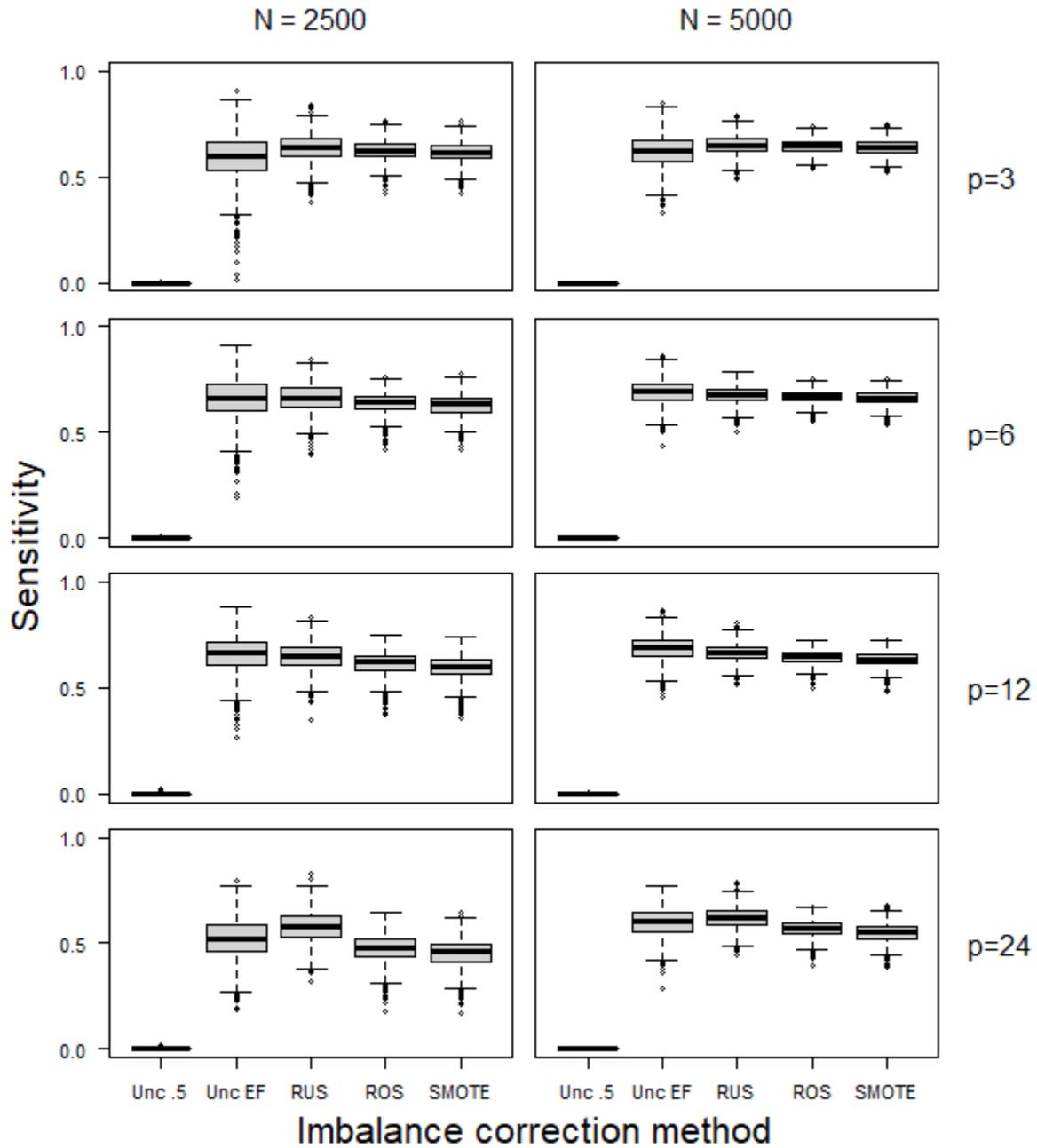

Figure S33. Test set sensitivity for the SLR models in the simulation scenarios with an event fraction of 10%. For uncorrected training sets, we used either the default threshold of 0.5 ("Unc .5") or a threshold based on the true event fraction ("Unc EF"). For RUS/ROS/SMOTE, the default threshold of 0.5 was used.

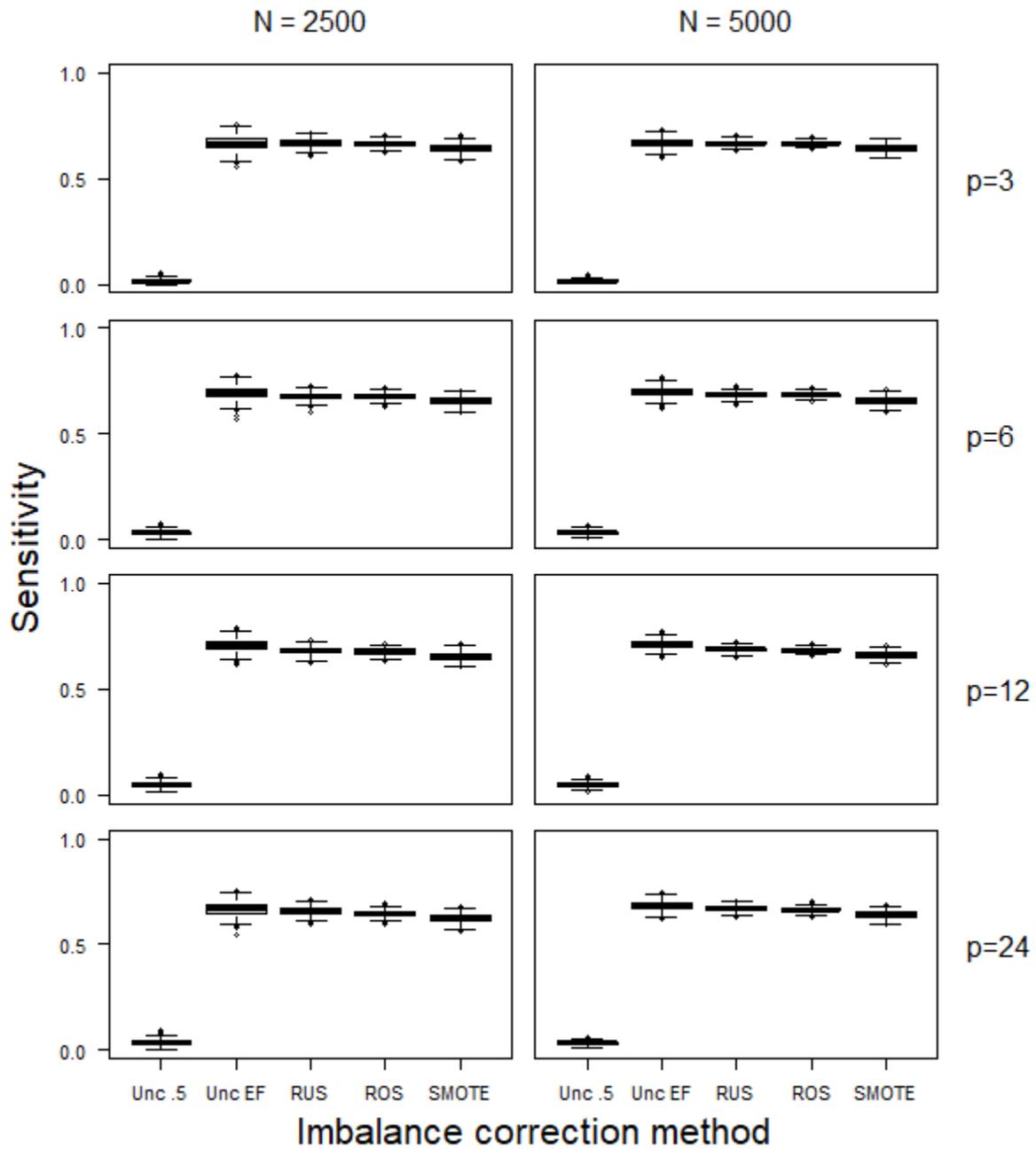

Figure S34. Test set sensitivity for the SLR models in the simulation scenarios with an event fraction of 30%. For uncorrected training sets, we used either the default threshold of 0.5 ("Unc .5") or a threshold based on the true event fraction ("Unc EF"). For RUS/ROS/SMOTE, the default threshold of 0.5 was used.

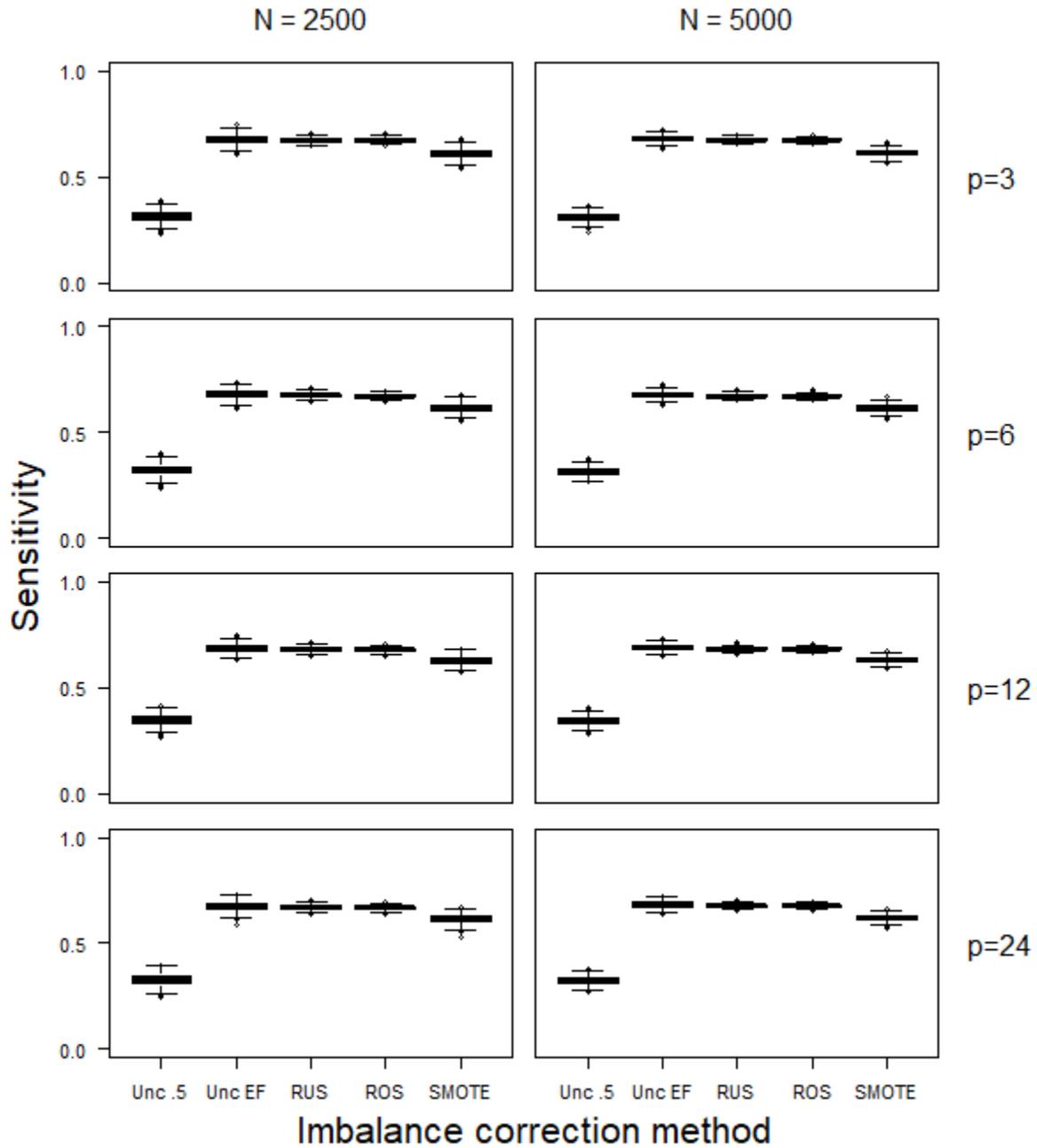

Figure S35. Test set specificity for the Ridge models in the simulation scenarios with an event fraction of 1%. For uncorrected training sets, we used either the default threshold of 0.5 ("Unc .5") or a threshold based on the true event fraction ("Unc EF"). For RUS/ROS/SMOTE, the default threshold of 0.5 was used.

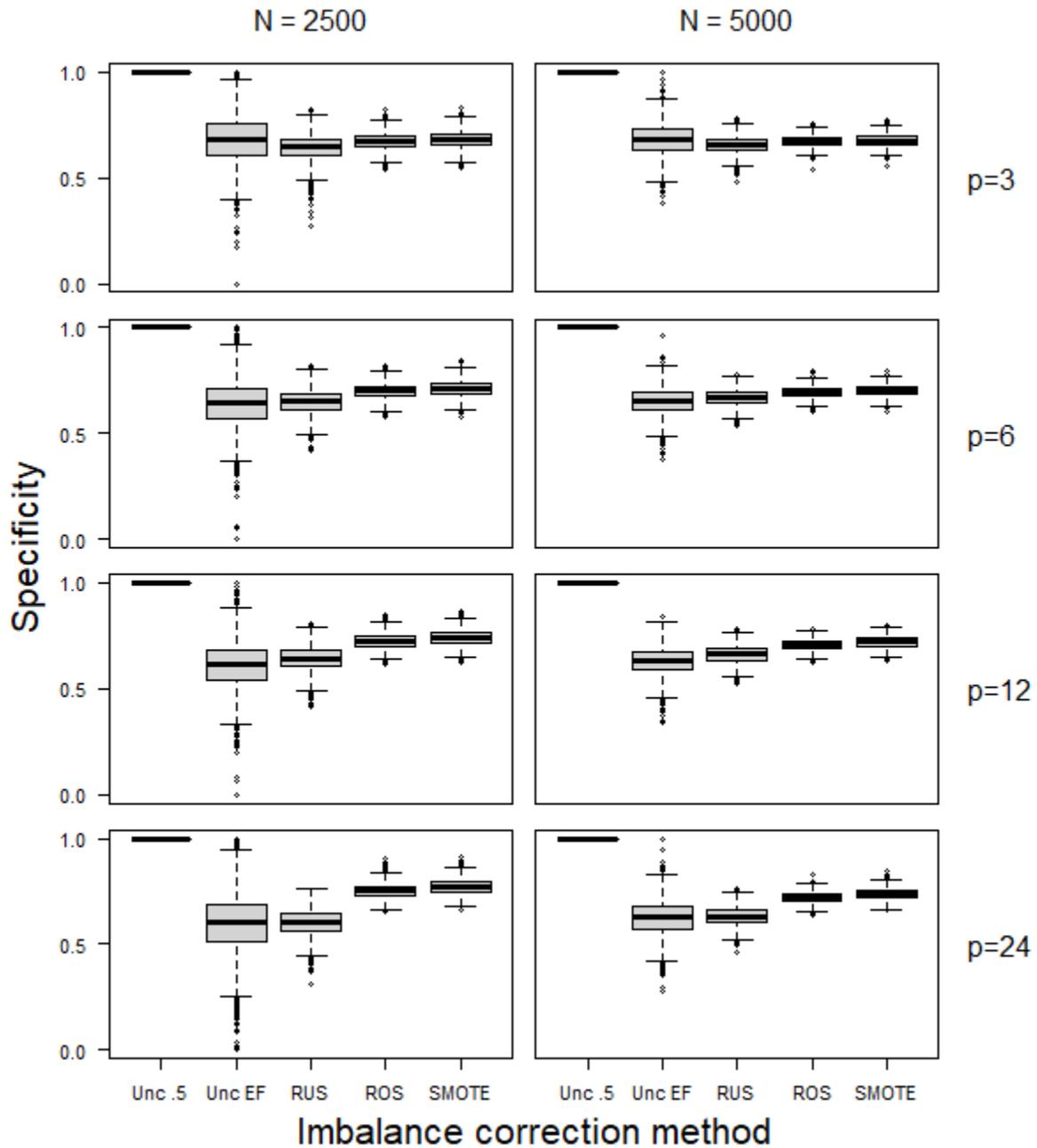

Figure S36. Test set specificity for the Ridge models in the simulation scenarios with an event fraction of 10%. For uncorrected training sets, we used either the default threshold of 0.5 ("Unc .5") or a threshold based on the true event fraction ("Unc EF"). For RUS/ROS/SMOTE, the default threshold of 0.5 was used.

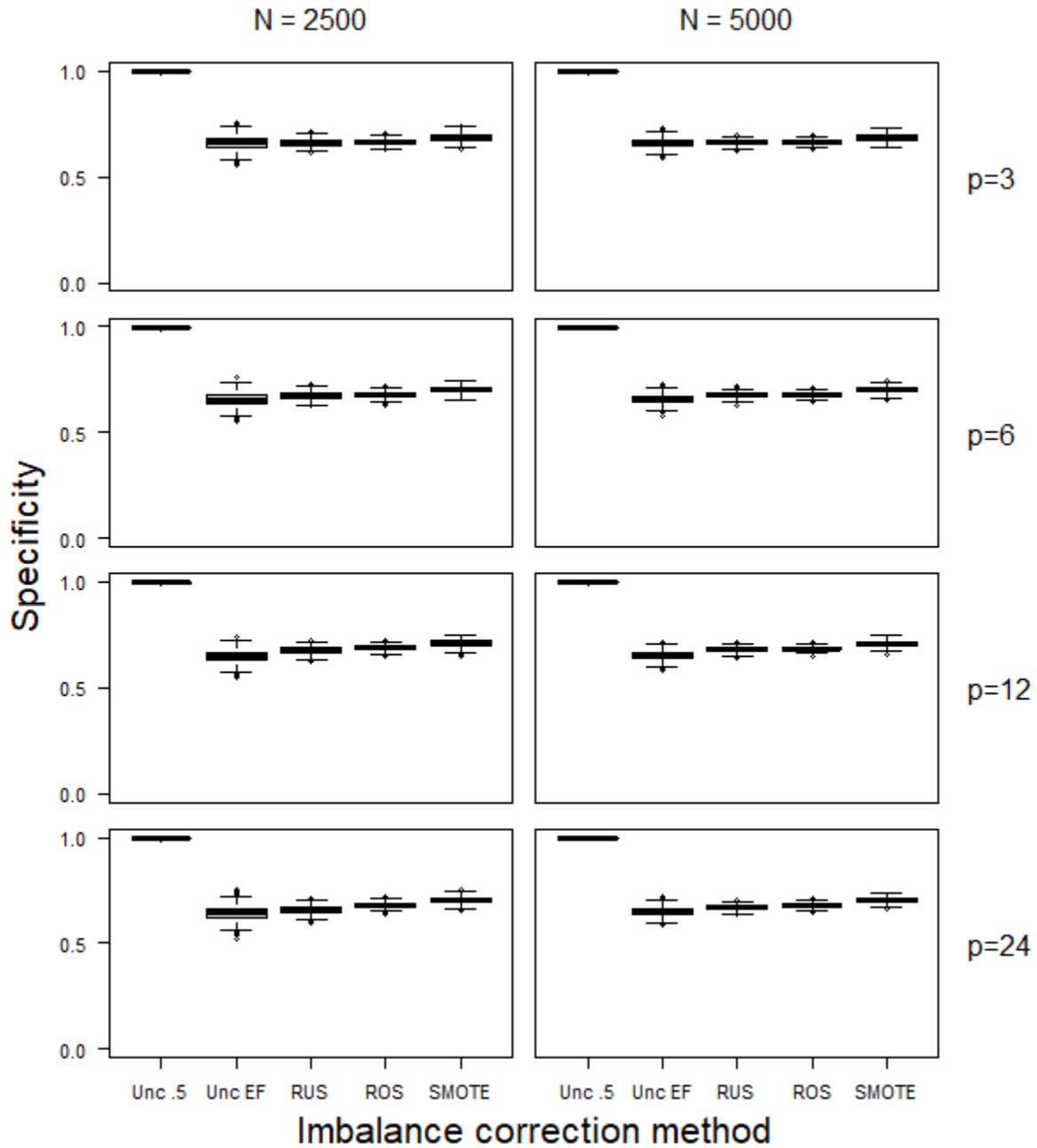

Figure S37. Test set specificity for the Ridge models in the simulation scenarios with an event fraction of 30%. For uncorrected training sets, we used either the default threshold of 0.5 ("Unc .5") or a threshold based on the true event fraction ("Unc EF"). For RUS/ROS/SMOTE, the default threshold of 0.5 was used.

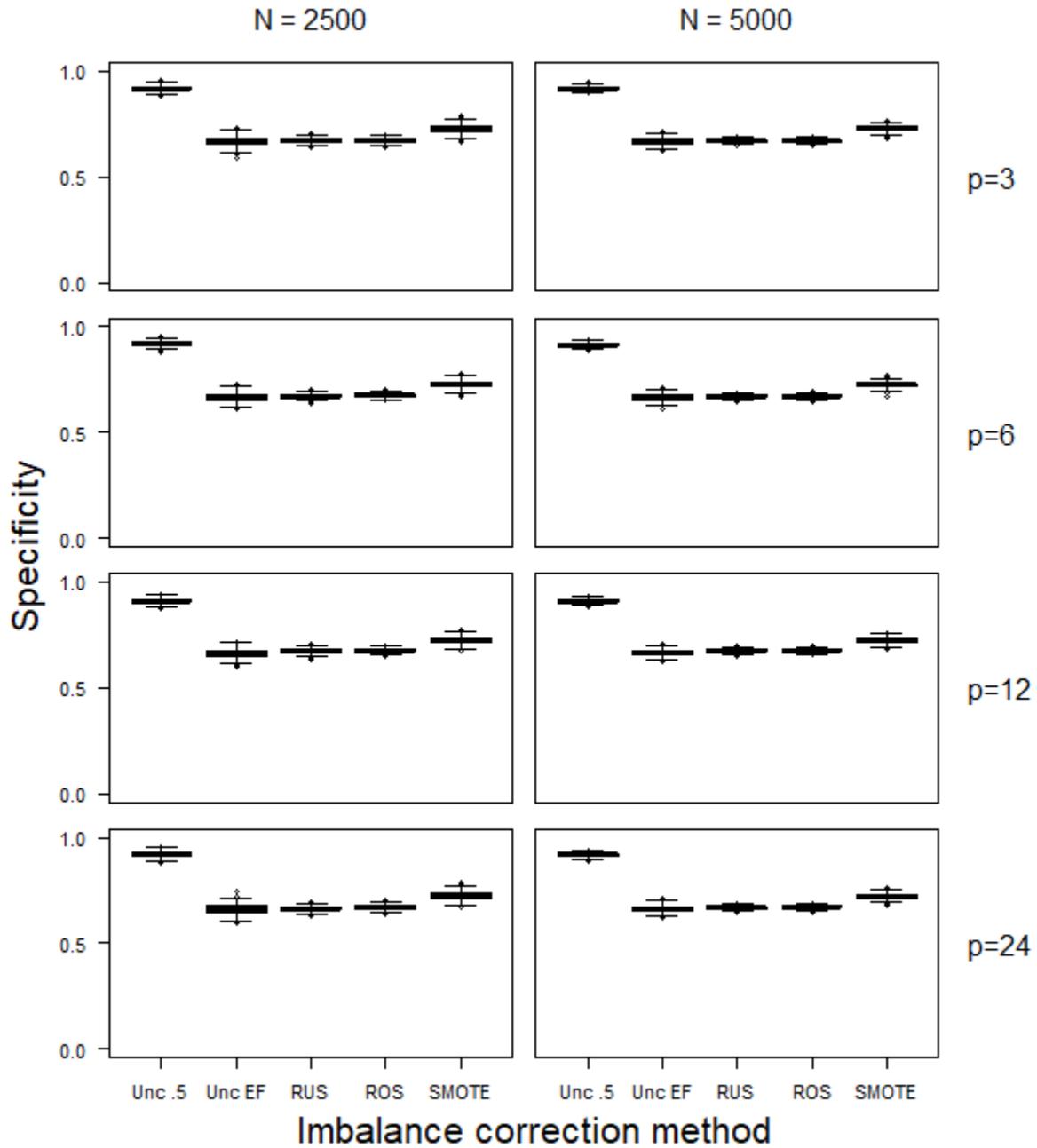

Figure S38. Test set specificity for the SLR models in the simulation scenarios with an event fraction of 1%. For uncorrected training sets, we used either the default threshold of 0.5 ("Unc .5") or a threshold based on the true event fraction ("Unc EF"). For RUS/ROS/SMOTE, the default threshold of 0.5 was used.

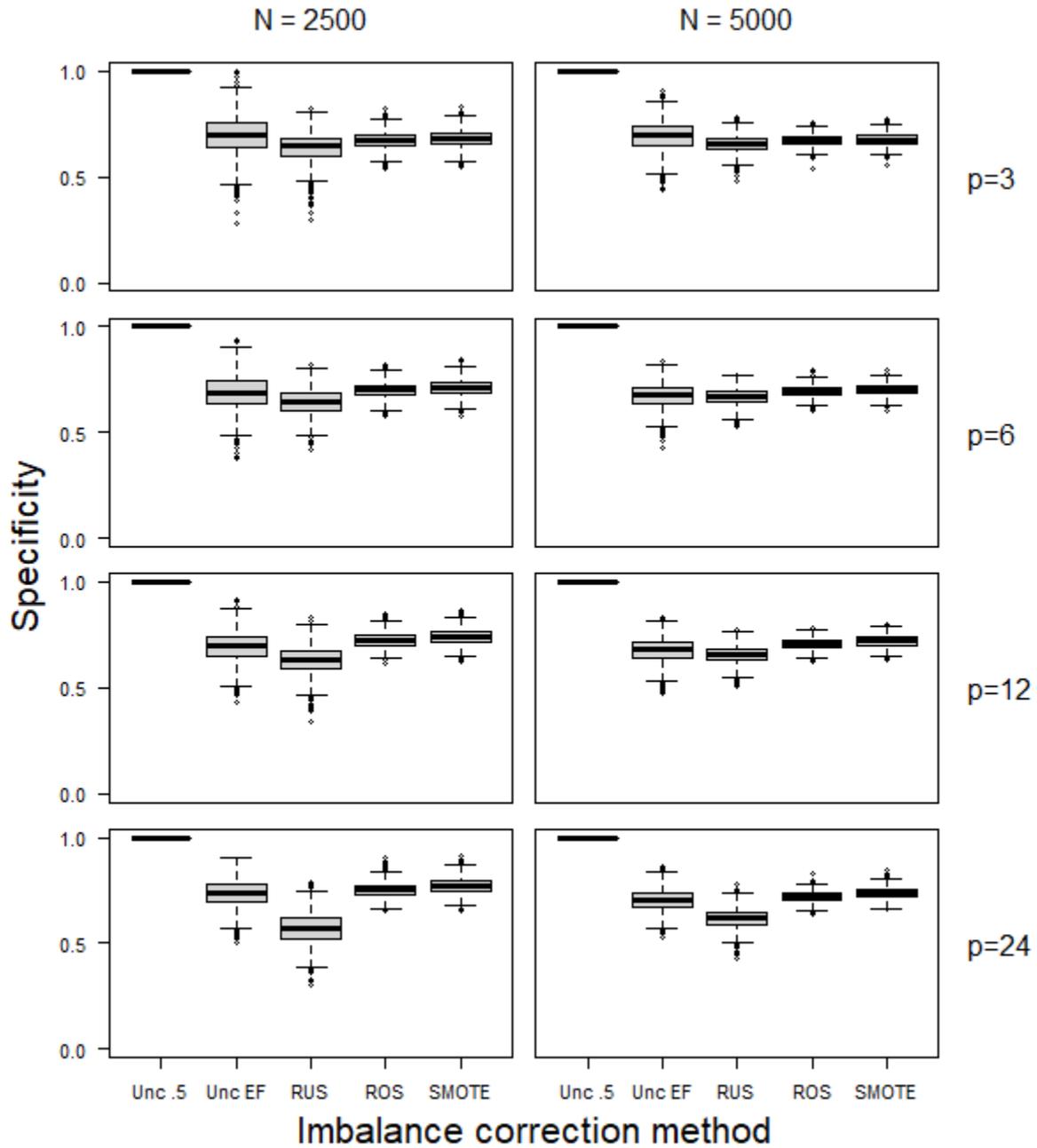

Figure S39. Test set specificity for the SLR models in the simulation scenarios with an event fraction of 10%. For uncorrected training sets, we used either the default threshold of 0.5 ("Unc .5") or a threshold based on the true event fraction ("Unc EF"). For RUS/ROS/SMOTE, the default threshold of 0.5 was used.

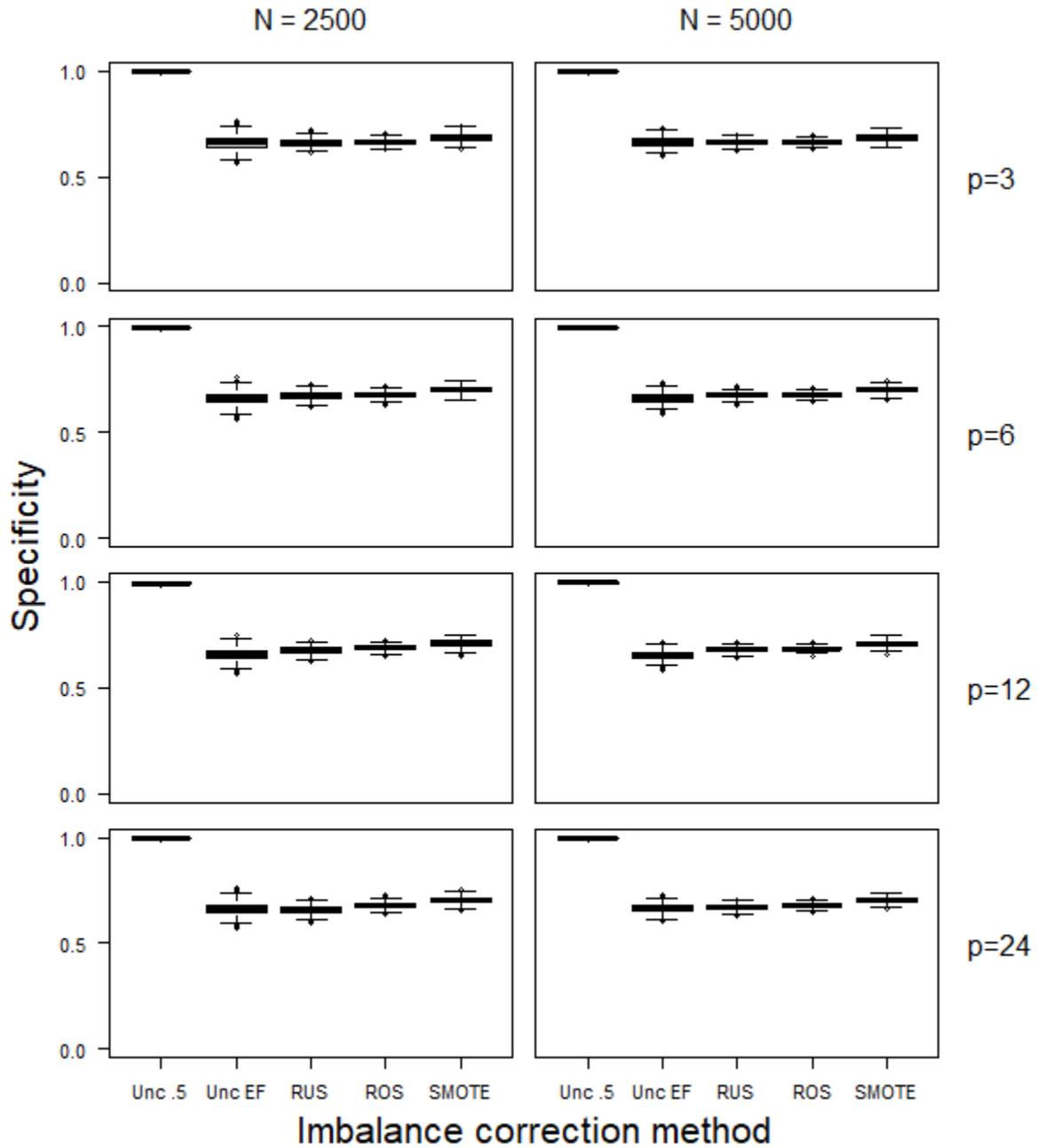

Figure S40. Test set specificity for the SLR models in the simulation scenarios with an event fraction of 30%. For uncorrected training sets, we used either the default threshold of 0.5 ("Unc .5") or a threshold based on the true event fraction ("Unc EF"). For RUS/ROS/SMOTE, the default threshold of 0.5 was used.

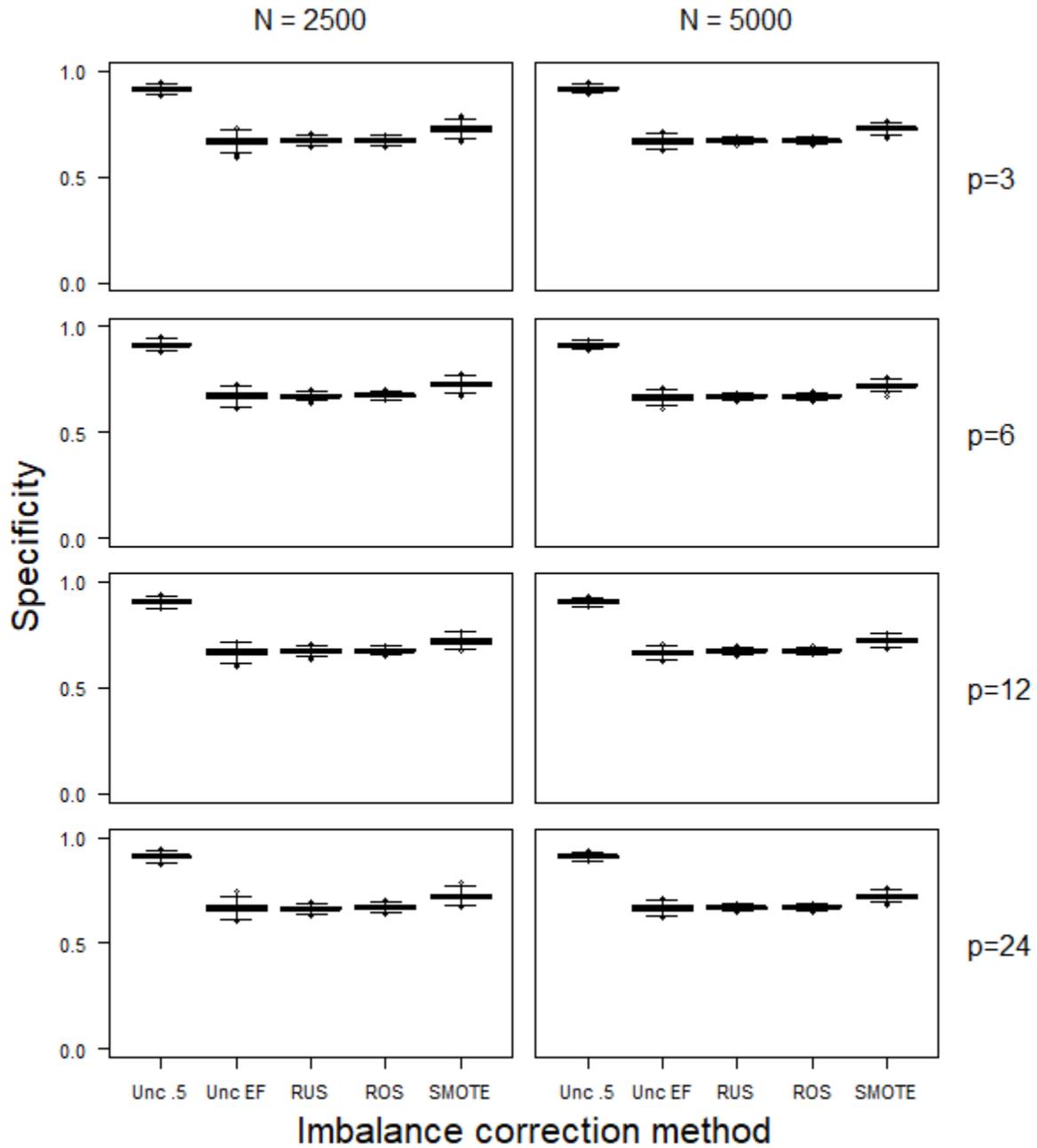